\documentclass{aastex631}

\usepackage{graphicx,amssymb,amsmath,times,xcolor}
\usepackage{longtable}
\usepackage{bm} 
\usepackage{url}

\usepackage[varg]{txfonts}
\usepackage[T1]{fontenc}
\usepackage{natbib}
\bibpunct{(}{)}{;}{a}{}{,} 
\usepackage[normalem]{ulem}

\def\simlt{\lower.5ex\hbox{\ltsima}}
\def\simgt{\lower.5ex\hbox{\gtsima}}

\def\g{{\rm\,g}}

\def \r {r$^{1/4}$ }

\def\AA{$\; \buildrel \circ \over {\rm A}$}

\def\tmaxo{$T_{max}^{opt}$} 
\def\tmeano{$T_{mean}^{opt}$} 
\def\tminv{$T_{min}^{RV}$} 
\def\tmeanv{$T_{mean}^{RV}$} 
\def\tminvfe{$T_{min}^{RV(Fe)}$} 
\def\tmeanvfe{$T_{mean}^{RV(Fe)}$} 
\def\tminvhb{$T_{min}^{RV(H\beta)}$} 
\def\tmeanvhb{$T_{mean}^{RV(H\beta)}$} 

\def\gtsim{\;\lower.6ex\hbox{$\sim$}\kern-6.7pt\raise.4ex\hbox{$>$}\;}
\def\ltsim{\;\lower.6ex\hbox{$\sim$}\kern-6.9pt\raise.4ex\hbox{$<$}\;}

\def\wcen{$\omega$ Cen~}

\shorttitle{Optical and radial velocity curve templates for RR Lyrae}
\shortauthors{Braga et al.}

\begin{document}

\title{On the Use of Field RR Lyrae as Galactic Probes. V. Optical and radial velocity curve templates}

\author[0000-0001-7511-2830]{V.~F.~Braga }
\affiliation{INAF-Osservatorio Astronomico di Roma, via Frascati 33, 00040 Monte Porzio Catone, Italy}
\affiliation{Space Science Data Center, via del Politecnico snc, 00133 Roma, Italy}
\author[0000-0001-8926-3496]{J.~Crestani}
\affiliation{INAF-Osservatorio Astronomico di Roma, via Frascati 33, 00040 Monte Porzio Catone, Italy}
\affiliation{Departamento de Astronomia, Universidade Federal do Rio Grande do Sul, Av. Bento Gon\c{c}alves 6500, Porto Alegre 91501-970, Brazil}
\affiliation{Dipartimento di Fisica, Universit\`a di Roma Tor Vergata, via della Ricerca Scientifica 1, 00133 Roma, Italy}
\author[0000-0001-5829-111X]{M.~Fabrizio }
\affiliation{INAF-Osservatorio Astronomico di Roma, via Frascati 33, 00040 Monte Porzio Catone, Italy}
\affiliation{Space Science Data Center, via del Politecnico snc, 00133 Roma, Italy}
\author[0000-0002-4896-8841]{G.~Bono }
\affiliation{INAF-Osservatorio Astronomico di Roma, via Frascati 33, 00040 Monte Porzio Catone, Italy}
\affiliation{Dipartimento di Fisica, Universit\`a di Roma Tor Vergata, via della Ricerca Scientifica 1, 00133 Roma, Italy}
\author[0000-0002-3456-5929]{C.~Sneden }
\affiliation{Department of Astronomy and McDonald Observatory, The University of Texas, Austin, TX 78712, USA}
\author{G.~W.~Preston}
\affiliation{The Observatories of the Carnegie Institution for Science, 813 Santa Barbara St., Pasadena, CA 91101, USA}
\author[0000-0002-8627-6096]{J.~Storm }
\affiliation{Leibniz-Institut f\"ur Astrophysik Potsdam, An der Sternwarte 16, 14482, Potsdam, Germany}
\author[0000-0001-6604-0505]{S.~Kamann}
\affiliation{Astrophysics Research Institute, Liverpool John Moores University, IC2, Liverpool Science Park, 146 Brownlow Hill, Liverpool,L3 5RF, UK}
\author[0000-0002-7547-6180]{M.~Latour}
\affiliation{Institute for Astrophysics, Georg-August-University G\:ottingen, Friedrich-Hund-Platz 1, D-37077 G\:ottingen, Germany}
\author{H.~Lala}
\affiliation{Astronomisches Rechen-Institut, Zentrum f\"ur Astronomie der Universit\"at Heidelberg, M\"onchhofstr. 12-14, D-69120 Heidelberg, Germany}
\author{B.~Lemasle}
\affiliation{Astronomisches Rechen-Institut, Zentrum f\"ur Astronomie der Universit\"at Heidelberg, M\"onchhofstr. 12-14, D-69120 Heidelberg, Germany}
\author[0000-0001-5497-5805]{Z.~Prudil}
\affiliation{Universit{\'e} de Nice Sophia-antipolis, CNRS, Observatoire de la C\^{o}te d'Azur, Laboratoire Lagrange, BP 4229, F-06304 Nice, France}
\author[0000-0002-9934-1352]{G.~Altavilla}
\affiliation{INAF-Osservatorio Astronomico di Roma, via Frascati 33, 00040 Monte Porzio Catone, Italy}
\affiliation{Space Science Data Center, via del Politecnico snc, 00133 Roma, Italy}
\author[0000-0003-3096-4161]{B.~Chaboyer }
\affiliation{Department of Physics and Astronomy, Dartmouth College, Hanover, NH 03755, USA}
\author[0000-0001-8209-0449]{M.~Dall'Ora }
\affiliation{INAF-Osservatorio Astronomico di Capodimonte, Salita Moiariello 16, 80131 Napoli, Italy}
\author[0000-0001-8514-7957]{I.~Ferraro }
\affiliation{INAF-Osservatorio Astronomico di Roma, via Frascati 33, 00040 Monte Porzio Catone, Italy}
\author[0000-0003-4510-0964]{C.~K.~Gilligan }
\affiliation{Department of Physics and Astronomy, Dartmouth College, Hanover, NH 03755, USA}
\author[0000-0003-0376-6928]{G.~Fiorentino }
\affiliation{INAF-Osservatorio Astronomico di Roma, via Frascati 33, 00040 Monte Porzio Catone, Italy}
\author[0000-0001-9816-5484]{G.~Iannicola }
\affiliation{INAF-Osservatorio Astronomico di Roma, via Frascati 33, 00040 Monte Porzio Catone, Italy}
\author[0000-0002-0271-2664]{L.~Inno }
\affiliation{Universit\`a degli Studi di Napoli ``Parthenope'', Via Amm. F. Acton, 38, 80133 Napoli, Italy}
\author{S.~Kwak}
\affiliation{Dipartimento di Fisica, Universit\`a di Roma Tor Vergata, via della Ricerca Scientifica 1, 00133 Roma, Italy}
\author[0000-0001-9910-9230]{M.~Marengo }
\affiliation{Department of Physics and Astronomy, Iowa State University, Ames, IA 50011, USA}
\author[0000-0001-7990-6849]{S.~Marinoni }
\affiliation{INAF-Osservatorio Astronomico di Roma, via Frascati 33, 00040 Monte Porzio Catone, Italy}
\affiliation{Space Science Data Center, via del Politecnico snc, 00133 Roma, Italy}
\author[0000-0002-8162-3810]{P.~M.~Marrese }
\affiliation{INAF-Osservatorio Astronomico di Roma, via Frascati 33, 00040 Monte Porzio Catone, Italy}
\affiliation{Space Science Data Center, via del Politecnico snc, 00133 Roma, Italy}
\author[0000-0002-9144-7726]{C.~E.~Mart{\'i}nez-V{\'a}zquez}
\affiliation{Cerro Tololo Inter-American Observatory, NSF's National Optical-Infrared Astronomy Research Laboratory, Casilla 603, La Serena, Chile}
\author[0000-0001-5292-6380]{M.~Monelli }
\affiliation{Instituto de Astrof\'isica de Canarias, Calle Via Lactea s/n, E38205 La Laguna, Tenerife, Spain}
\author{J.~P.~Mullen}
\affiliation{Department of Physics and Astronomy, Iowa State University, Ames, IA 50011, USA}
\author{N.~Matsunaga }
\affiliation{Department of Astronomy, The University of Tokyo, 7-3-1 Hongo, Bunkyo-ku, Tokyo 113-0033, Japan}
\author[0000-0002-8894-836X]{J.~Neeley }
\affiliation{Department of Physics, Florida Atlantic University, 777 Glades Rd, Boca Raton, FL 33431 USA}
\author[0000-0001-6074-6830]{P.~B.~Stetson }
\affiliation{Herzberg Astronomy and Astrophysics, National Research Council, 5071 West Saanich Road, Victoria, British Columbia V9E 2E7, Canada}
\author[0000-0002-6092-7145]{E.~Valenti }
\affiliation{European Southern Observatory, Karl-Schwarzschild-Str. 2, 85748 Garching bei Munchen, Germany}
\author[0000-0002-5829-2267]{M.~Zoccali }
\affiliation{Instituto Milenio de Astrof{\'i}sica, Santiago, Chile}
\affiliation{Pontificia Universidad Catolica de Chile, Instituto de Astrofisica, Av. Vicu\~na Mackenna 4860, Santiago, Chile}

\begin{abstract}
We collected the largest spectroscopic catalog of RR Lyrae (RRLs) 
including $\approx$20,000 high-, medium- and low-resolution spectra 
for $\approx$10,000 RRLs. We provide the analytical 
forms of radial velocity curve (RVC) templates. 
These were built using 36 RRLs (31 fundamental---split into three 
period bins---and 5 first overtone pulsators) 
with well-sampled RVCs based on three groups 
of metallic lines (Fe, Mg, Na) and four Balmer lines
(H$_\alpha$, H$_\beta$, H$_\gamma$, H$_\delta$).
We tackled the long-standing problem of the reference epoch to anchor 
light curve and RVC templates. For the $V$-band, we found that the 
residuals of the templates anchored to the phase of the mean magnitude 
along the rising branch are $\sim$35\% to $\sim$45\% smaller than those anchored 
to the phase of maximum light. For the RVC, we used two independent 
reference epochs for metallic and Balmer lines and we verified that 
the residuals of the RVC templates anchored to the phase of mean RV 
are from 30\% (metallic lines) up to 45\% (Balmer lines) smaller than 
those anchored to the phase of minimum RV.
We validated our RVC templates by using both the single- and the 
three-phase points approach. We found that barycentric velocities 
based on our RVC templates are two-three times more accurate than 
those available in the literature.
We applied the current RVC templates to Balmer lines 
RVs of RRLs in the globular NGC~3201 collected with 
MUSE at VLT. We found the cluster barycentric RV of 
$V_{\gamma}$=496.89$\pm$8.37(error)$\pm$3.43 (standard deviation) 
km/s, which agrees well with literature estimates.
\end{abstract}


\section{Introduction} \label{chapt_intro}

Pulsating variables are behind numerous breakthroughs in 
astrophysics. Classical Cepheids (CCs) were used 
to estimate the distance to M31 and solve the Great 
Debate concerning the extragalactic nature of the 
so-called Nebulae \citep{hubble1926} and to trace, for the 
first time, the rotation of the Galactic thin disc 
\citep{oort1927,joy1939}. The size 
and the age of the Universe were revolutionized thanks 
to the discovery of the difference between 
CCs, RRLs, and type II Cepheids \citep[TIICs,][]{baade56}. Indeed, while they 
were previously thought to represent the same type of 
variable stars, it became clear that they represented 
very distinct populations, with the RRLs and TIICs being 
very old (t$\ge$10 Gyr), and the CCs very young (t$\le$300 Myr). 

Nowadays, CCs are among the most popular calibrators of
the extragalactic distance scale \citep{riess2019}. 
RRLs, albeit fainter, are excellent standard candles that
can provide robust, independent distance measurements even 
for stellar populations where the young CCs are absent.
RRLs obey the well-defined Period-Luminosity-Metallicity (PLZ)
relations for wavelengths longer than the $R$-band 
\citep{bono03c,catelan09}. As tracers of purely old stellar 
populations, they can be used to investigate the early formation 
and evolution of both the Galactic Halo 
\citep{fiorentino15a,belokurov2018,fabrizio2019}
and Bulge \citep{pietrukowicz2015,braga2018b}.

It is noteworthy that we lack general consensus on the Galactic Halo
structure, in part because different stellar tracers provide different views
concerning its spatial structure and the timescale for its formation. 
Indeed, \citet{carollo07} by using Main Sequence, subgiants and RGs and 
\citet{kinman12} by using RRLs, suggested that the outer halo is
more spherical and its density profile is shallower when compared 
with the inner halo. In contrast, \citet{keller2008}
by using RRLs and \citet{sesar2010,sesar2011} by using RRLs plus
Main Sequence stars suggested that the outer halo has a steeper
density profile when compared with the inner halo. \citet{deason2011} 
by using Blue Horizontal Branch stars and
Blue Stragglers found no change in the flattening as a function
of the Galactocentric distance \citep{sesar2011}. More recently,
\citet{xue2015}, by adopting a global ellipsoidal stellar density
model with Einasto profile found that the models with constant
flattening provide a good fit to the entire Halo.

The tension between different measurements may be due to the sample
selection of each study. On the one hand, the ages of the RRLs cover 
a narrow range from $\sim$10 to $\sim$13 Gyrs. There is evidence that a few 
RRLs---or stars that mimic RRLs, see \citet{smolec2013}---are the aftermath of binary evolution, but
they only represent a few percent of the populations 
\citep{bono97b,pietrzynski2012,kervella2019}. On the other hand,
Red giants (RGs) and main sequence (MS) stars, typically used to investigate 
the Halo, have only very weak age constraints \citep{conroy2021}. Indeed, all stellar 
structures less massive than 2M$_\odot$ (older than $\sim$0.5-1.0 Gyr) experience 
a RG phase and MS stars also cover a broad range in stellar
masses/ages. This means that if the Halo is the result of an in-
tense disruption and merging activity \citep{monachesi2019}
RG and MS stars are far from being optimal tracers of the early
formation, because they are a mixed bag concerning the age dis-
tribution. 

Field RRLs are less numerous when compared
with RG and MS stars, but their narrow age distribution makes them
uniquely suited for Galactic archaeology. They probe a significant Halo fraction
(Galactocentric distance $\le$150 kpc) with high accuracy.
Their individual distances have uncertainties on average 
smaller than 3-5\% and their accuracy improves
when moving from optical to NIR \citep{longmore86,catelan04,braga15}. 
This is a key advantage even in the Gaia era: Gaia EDR3 has an accuracy of 3\%
for Halo RRLs (G$\le$15 mag) at 1 kpc and this accuracy will be
extended to 2 kpc at the end of the mission \citep{clementini2019}.
RRLs are also valuable targets from the kinematical point of
view. In fact, by measuring their velocities, one gets 
information on the kinematical state of the old population
(Halo, Globular Clusters, Bulge). The pioneering work by 
\citet{layden94,layden1995}, based on 302 RRLs, pointed towards a 
non-steady formation of the Halo, favouring a fragmented accretion scenario
\citep{searle78}. More recently, \citet{zinn2020} were
able to pinpoint the membership of several Halo RRLs to past merger events 
\citep[Gaia-Enceladus and the Helmi streams,][]{helmi1999,myeong2018,helmi2018}. 
A few Halo RRLs were also associated with the Orphan stream by \citet{prudil2021}, 
leading to more solid constraints on the origin of the
stream itself. Concerning the Bulge, the kinematic properties of
RRLs display a duality, with one group of stars associated with the
spheroidal component and the other with the Galactic bar \citep{kunder2020}.

The number of identified RRL is rapidly growing thanks to the 
enhancements in telescope collecting areas and instrument 
efficiency. Thanks to long-term optical (Catalina, \citealt{drake2009,drake2017}; 
ASAS, \citealt{pojmanski1997}; ASAS-SN, \citealt{jayasinghe2019}; 
 DES, \citealt{stringer2019}; \textit{Gaia}, \citealt{clementini2019}; OGLE, 
 \citealt{soszynski2019}; Pan-STARRS \citealt{sesar2017b})
near-infrared \citep[VVV, VVV-X,][]{minniti2011} and mid-infrared
\citep[neo-WISE,][]{wright2010} surveys, more
than 200,000 RRLs were identified in the Galactic spheroid. 
However, RRLs are demanding targets from an observational
point of view. Well-sampled time series, meaning at least a dozen, properly sampled, photometric measurements, are
required for a solid identification and an accurate characterization. 
The same limitation applies to the measurement of the
RRL barycentric radial velocity (V$_\gamma$), because it requires 
multiple measurements to trace the radial velocity (RV) variation along the
pulsation cycle. To overcome this limitation, several authors have used 
the radial velocity curve (RVC) of X Ari, observed more than half a 
century ago by \citet{oke1966}, as 
a pseudo-template. More recently, RVC
templates have been developed for fundamental (RRab) RRLs 
\citep[][henceforth, S12]{liu1991,sesar2012}. They allow to estimate V$_\gamma$
even with a small number of velocity measurements, provided
that the $V$-band pulsation properties are known. The 
current RVC templates are affected by several limitations: despite 
being based on 22 RRab stars with periods between 0.37 and 0.71 days,
the template of \citet{liu1991} was derived from RVCs with---at
most---a few tens of points each. These points are
velocities obtained from a heterogeneous set of unidentified
metallic lines, since they were collected from several different
papers. S12 provided templates for both metallic and 
H$_\alpha$, H$_\beta$ and H$_\gamma$ lines with a few hundreds of RV
measurements. However, their Balmer templates do not cover the 
steep decreasing branch and, even more
importantly, the templates were based on only six RRab with
periods in a very narrow range (0.56-0.59 days). Finally, 
no RVC templates are available for first-overtone RRLs (RRc).

This work aims at providing new RVC templates
for both RRab and RRc variables by addressing all the limitations
described above. We adopted a wide set of specific and well-identified 
metallic and Balmer lines for both RRab and RRc stars and hundreds of 
velocity measurement for each template. As the velocity curves of the
RRab display some peculiar variations among themselves, we also separated
them into three bins according to their specific shape and pulsation
period. Thus, we can provide uniquely precise templates that
cover a wide range of intrinsic parameters of these variable stars.

The paper is structured as follows. In section 2 we investigate the phasing of the 
optical light curve and discuss on a quantitative basis the difference between the 
reference epoch anchored to the luminosity maximum and to the mean magnitude along the 
rising branch. We present the spectroscopic dataset in Section~\ref{sect:data} 
and provide new RVCs and their properties in Section~\ref{sect:rvcurves}. 
We put together the RVCs and derive the analytical form of the RVC templates in
Section~\ref{chapt_template}, discuss the reference epoch to be used to apply
the templates in Section~\ref{sect:deltaphi_opt_rv}
and validate them in Section~\ref{chapt_validation}.
We provide a practical example of how to use the RVC template on
spectroscopic observations of NGC 3201 in Section~\ref{chapt_3201}.
Finally, in Section~\ref{chapt_discussion}, we summarize the current results 
and outline future developments of this project.

\section{Optical light curve templates}\label{sect:template_vband}

Light curve templates are powerful tools that model the light curve
of a periodic variable star. The templates are parametrized with the
properties of the variable stars (pulsation mode, period and amplitude). 
These come in hand, e.g., to estimate the pulsation
properties with a few available data \citep{stringer2019}, to obtain $O-C$ diagrams
that trace the rate of period change \citep{hajdu2021}, to predict the luminosity
of the star at a given phase, and for various other purposes.

The use of both luminosity and RVC templates relies on 
the use of a reference epoch. This means that the phase zero of the RVC template has 
to be anchored to a specific feature of the luminosity/RV curve. 
The most common reference epoch adopted in the field of pulsating variable 
stars is the time of maximum light in the optical (\tmaxo).
The RVC templates available in the 
literature are also anchored to \tmaxo~ because it matches, within the 
uncertainties, the time of minimum in the RVC \tminv. Note that, by ``minimum''
in the RVC, we mean the numerical minimum, i.e., the epoch of maximum blueshift.
This is an approximate choice due 
to the well-known phase-lag between light and RVC \citep{castor1971}. 

Our group introduced a new reference epoch, namely, the epoch at which the magnitude 
along the rising branch of the $V$-band light curve---that is, the section 
of the light curve where brightness changes from minimum to maximum---becomes equal to the mean 
$V$-band magnitude \citep[\tmeano,][]{inno15,braga2019}. We thoroughly discussed 
the advantages of using \tmeano~ versus \tmaxo~ in the context of NIR light curve 
templates for both CCs and RRLs. The reader interested in a detailed 
discussion is referred to the quoted papers. Here, we summarize the 
key advantages in adopting \tmeano~ for RRL variables. 
{\it i)} RRab variables with large amplitudes have RVCs with a 
``sawtooth'' shape, where the maximum can be misidentified
by an automatic analytical fit if the phase coverage is not 
optimal. The rising branch, however, can be more easily fitted.
{\it ii)} A significant fraction of RRc variables displays a 
well-defined bump/dip before the maximum in luminosity. A clear 
separation between the two maxima is not trivial if the phase 
coverage is not optimal.
{\it iii)} The estimate of \tmaxo~ is more prone to possible 
systematics, even with well-sampled light curves, because several 
RRc and long-period RRab variables display flat-topped light curves 
i.e. light curves in which the maximum is almost flat for a 
relatively broad fraction of the phase cycle ($\sim$0.10).
{\it iv)} \tmaxo~ is typically estimated either as the top value 
of the fit of the light cure or the brightest observed point, 
when the sampling is optimal (e.g., ASAS-SN). This means that 
\tmaxo~ is affected by the intrinsic dispersion of the observations 
and by the time resolution of photometric data. Meanwhile, \tmeano~ 
is estimated by interpolating the analytical fit the mean magnitude
(see Appendix~\ref{chapt_howto_tmeano}),
which is a very robust property of the star. Therefore, \tmeano~ 
is intrinsically more robust because its precision is less dependent
of sampling.

In the following, we address on a more quantitative basis these key 
issues in the context of optical light curves. For this purpose, 
we take advantage of a homogeneous and complete sample of 
$V$-band light curves for cluster and field RRL variables. In particular, 
we use visual light curves for RRLs in M4 \citep{stetson14a} and in 
\wcen \citep{braga16} together with literature observations for 
RRLs with Baade-Wesselink (BW) analysis, 
\citep[][and references therein]{braga2019}. The
RRLs in M4 and in $\omega$~Cen have well-sampled light curves, 
with the number of phase points ranging from hundreds to more than one thousand.

\subsection{Phasing of optical light curves}\label{sect:t0tmax}

\begin{figure*}[!htbp]
\centering
\includegraphics[width=10cm]{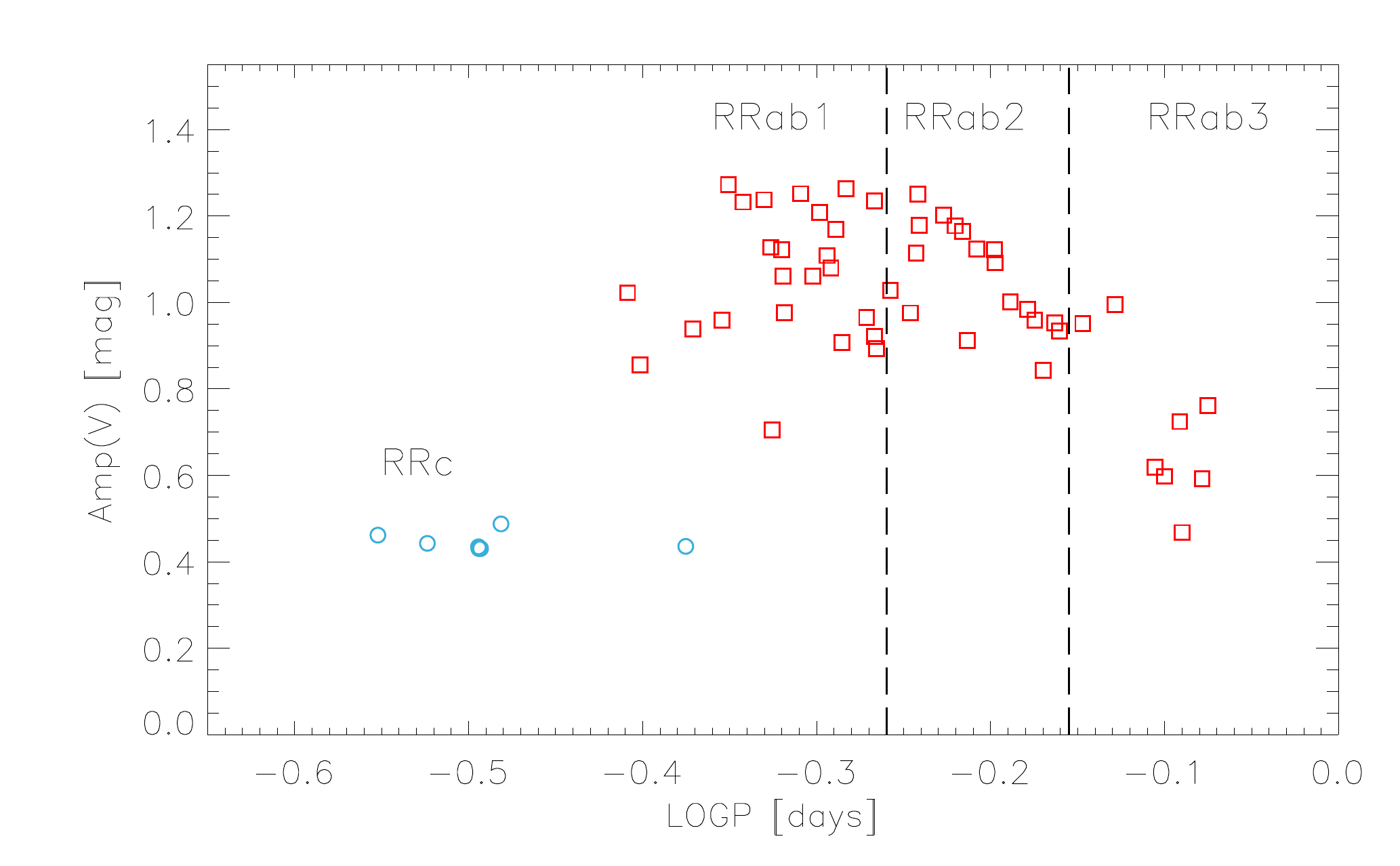}
\caption{Bailey diagram ($V$-band amplitude versus logarithmic period)
for the RRLs adopted to build the $V$-band light curve templates. The RRc 
variables are marked with blue circles and the RRab variables with red squares. 
The template bins are labeled and the period thresholds are displayed as 
vertical dashed lines.}
\label{fig:bailey_vband}
\end{figure*}

We selected 57 RRLs (7 RRc, 50 RRab) from our M4, \wcen and BW RRLs and separated them into 
four period bins (See Fig.~\ref{fig:bailey_vband}). Following \citet{braga2019}
the thresholds are the following: 
``{\it RRc}'', ``{\it RRab1}'' (RRab with periods shorter than 0.55 days), 
``{\it RRab2}'' (RRab with periods between 0.55 and 0.70 days) and
``{\it RRab3}'' (RRab with periods longer than 0.70 days).
See Section~\ref{chapt_bins} for a more detailed discussion.

We normalized all the light curves by subtracting
the mean magnitude and dividing them by their 
peak-to-peak amplitude $A(V)$, and estimated
\tmeano~ and \tmaxo~ for the entire sample and the individual values are listed 
in columns 5 and 6 of Table~\ref{tab:tmeantmax}. \tmeano~ was estimated as 
described in Appendix~\ref{chapt_howto_tmeano} and \tmaxo~ was estimated by 
converting the phase of maximum light of the model light curve, into an Heliocentric
Julian Date.

\begin{deluxetable*}{l c c c c c}
\tablenum{1}
\tabletypesize{\scriptsize}
\tablecaption{Photometric properties of the RRLs adopted to evaluate the optical 
light curve template. From left to right, the columns give the ID of the variable, 
the pulsation period, the mean visual magnitude, the $V$-band amplitude and the 
two reference epochs.}
 \tablehead{\colhead{Name} & \colhead{Period} & \colhead{$<V>$} & \colhead{$Amp(V)$} & \colhead{\tmeano} & \colhead{\tmaxo}\\
\label{tab:tmeantmax}
 & (days) & \multicolumn{2}{c}{(mag)} & \multicolumn{2}{c}{HJD-2,400,000 (days)}}
\startdata
\multicolumn{6}{c}{---RRc---}\\
\wcen~V16        & 0.3301961 & 14.558 &  0.487 & 51766.7639 & 51766.5041\\
\wcen~V19        & 0.2995517 & 14.829 &  0.442 & 49869.6627 & 55715.4767\\
\wcen~V98        & 0.2805656 & 14.773 &  0.461 & 55715.6665 & 51693.5392\\
\wcen~V117       & 0.4216425 & 14.444 &  0.435 & 51277.1910 & 50985.5294\\
\wcen~V264       & 0.3213933 & 14.703 &  0.430 & 54705.4664 & 52743.7623\\
M4~V6            & 0.3205151 & 13.454 &  0.434 & 55412.8765 & 49469.6457\\
M4~V43           & 0.3206600 & 13.082 &  0.430 & 43724.9882 & 43681.1370\\
\multicolumn{6}{c}{---RRab1---}\\
\wcen~V8         & 0.5213259 & 14.671 &  1.263 & 49824.5018 & 52443.1571\\
\wcen~V23        & 0.5108703 & 14.821 &  1.079 & 49866.6429 & 54705.6430\\
\wcen~V59        & 0.5185514 & 14.674 &  0.907 & 51277.1348 & 51370.5255\\
\wcen~V74        & 0.5032142 & 14.620 &  1.208 & 55711.7447 & 55715.3000\\
\wcen~V107       & 0.5141038 & 14.753 &  1.169 & 49860.6035 & 53865.5023\\
M4~V2            & 0.5356819 & 13.411 &  0.965 & 55412.1842 & 52087.7853\\
M4~V7            & 0.4987872 & 13.415 &  1.061 & 55412.2209 & 50601.4511\\
M4~V8            & 0.5082236 & 13.323 &  1.108 & 55412.5025 & 50601.6924\\
M4~V10           & 0.4907175 & 13.327 &  1.251 & 55412.2533 & 50601.2890\\
M4~V12           & 0.4461098 & 13.578 &  1.272 & 55412.7833 & 50601.5210\\
M4~V16           & 0.5425483 & 13.344 &  0.893 & 55412.2979 & 50601.5688\\
M4~V18           & 0.4787920 & 13.358 &  1.121 & 55412.8648 & 50601.5170\\
M4~V19           & 0.4678111 & 13.376 &  1.237 & 55412.3131 & 50601.3741\\
M4~V21           & 0.4720074 & 13.190 &  1.127 & 55412.6133 & 50601.4735\\
M4~V26           & 0.5412174 & 13.247 &  1.235 & 55412.4631 & 50552.3694\\
M4~V36           & 0.5413092 & 13.424 &  0.921 & 55412.8657 & 52088.7312\\
M4~C303          & 0.4548026 & 16.037 &  1.232 & 55412.5626 & 50601.6896\\
AR Per           & 0.4255489 & 10.452 &  0.938 & 47123.6655 & 46773.4731\\
AV Peg           & 0.3903912 & 10.452 &  0.938 & 47123.7076 & 47116.3202\\
BB Pup           & 0.4805437 & 10.492 &  1.022 & 47193.3909 & 47193.4293\\
DX Del           & 0.4726174 & 10.492 &  1.022 & 43689.8611 & 30950.5060\\
SW And           & 0.4422660 & 12.164 &  0.976 & 47065.7327 & 47116.1847\\
V445 Oph         & 0.3970227 & 12.164 &  0.976 & 46981.3385 & 46868.6233\\
V Ind            & 0.4796012 &  9.937 &  0.704 & 47815.0317 & 47812.6680\\
\multicolumn{6}{c}{---RRab2---}\\
\wcen~V13        & 0.6690484 & 14.471 &  0.959 & 51316.5671 & 51314.6124\\
\wcen~V33        & 0.6023333 & 14.538 &  1.177 & 51285.7634 & 52446.5015\\
\wcen~V40        & 0.6340978 & 14.511 &  1.121 & 49863.7202 & 54705.7382\\
\wcen~V41        & 0.6629338 & 14.505 &  0.983 & 52743.9786 & 52447.0363\\
\wcen~V44        & 0.5675378 & 14.709 &  0.975 & 50971.6089 & 50971.6529\\
\wcen~V46        & 0.6869624 & 14.501 &  0.952 & 49821.6201 & 55715.8201\\
\wcen~V51        & 0.5741424 & 14.511 &  1.178 & 51276.8553 & 50984.6520\\
\wcen~V62        & 0.6197964 & 14.423 &  1.123 & 50984.4926 & 53860.3888\\
\wcen~V79        & 0.6082869 & 14.596 &  1.164 & 49922.5029 & 50165.8572\\
\wcen~V86        & 0.6478414 & 14.509 &  1.001 & 50978.5945 & 52743.3654\\
\wcen~V100       & 0.5527477 & 14.638 &  1.028 & 50975.6290 & 50975.6676\\
\wcen~V102       & 0.6913961 & 14.519 &  0.933 & 50975.5249 & 53864.9282\\
\wcen~V113       & 0.5733764 & 14.596 &  1.250 & 50978.5866 & 52743.4734\\
\wcen~V122       & 0.6349212 & 14.520 &  1.091 & 54705.4856 & 53870.6116\\
\wcen~V125       & 0.5928780 & 14.587 &  1.202 & 49116.6901 & 51600.8905\\
\wcen~V139       & 0.6768713 & 14.324 &  0.843 & 50972.5424 & 51276.5148\\
M4~V9            & 0.5718945 & 13.303 &  1.114 & 55412.7595 & 50601.4580\\
M4~V27           & 0.6120183 & 13.214 &  0.911 & 55412.7165 & 50601.6926\\
\multicolumn{6}{c}{---RRab3---}\\
\wcen~V3         & 0.8412616 & 14.391 &  0.761 & 52743.3051 & 55715.5717\\
\wcen~V7         & 0.7130342 & 14.594 &  0.950 & 49082.5766 & 49191.0218\\
\wcen~V15        & 0.8106543 & 14.368 &  0.724 & 54705.5137 & 54705.6080\\
\wcen~V26        & 0.7847215 & 14.470 &  0.618 & 50978.6516 & 54705.3909\\
\wcen~V57        & 0.7944223 & 14.469 &  0.597 & 51766.3964 & 49876.5654\\
\wcen~V109       & 0.7440992 & 14.426 &  0.995 & 50984.5494 & 52743.6624\\
\wcen~V127       & 0.8349918 & 14.341 &  0.591 & 54705.2972 & 50984.6784\\
\wcen~V268       & 0.8129334 & 14.544 &  0.467 & 51305.5583 & 2451336.5593\\
\enddata
\tablecomments{Tha table lists a 
few RRLs in common with those used for the RVC template. Periods might be slightly different, because we 
adopted different datasets for these two analyses.}
\end{deluxetable*}

\begin{longrotatetable}
\begin{deluxetable*}{l r c ccc ccc ccc ccc ccc c}
\tablenum{2}
 \tabletypesize{\scriptsize}
 \tablecaption{Coefficients and standard deviations of the PEGASUS fits to the 
$V$-band light curve template.}
\tablehead{
\colhead{Template bin} & \colhead{N} & \colhead{$A_0$} & \colhead{$A_1$} & \colhead{$\phi_1$} & \colhead{$\sigma_1$} & \colhead{$A_2$} & \colhead{$\phi_2$} & \colhead{$\sigma_2$} & \colhead{$A_3$} & \colhead{$\phi_3$} & \colhead{$\sigma_3$} & \colhead{$A_4$} & \colhead{$\phi_4$} & \colhead{$\sigma_4$} & \colhead{$A_5$} & \colhead{$\phi_5$} & \colhead{$\sigma_5$} & \\
\multicolumn{2}{c}{} & & \colhead{$A_6$} & \colhead{$\phi_6$} & \colhead{$\sigma_6$} & \colhead{$A_7$} & \colhead{$\phi_7$} & \colhead{$\sigma_7$} & \colhead{$A_8$} & \colhead{$\phi_8$} & \colhead{$\sigma_8$} & \colhead{$A_9$} & \colhead{$\phi_9$} & \colhead{$\sigma_9$} & & & & \colhead{$\sigma$ }}
 \startdata 
            RRc (\tmeano) &  8387 & --0.5307 &     1.5916 &     8.0136 &     0.2750 &   --0.9843 &     0.5681 &     2.4905 &   --3.0543 &     0.0136 &     0.2443 &     1.8137 &     0.6700 &     1.1791 &     1.4625 &     0.0107 &     0.2233 &  \\
\multicolumn{2}{c}{} &  &   0.4535 &   --0.1160 &     0.5254 &      \ldots  &      \ldots  &      \ldots  &      \ldots  &      \ldots  &      \ldots  &      \ldots  &      \ldots  &      \ldots  & \multicolumn{3}{c}{} &     0.049 \\
 \hline 
           RRab1 (\tmeano) & 25956 &  --0.0537 &     0.0241 &     2.4686 &   --0.3268 &   --0.4234 &     0.1043 &     0.5060 &   --0.6608 &     0.0204 &     0.2749 &     0.2908 &     0.6227 &     0.6791 &     0.3389 &   --0.0153 &   --0.1522 &  \\
\multicolumn{2}{c}{} &  &   0.9761 &   --0.0829 &     0.4306 &   --0.4576 &   --0.1291 &     0.2920 &      \ldots  &      \ldots  &      \ldots  &      \ldots  &      \ldots  &      \ldots  & \multicolumn{3}{c}{} &     0.035 \\
 \hline 
           RRab2 (\tmeano) &  26029 & --0.3753 &     0.5372 &     3.5621 &     0.9021 &   --0.4391 &     0.0470 &     0.1680 &   --0.1191 &     0.0124 &     0.0657 &     0.1553 &     0.7271 &     0.3803 &     0.0485 &     0.0239 &   --0.0506 &  \\
\multicolumn{2}{c}{} &  &   0.6073 &   --0.0445 &     0.5040 &   --0.2475 &     0.0976 &     0.2170 &   --0.2263 &     0.1454 &     0.3674 &      \ldots  &      \ldots  &      \ldots  & \multicolumn{3}{c}{} &     0.028 \\
 \hline 
           RRab3 (\tmeano) &  9787 & --2.3494 &   --0.6802 &     4.0938 &     0.3433 &   --0.4498 &     0.2080 &     0.4693 &   --0.2342 &     0.0167 &     0.1190 &     0.6038 &     0.5839 &     0.7883 &   --0.0828 &     0.0854 &     0.1449 &  \\
\multicolumn{2}{c}{} &  &   2.8333 &     0.0219 &     1.5893 &      \ldots  &      \ldots  &      \ldots  &      \ldots  &      \ldots  &      \ldots  &      \ldots  &      \ldots  &      \ldots  & \multicolumn{3}{c}{} &     0.036 \\
 \hline 
            RRc (\tmaxo) & 8387 &  --4.7750 &   --0.6122 &     2.8134 &     0.5686 &     0.0666 &     0.6401 &     0.2090 &     3.5226 &     0.0897 &     0.4441 &   --3.9216 &     1.0871 &     0.4602 &     5.4369 &   --0.3928 &     3.0508 &  \\
\multicolumn{2}{c}{} & &     \ldots  &      \ldots  &      \ldots  &      \ldots  &      \ldots  &      \ldots  &      \ldots  &      \ldots  &      \ldots  &      \ldots  &      \ldots  &      \ldots  & \multicolumn{3}{c}{} &     0.069 \\
 \hline 
           RRab1 (\tmaxo) & 25956 &  --0.4586 &   --0.5500 &     1.1767 &     0.4577 &     0.0221 &     0.8182 &   --0.1123 &   --1.5880 &     0.9838 &     0.4772 &     0.9760 &     0.9329 &     0.3396 &   --0.2869 &     1.3225 &     0.4907 &  \\
\multicolumn{2}{c}{} &  & --0.4888 &   --1.0239 &   --0.1842 &     1.3798 &     1.0799 &   --1.3487 &      \ldots  &      \ldots  &      \ldots  &      \ldots  &      \ldots  &      \ldots  & \multicolumn{3}{c}{} &     0.043 \\
 \hline 
           RRab2 (\tmaxo) &  26029 &   0.4326 &     0.3797 &     0.5056 &     2.2368 &   --1.3584 &   --0.0486 &     1.3392 &   --2.8021 &     0.0117 &     0.2499 &     0.4585 &     0.7517 &     0.7523 &     2.1228 &     0.0195 &     0.2295 &  \\
\multicolumn{2}{c}{} &   &  0.7873 &   --0.0764 &     0.4384 &      \ldots  &      \ldots  &      \ldots  &      \ldots  &      \ldots  &      \ldots  &      \ldots  &      \ldots  &      \ldots  & \multicolumn{3}{c}{} &     0.035 \\
 \hline 
           RRab3 (\tmaxo) &  9787 & --1.5356 &   --0.2273 &     1.0032 &     0.3180 &     0.1744 &     0.7173 &     0.5384 &   --0.0470 &   --0.0120 &     0.0916 &     1.7028 &     0.5273 &     1.7784 &     0.3626 &     0.7872 &     0.3318 &  \\
\multicolumn{2}{c}{} &  &   0.1686 &   --0.1563 &     0.1623 &      \ldots  &      \ldots  &      \ldots  &      \ldots  &      \ldots  &      \ldots  &      \ldots  &      \ldots  &      \ldots  & \multicolumn{3}{c}{} &     0.055 \\
\enddata
\label{tab:coeff_vband}
 \end{deluxetable*} 
\end{longrotatetable}

\begin{figure*}[!htbp]
\centering
\includegraphics[width=8.5cm]{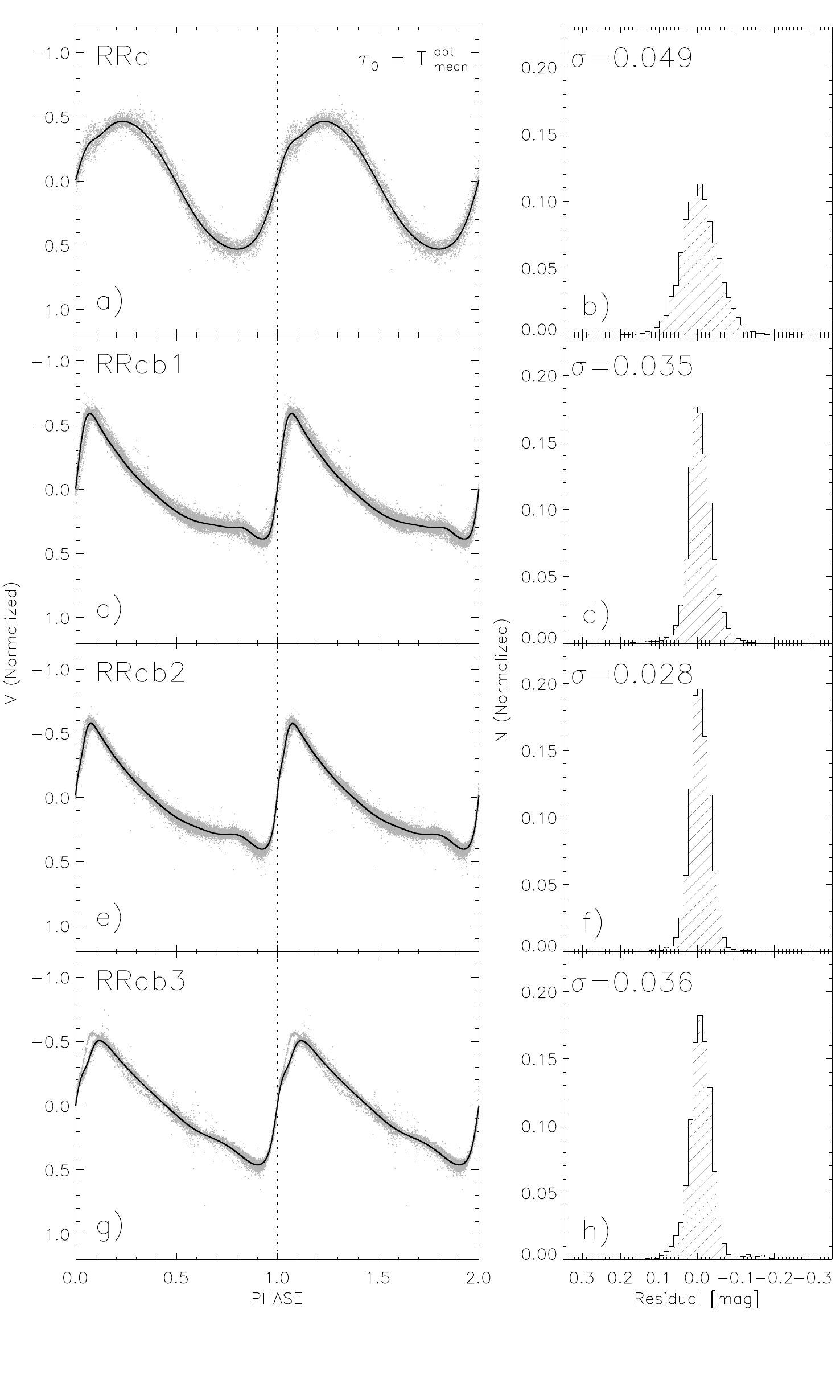}\quad\quad\quad
\includegraphics[width=8.5cm]{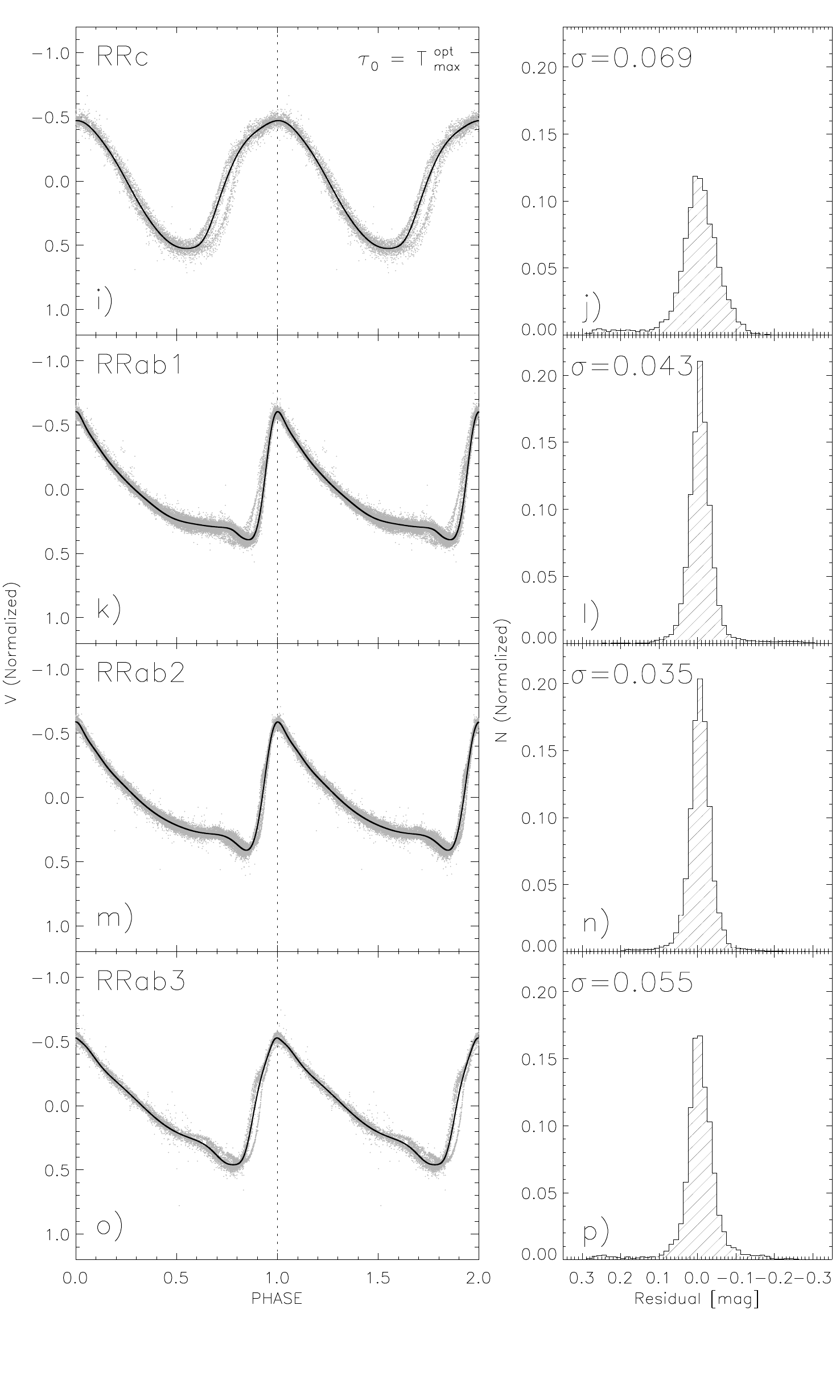}
\vspace{-1truecm}
\caption{Panels a) to h): the left panels---a), c), e) and g)---display, 
from top to bottom, cumulative and 
normalized $V$-band light curves for the four different period bins phased 
by assuming as reference epoch $\tau_0$=\tmeano. The individual measurements 
are marked with light grey dots, while the black solid line shows the analytic 
fit of the light curve template and the vertical dotted line the reference 
epoch.  
The right panels---b), d), f) and h)---display the histogram of 
the residuals of individual phase 
points with respect to the analytical fit. The standard deviations of the
distribution of the residuals are also labeled.
Panels i) to p): Same as a) to h), but the $V$-band light curves were phased by assuming 
as reference epoch $\tau_0$=\tmaxo.
}
\label{fig:vbandtemplate}
\end{figure*}

We visually inspected all the reference epochs derived in this work (see 
Figure~\ref{fig:vbandtemplate}). To 
overcome thorny problems in the phasing of light curves, we manually selected
the best value of \tmaxo~ as the HJD of the phase point closest to the
maximum for the variables where the fit does not follow
closely the data around maximum light. In contrast, no manual selection
of \tmeano~ was needed because its estimation is, by its
own nature, based on a more robust approach. 
We anchored the phases to both \tmaxo~ and \tmeano~ 
and we piled up the light curves into four period bins. 
We ended up with eight cumulative and normalized light curves: four with $\tau_0$ 
anchored to \tmaxo~ and four anchored to \tmeano.

Finally, we adopted the PEGASUS (PEriodic GAuSsian Uniform and Smooth) 
function \citep[a series of multiple periodic 
Gaussians,][]{inno15} to fit the cumulative and normalized light curves. 
The form of the PEGASUS fit is:

\begin{equation}\label{eq_pegasus}
P(\phi) = A_0 + \Sigma_i A_i \exp{\Big(-\sin{\Big(\dfrac{\pi (\phi - \phi_i)}{\sigma_i}\Big)^2}\Big)}
\end{equation}

where $A_0$ and $A_i$ are the zero points and the amplitudes of the Gaussians, 
while $\phi_i$ and $\sigma_i$ are the centers and the $\sigma$ of the Gaussians.

Figure~\ref{fig:vbandtemplate} displays the cumulative and normalized light 
curves of the four period bins. The black solid lines plotted in left and right 
panels show the analytical fits of the cumulative and normalized $V$-band light 
curves with PEGASUS functions (see Eq.~\ref{eq_pegasusfit}). 
The coefficients of the PEGASUS fits are listed in Table~\ref{tab:coeff_vband}. 
The standard deviations plotted to the right of the light curves 
(see also the last column in Table~\ref{tab:coeff_vband}) bring 
forward two interesting results. 
{\em i)}-- The standard deviations of the 
light curves phased by using \tmeano~ are systematically smaller than 
those phased using \tmaxo. The difference for the period bins in which 
the light curves display a cuspy maximum (RRab1, RRab2) is $\sim$37\% smaller, 
but it becomes $\sim$45\% smaller for the RRc and the RRab3 period bins, because 
they are characterized by flat-topped light curves.   
{\em ii)}-- The cumulative light curves for the RRc and RRab3 period bins phased 
using \tmaxo~ show offsets along the rising branch. This mismatch 
could lead to systematic offsets of $\sim$30\% in $Amp(V)$ adopted 
to estimate the mean $<V>$ magnitude. Meanwhile, the cumulative light 
curves phased using \tmeano~ overlap better with each other over the entire pulsation 
cycle. There is one exception: the RRab3 period bin shows a marginal difference 
across the phases of maximum in luminosity, but the error in the adopted 
$Amp(V)$ is on average a factor of two smaller ($\sim$15\%) than those obtained
by using \tmaxo~ as anchor. 

The current circumstantial evidence, based on the same photometric data, indicates 
that the use of a reference epoch anchored to the phase of mean magnitude along 
the rising branch allows a more accurate phasing with respect to the 
phase of the maximum in luminosity.

\subsection{Phase offset between \tmaxo~ and \tmeano}\label{chapt_deltaphi}

\begin{figure*}[htbp]
\centering
\label{fig:deltaphi}
\includegraphics[width=10cm]{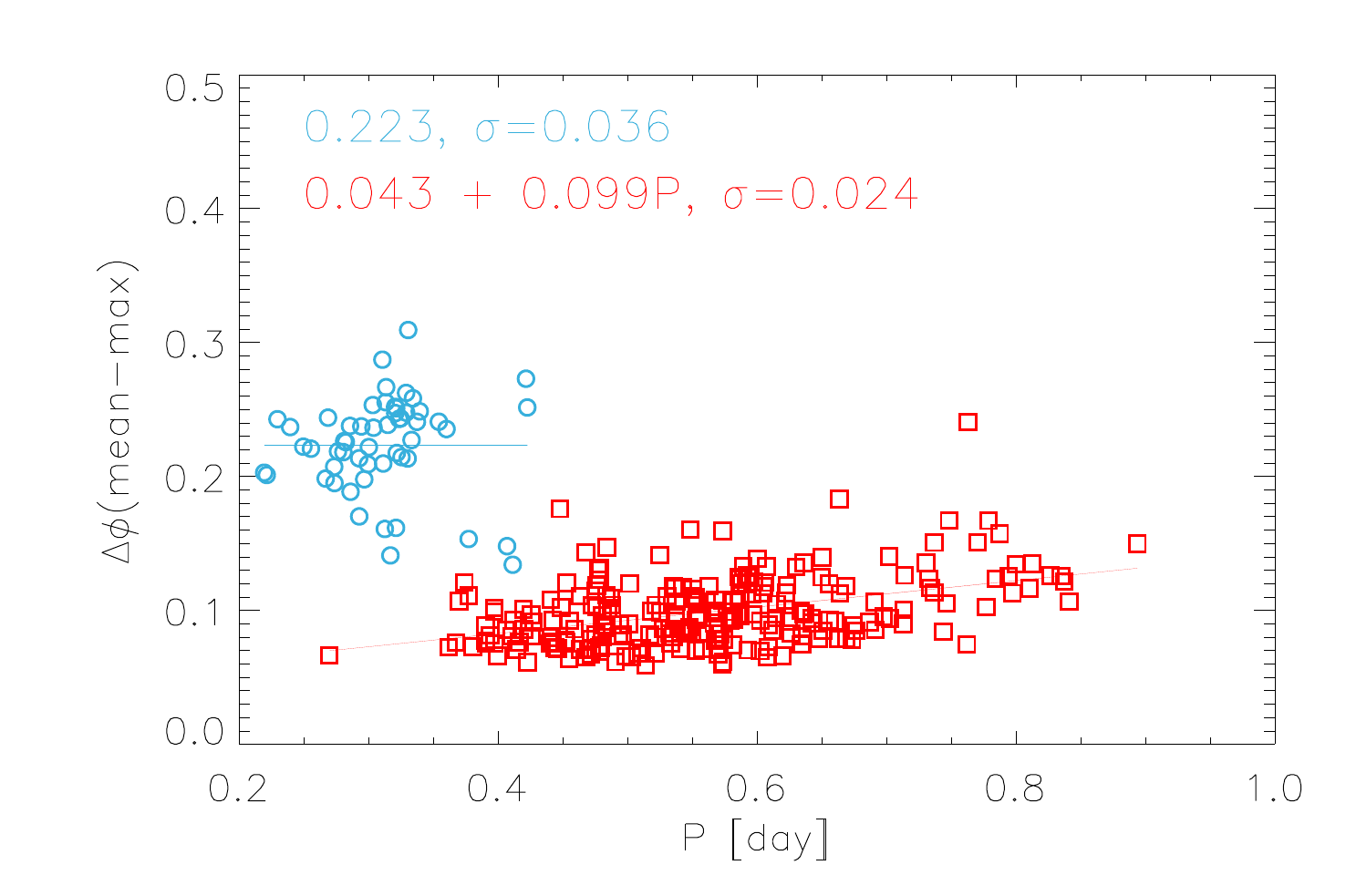}
\caption{$V$-band $\Delta\Phi$ versus pulsation period for RRc (blue) and 
RRab (red) variables. The constant offset for RRc variables and the linear 
relation for RRab variables are displayed as solid lines. The offset and the 
linear relation are labeled at the top-left corner together with their 
standard deviations.}
\end{figure*}

We are aware that large photometric surveys---but also smaller projects focused 
on variable stars---provide, as reference epoch, \tmaxo. To overcome this 
difficulty and to provide a homogeneous empirical framework, we investigated the 
phase offset between \tmeano~ and \tmaxo. In particular, we defined the phase 
difference 

$$\Delta\Phi = \dfrac{(T_{max}^{opt}-T_{mean}^{opt})}{P} \bmod{1} $$

where $\bmod$ is the remainder operator. For this purpose, we could
adopt a larger sample of visual light curves of 291 RRLs
(54~RRc and 237~RRab) from large photometric surveys 
({\it Gaia}, ASAS, ASAS-SN and Catalina), from our own photometry of
globular clusters 
($\omega$~Cen, M4), and from the literature (BW sample, see caption of 
Table~\ref{tbl:deltaphase}). We found that the phase difference shows, 
as expected, a trend with the pulsation period (see Figure~\ref{fig:deltaphi}).

In particular, the RRab variables show a quite clear linear trend of phase 
offset with period
($\Delta\Phi = 0.043 + 0.099 \cdot P$), with an intrinsic dispersion of 
0.024. The standard deviation for RRc variables is larger, but there is no 
clear sign of a period dependency. Therefore, we assume a 
constant phase difference ($\Delta\Phi=$0.223$\pm$0.036) for RRc variables.
We also investigated a possible correlation of phase offset with metallicity 
by adopting the estimates recently provided by \citet{crestani2021a}, but we found none.

\section{Radial velocity database}\label{sect:data}

To provide new RVC templates we performed a large spectroscopic  
campaign aimed at providing RV measurements for both field and 
cluster RRLs. We reduced and analyzed a large sample of high-, medium- and 
low-resolution (HR, MR, LR) spectra. This mix of proprietary data and data 
retrieved from public science archives was supplemented with 
RVCs of RRLs available in the literature.

\subsection{Spectroscopic catalog}

\begin{figure*}[htbp]
\centering
\label{fig:spectroscopic_catalog}
\includegraphics[width=14cm]{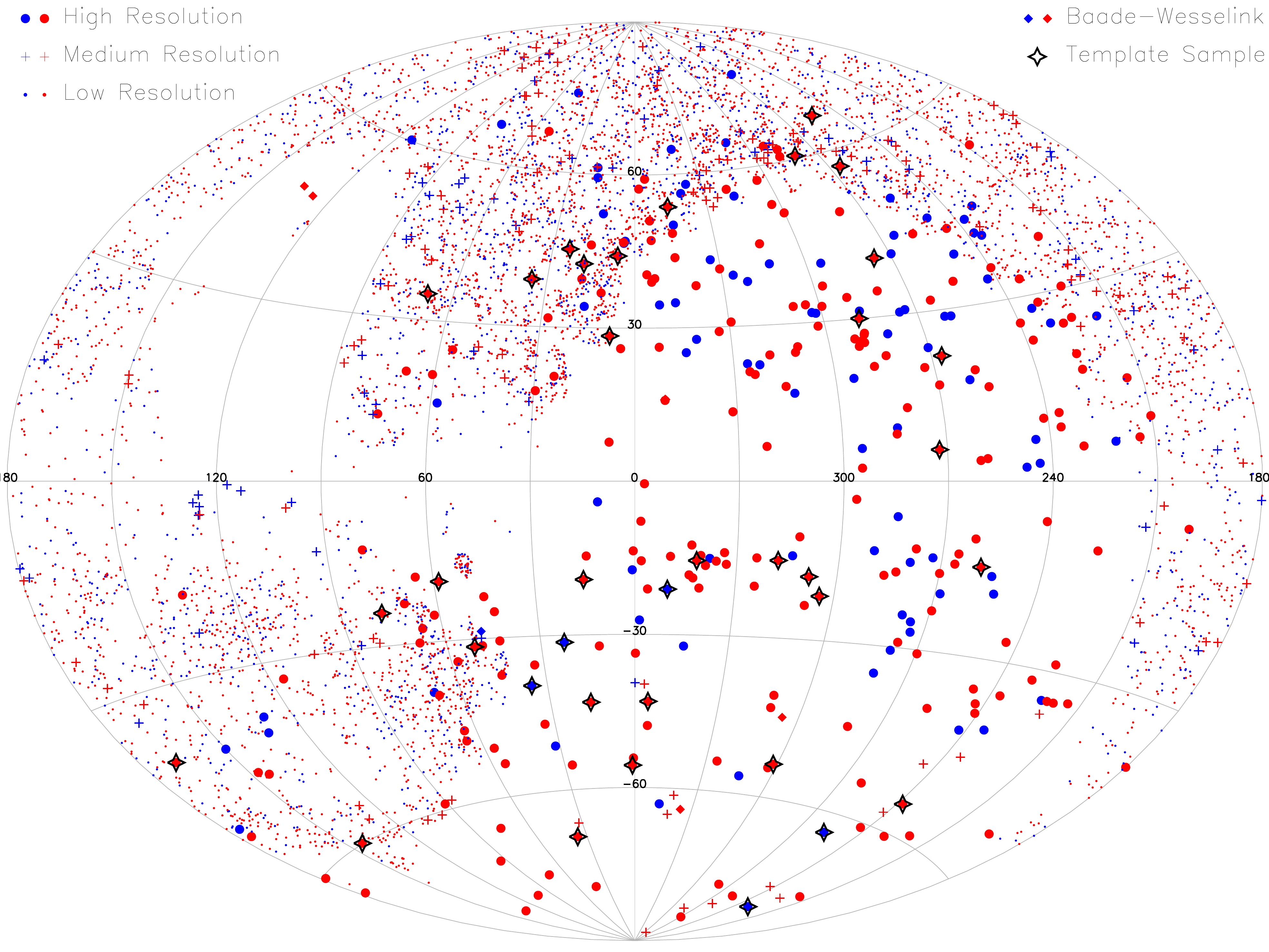}
\caption{Aitoff projection in Galactic coordinates ($l,b$) of the RRL spectroscopic 
dataset. Blue and red symbols display RRc and RRab variables. 
High-, medium-, and low-resolution spectra are marked with large circles, 
crosses, and small circles, respectively. RRL for which we only have radial velocity 
measurements from the literature are plotted as diamonds (Baade-Wesselink sample, 
see Section~\ref{sect:baadewess}). 
Black stars indicate the RRLs used to build the RVC templates.}
\end{figure*}

We collected the largest spectroscopic dataset---both proprietary and 
public---for RRLs. Preliminary versions of this spectroscopic catalog were
already used in studies focused on chemical abundances 
\citep{fabrizio2019,crestani2021a,crestani2021b} and 
RV \citep{bono2020}. 
In this investigation, we added new spectroscopic data and discuss in detail 
the spectra used for RV measurements. We ended up with 23,865 
spectra for 10,413 RRLs. Figure~\ref{fig:spectroscopic_catalog} shows that 
the distribution of the RRLs is well-spread over the Galactic Halo. 
The key properties of the spectra (spectral resolution and signal-to-noise ratio), 
the spectrographs and the spectroscopic sample are summarized in 
Table~\ref{tab:spectra}.

\begin{deluxetable*}{lrrrrr}
\tablenum{3}
\tabletypesize{\scriptsize}
\tablecaption{Key properties of the different spectroscopic datasets adopted in this investigation.}
\label{tab:spectra}
\tablehead{
 \colhead{Instrument} & \colhead{N$_{spectra}$} & \colhead{N$_{RRab}$} & \colhead{N$_{RRc}$} & \colhead{R} & \colhead{SNR} }
\startdata
\multicolumn{6}{c}{---High resolution---} \\
du~Pont & 6208 & 114 &  76 & 35,000 & 40        \\
FEROS@2.2m & 55 & 3 &  0 & 48,000 & 13         \\
HARPS-N@TNG & 10 & 0 &  4 & 115,000 & 40      \\
HARPS@3.6m & 320 & 19 &  6 & 115,000 & 10      \\
HRS@SALT & 81 & 64 &  5 & 40,000 & 50         \\
SES@STELLA & 100 & 0 &  8 & 55,000 & 35        \\
HDS@Subaru & 34 & 23 &  2 & 60,000 & 35        \\
UVES@VLT & 277 & 62 &  8 & 34,540-107,200 & 20  \\
\multicolumn{6}{c}{---Medium resolution---} \\
X-Shooter@VLT & 121 & 16 &  2 & 4,300–18,000 & 45 \\
LAMOST-MR & 1271 & 106 & 66 & 7,500 & 22 \\
\multicolumn{6}{c}{---Low resolution---} \\
LAMOST-LR & 9099 & 4275 & 1935 & 2,000 & 22 \\
SEGUE-SDSS & 5110 & 2487 & 1197 & 2,000 & 21 \\
\multicolumn{6}{c}{---Total---} \\
 & 23865 & 7070 & 3343 & &  \\
\enddata
\tablecomments{Each row gives either the spectrograph or the spectroscopic 
dataset (column 1), the total number of spectra (column 2),
the number of RRab and RRc variables (column 3 and 4), the typical spectral 
resolution (column 5) and the typical SNR@3950 \AA (column 6).}
\end{deluxetable*}

The HR sample mainly includes spectra collected with  
the Las Campanas Observatory du~Pont echelle spectrograph
(du~Pont, 6,208 spectra), plus HR spectra collected from
ESO telescopes (277 from UVES@VLT, 320 from HARPS@3.6m, 
55 from FEROS@2.2m MPG). We also have 100 HR spectra 
from SES@STELLA, 81 from HRS@SALT, 10 from HARPS-N@TNG and 
34 from HDS@Subaru. We collected MR spectra from both X-Shooter@VLT (121 spectra) 
and the LAMOST MR survey (1271 spectra). Finally, our spectroscopic 
dataset includes LR spectra from the LAMOST (9,099 spectra) and 
from the SDSS-SEGUE (6,289 spectra) surveys.

\section{Radial velocity curves}\label{sect:rvcurves}

The main aim of this investigation is to provide RVC templates that can be used 
to provide V$_\gamma$ for RRLs from a few random RV
measurements based on a wide variety of spectra. For this purpose, we 
selected a broad range of strong and weak spectroscopic diagnostics.

\subsection{Radial velocity spectroscopic diagnostics}\label{sect:rvdiagnostics}
The decision to use multiple spectroscopic diagnostics was made because 
different lines form at different atmospheric layers. As the RRL are pulsating 
stars, different lines may trace very different kinematics even 
when observed at the same phase. The resulting velocity curves for different 
lines, therefore, may have different shapes and amplitudes. Consequently, 
combining different hydrogen and/or metallic lines for a single velocity
determination would blur the fine detail of the velocity curves and decrease the 
accuracy of the V$_\gamma$ estimate. With this in mind, we performed RV measurements separately with the following diagnostics: four 
Balmer lines (H$_\alpha$, H$_\beta$, H$_\gamma$ and H$_\delta$), the 
Na doublet (D1 and D2), the Mg I b triplet (Mg b$_1$, Mg b$_2$ and Mg b$_3$) 
and a set of Fe and Sr lines 
\citep[three lines of the Fe I multiplet 43 and a resonant Sr II line,][]{moore1972}.

\begin{figure*}[htbp]
\centering
\label{fig:lines}
\includegraphics[width=15cm]{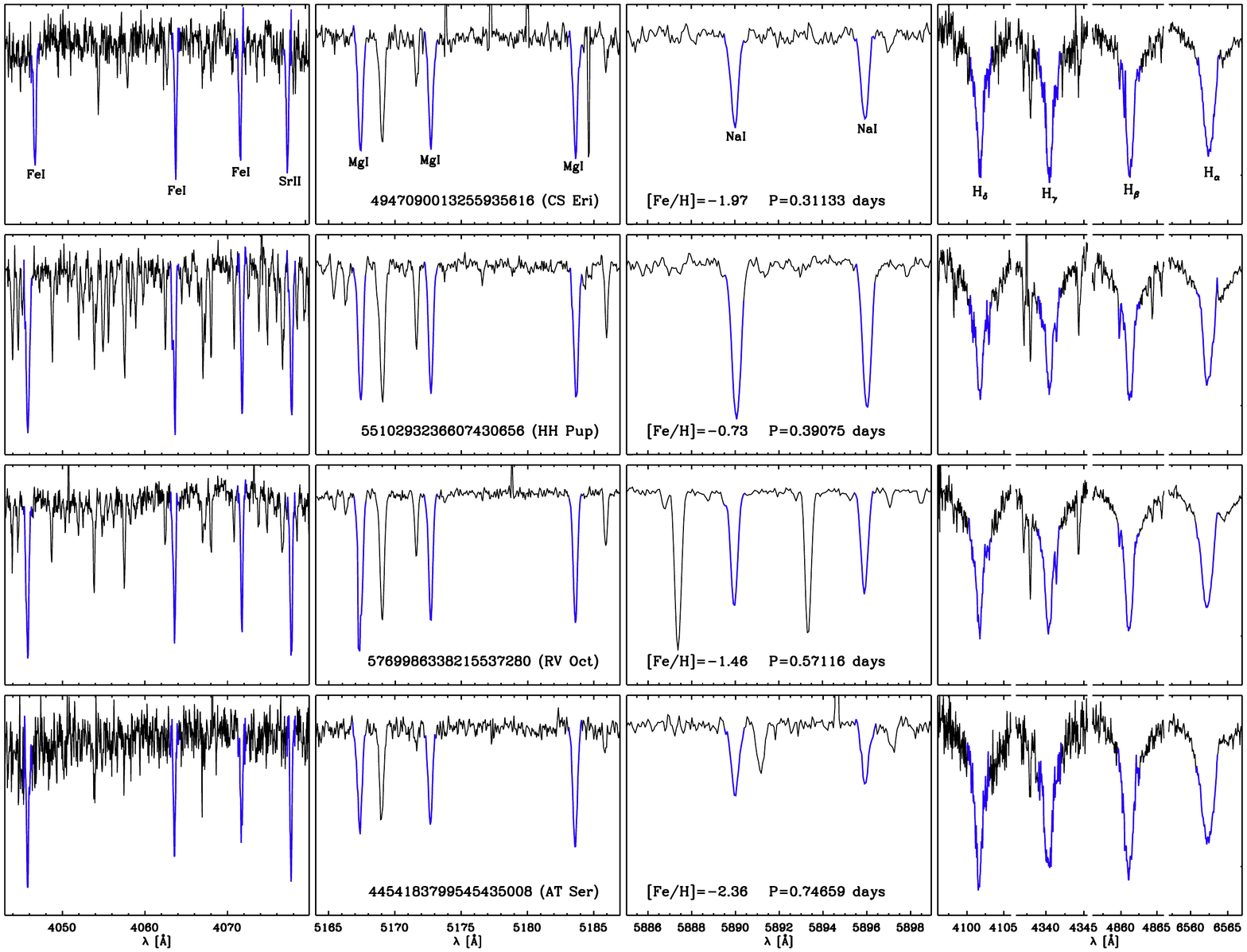}
\caption{From top to bottom, HR spectra collected with 
du~Pont for RRLs adopted in the four bins of the RVC template. 
The flux units are arbitrary.
The Gaia ID (alternative name in parentheses), iron abundance ([Fe/H]) and 
the pulsation period are labeled. The four portions of the spectrum display, 
from left to right, the metallic lines, Mg I b triplet, Na doublet 
and the Balmer lines. Each line is marked in blue and labeled.
All the spectra presented in this figure were taken at phases near 
one third of the rising branch of the RV curve
and are only minimally affected by nonlinear phenomena. 
The first row shows the spectrum of the RRc variable CS Eri.  
Second row: same as the top, but for the RRab HH Pup.
Third row: same as the top but, for the RRab RV Oct.
Fourth row: same as the top but, for the RRab AT Ser.
}
\end{figure*}

The laboratory wavelengths of the
quoted absorption lines are listed in Table~\ref{tab:wav}. 
Figure~\ref{fig:lines} displays 
the regions of the spectrum of four RRLs where the quoted lines
are located. The four RRLs were selected in order to have one RRL 
for each period bin of the RVC template (see Section~\ref{chapt_bins}).
To measure the RVs for the quoted diagnostics we performed a Lorentian fit to the absorption 
lines by using an automated procedure written in IDL. The wavelength range adopted by the 
fitting algorithm is fixed according to the spectral resolution of the different 
spectrographs. Typically,  we selected a range in wavelength that is ten Full Width at 
Half Maximum (FWHM) to the left and ten to the right. 
The FWHM was estimated as FWHM=$2.355\times\dfrac{\lambda_{obs}}{R}$, 
where $\lambda$ is the wavelength of the diagnostic and R is the spectral resolution.

\begin{deluxetable}{rrr}
\tablenum{4}
\tabletypesize{\footnotesize}
\tablecaption{Wavelengths of the lines adopted for radial velocity measurements.}
\label{tab:wav}
\tablehead{ \colhead{Species} & \colhead{Line ID} & \colhead{$\lambda$ [\AA]}}
\startdata
\multicolumn{3}{c}{---Balmer lines---} \\
H$_\alpha$ &H$_\alpha$ & 6562.80 \\
H$_\beta$  &H$_\beta$  & 4861.36 \\
H$_\gamma$ &H$_\gamma$ & 4340.46 \\
H$_\delta$ &H$_\delta$ & 4101.74 \\
\multicolumn{3}{c}{---Fe group---} \\
Fe I      & Fe1      & 4045.81 \\
Fe I      & Fe2      & 4063.59 \\
Fe I      & Fe3      & 4071.74 \\
Sr II     & Sr       & 4077.71 \\
\multicolumn{3}{c}{---Mg group---} \\
Mg I      & Mg b$_1$      & 5167.32 \\
Mg I      & Mg b$_2$      & 5172.68 \\
Mg I      & Mg b$_3$      & 5183.60 \\
\multicolumn{3}{c}{---Na group---} \\
Na I      & D1      & 5889.95 \\
Na I      & D2      & 5895.92 \\
\enddata
\end{deluxetable}

The median uncertainties of the single RV estimates for the adopted 
spectroscopic diagnostics and the standard deviations of the different 
datasets are listed in Table~\ref{tab:errors}. Note that the different 
datasets have median uncertainties, on average, smaller than 1.5 km/s.

\begin{deluxetable*}{l cc cc cc cc cc cc cc}
\tablenum{5}
\tabletypesize{\scriptsize}
\tablecaption{Typical uncertainties in radial velocity measurements for the adopted 
diagnostics in the different spectroscopic datasets.}
\label{tab:errors}
\tablehead{ & \multicolumn{2}{c}{eRV(Fe) } & \multicolumn{2}{c}{eRV(Mg) } & \multicolumn{2}{c}{eRV(Na) } & \multicolumn{2}{c}{eRV(H$_\alpha$) } & \multicolumn{2}{c}{eRV(H$_\beta$) } & \multicolumn{2}{c}{eRV(H$_\gamma$) } & \multicolumn{2}{c}{eRV(H$_\delta$) } \\
 Instrument & mdn & $\sigma$& mdn & $\sigma$& mdn & $\sigma$& mdn & $\sigma$& mdn & $\sigma$& mdn & $\sigma$& mdn & $\sigma$\\
 & \multicolumn{2}{c}{(km/s)}& \multicolumn{2}{c}{(km/s)}& \multicolumn{2}{c}{(km/s)}& \multicolumn{2}{c}{(km/s)}& \multicolumn{2}{c}{(km/s)}& \multicolumn{2}{c}{(km/s)}& \multicolumn{2}{c}{(km/s)}}
\startdata
\multicolumn{15}{c}{---High resolution---} \\
du Pont  &  0.168  &   2.154  &  0.217  & 1.592 & 0.382 & 0.257 & 0.966  & 0.488 &  0.944 &   0.510 &   1.168 &   1.173 &   1.200 &   1.446  \\
FEROS@2.2m &   0.145  &   0.023  &  0.166  & 0.028 & 0.161 & 0.033 & 1.027  & 0.162 &  0.819 &   0.130 &   1.008 &   0.110 &   0.991 &   0.063 \\
HARPS-N@TNG &   0.094  &   0.032  &  0.133  & 0.030 & 0.160 & 0.033 & 0.938  & 0.119 &  0.923 &   0.090 &   1.010 &   0.088 &   1.037 &   0.110 \\
HARPS@3.6m &   0.180  &   0.054  &  0.203  & 0.059 & 0.160 & 0.154 & 1.238  & 0.246 &  0.897 &   0.558 &   1.106 &   0.648 &   1.045 &   0.613 \\
HRS@SALT &   0.209  &   2.186  &  0.214  & 0.811 & 0.229 & 0.270 & 1.181  & 0.407 &  1.105 &   0.387 &   1.851 &   0.742 &   2.948 &   2.174 \\
SES@STELLA &   0.329  &   1.094  &  0.178  & 0.411 & 0.000 & 1.491 & 0.808  & 3.193 &  0.677 &   0.278 &   0.785 &   1.323 &   0.801 &   1.379 \\
HDS@Subaru & \ldots     & \ldots     &  0.185  & 0.028 & 0.200 & 0.071 & 0.790  & 0.192 &\ldots & \ldots    & \ldots    & \ldots    & \ldots    & \ldots    \\
UVES@VLT &   0.167  &   0.041  &  0.196  & 0.047 & 0.219 & 0.075 & 1.089  & 0.344 &  0.912 &   0.252 &   1.106 &   0.218 &   1.045 &   0.358 \\
\multicolumn{15}{c}{---Medium resolution---} \\
X-Shooter@VLT &   0.259  &   0.003  &  0.329  & 0.011 & 0.263 & 0.136 & 1.189  & 0.232 &  1.071 &   0.181 &   1.106 &   0.100 &   1.045 &   0.039 \\
LAMOST-MR &   0.259  &   0.058  &  0.329  & 0.046 & 0.375 & 0.129 & 1.494  & 0.410 &  1.239 &   0.221 &   1.106 &   0.168 &   1.045 &   0.095  \\
\multicolumn{15}{c}{---Low resolution---} \\
LAMOST-LR &   2.863  &   1.495  &  3.657  & 1.214 & 4.168 & 1.874 & 4.645  & 1.182 &  3.440 &   0.528 &   3.072 &   0.421 &   1.983 &   0.492 \\
SEGUE-SDSS & \ldots     & \ldots     & \ldots     & \ldots     & \ldots     & \ldots  &     4.180  & 0.572 &     3.096  & 0.411 &     2.765  & 0.331 &    2.613  & 0.209 \\
\enddata
\tablecomments{Medians (mdn) and standard deviations ($\sigma$) of 
the uncertainties on the RV measurements for the Fe, Mg, Na,
H$_\alpha$, H$_\beta$, H$_\gamma$ and H$_\delta$ lines}
\end{deluxetable*}

\subsection{Radial velocity curves from the literature}\label{sect:baadewess}

To complement our dataset, we collected RVCs of RRLs from the literature
\citep{liujanes89,skillen93a,skillen93b,jones88a,jones88b,cacciari87,jones87b,clementini90,fernley90b,jones87a}. 
During the 80s and 90s of the previous century, several 
bright RRLs were observed both photometrically (optical and NIR) and
spectroscopically (velocities from metallic lines) to apply the 
Baade-Wesselink method \citep[BW,][]{baade1926,wesselink1946} 
in order to obtain accurate distance determinations. 
Therefore, we label the set of RVCs from
these works as the BW sample. Unfortunately, it was not possible to collect 
the spectra, therefore we adopted the RV estimates as provided in the quoted papers.
Overall, the BW sample includes 2,725 RV measurements for 36 RRab and 3 RRc. 

Although this dataset is inhomogeneous 
and based on a mix of weak metallic lines, it is extremely useful to complement our 
own measurements. Some of the works mentioned above included optical light curves, which we
used to validate the robustness of the reference epoch used in the phasing 
of the RVC template.

\subsection{Estimate of barycentric velocities, radial velocity amplitudes and 
reference epochs}

To derive the analytic form of the RVC templates, 
it is necessary to know the pulsation period (P), the reference epoch,
$V_{\gamma}$ and the RV amplitude ($Amp(RV)$) of 
the RRLs with a well-sampled RV curve. The former two are
needed to convert epochs into phases and the latter two are used 
for the normalization of the RV curve. The normalization is 
a crucial step because the RVC templates have to be provided 
as normalized curves, with zero mean and unit amplitude.

Our data set includes RV measurements for more than 10,000 RRLs, but
only 74 of them have a well-sampled pulsation cycle. A good phase coverage
is necessary for the determination of the pulsation properties required for the
creation of the RVC template. Reference epochs and $Amp(RV)$ are particularly 
sensitive to the quality of the pulsation cycle sampling.
We neglected all the RRLs displaying a clear Blazhko 
effect (a modulation of the pulsation amplitude, both in light and in RV) 
that would introduce a large intrinsic spread in the RVC templates. 
Because of this exacting quality control, we derived the analytic form
of the RVC template using only a subset of three dozen RRL in the spectroscopic 
template (31 RRab, 5 RRc). They cover a broad range in pulsation periods 
(0.27-0.84 days) and iron abundances (--2.6 $\le$ [Fe/H]$\le$--0.2). 
We label these stars with the name ``Template Sample'' (TS) 
and their properties are listed in Table~\ref{tab:rrls}. The
individual RV measurements for the TS variables are given in
Table~\ref{tab:allrvcs}.

\begin{deluxetable*}{rr c c c cc cc c c}
\tablenum{6}
\setlength{\tabcolsep}{4pt}
\tabletypesize{\scriptsize}
\tablecaption{Calibrating RRLs used to derive the RVC templates. From left to 
right the columns list the Gaia EDR3 ID, the alternative ID, pulsation period, mean 
visual magnitude, visual amplitude, reference epochs
(\tmeanv~ and \tminv, both for Fe and H$_\beta$ RVCs), iron abundance and its error.}
\label{tab:rrls}
\tablehead{ \colhead{Gaia EDR3 ID} & \colhead{name} & \colhead{Period} & \colhead{$V$} & \colhead{$Amp(V)$} & \colhead{\tmeanvfe} & \colhead{\tminvfe} & \colhead{\tmeanvhb} & \colhead{\tminvhb} & \colhead{[Fe/H]}\tablenotemark{a}  & \colhead{e[Fe/H]}\tablenotemark{a}  \\ 
 & & (days) & \multicolumn{2}{c}{(mag)} & \multicolumn{4}{c}{HJD-2,400,000 (days)}& & }
\startdata
\multicolumn{11}{c}{---RRc---} \\
 6884361748289023488 &                    YZ Cap & 0.2734529  & 11.275 & 0.490 & 55461.3137 & 55461.3460 & 58320.2616 & 58320.0357 & --1.50  &  0.02  \\
 6856027093125912064 &       ASAS J203145-2158.7 & 0.3107106  & 11.379 & 0.370 & 56915.4305 & 56915.1735 & 56915.1343 & 56915.1813 & --1.17  &  0.03  \\
 4947090013255935616 &                    CS Eri & 0.3113302  &  8.973 & 0.520 & 56919.7069 & 56919.4373 & 56919.3975 & 56919.4442 & --1.89  &  0.02  \\
 6662886605712648832 &                    MT Tel & 0.31689945 &  8.962 & 0.560 & 56919.3010 & 56919.3476 & 58574.4677 & 58574.5252 & --2.58  &  0.03  \\
 5022411786734718208 &                    SV Scl & 0.3773586  & 11.350 & 0.530 & 56916.2869 & 56916.3304 & 56916.2943 & 56916.3480 & --2.28  &  0.04  \\
\multicolumn{11}{c}{---RRab1---} \\                                                                     
 1793460115244988800 &                    AV Peg & 0.3903809  & 10.561 & 1.022 & 56531.4845 & 56531.1300 & 56531.0981 & 56531.1245 & --0.18  &  0.10  \\
 5510293236607430656 &                    HH Pup & 0.3908119  & 11.345 & 1.240 & 55962.2982 & 58472.9012 & 58472.8823 & 55959.5915 & --0.93  &  0.15  \\
 4352084489819078784 &                 V0445 Oph & 0.397026   & 10.855 & 0.810 & 56530.9768 & 56531.0228 & 56530.9811 & 56531.0078 & --0.01  &  0.15  \\
 3652665558338018048 &                    ST Vir & 0.41080754 & 11.773 & 1.180 & 56468.9303 & 56468.9685 & 58322.4995 & 58322.5333 & --0.86  &  0.15  \\
 3546458301374134528 &                     W Crt & 0.4120119  & 11.517 & 1.294 & 56076.2014 & 56076.2294 & 58620.7984 & 58620.8240 & --0.75  &  0.15  \\
 4467433017738606080 &                    VX Her & 0.4551803  & 10.791 & 1.200 & 56472.1573 & 57880.9753 & 57880.9507 & 57880.9885 & --1.42  &  0.17  \\
 2689556491246048896 &                    SW Aqr & 0.4593007  & 11.199 & 1.281 & 56175.1772 & 56175.2107 & 56175.1868 & 55815.5797 & --1.38  &  0.15  \\
 1760981190300823808 &                    DX Del & 0.47261773 &  9.898 & 0.700 & 56472.2987 & 56472.3478 & 56472.3059 & 56472.3422 & --0.40  &  0.10  \\
 3698725337376560512 &                    UU Vir & 0.47560267 & 10.533 & 1.127 & 56471.7854 & 56471.8215 & 58573.9638 & 58574.0025 & --0.81  &  0.10  \\
 6771307454464848768 &                 V0440 Sgr & 0.4775     & 10.269 & 1.101 & 54305.4643 & 54305.0241 & 54304.9903 & 54305.0308 & --1.15  &  0.10  \\
 3915998558830693888 &                    ST Leo & 0.47797595 & 11.585 & 1.190 & 56466.8416 & 56466.8722 & 56466.8559 & 56466.8884 & --1.31  &  0.15  \\
 6483680332235888896 &                     V Ind & 0.47959915 &  9.920 & 1.060 & 57620.0005 & 57620.0375 & 57620.0144 & 57620.0519 & --1.46  &  0.14  \\
 1191510003353849472 &                    AN Ser & 0.52207295 & 10.922 & 1.010 & 56468.9009 & 57880.6326 & 57880.5948 & 57880.6280 & --0.05  &  0.15  \\
\multicolumn{11}{c}{---RRab2---} \\                                                                     
 2558296724402139392 &                    RR Cet & 0.55302505 &  9.704 & 0.938 & 56171.2955 & 56171.3458 & 56171.3170 & 56171.3585 & --1.41  &  0.03  \\
 3479598373678136832 &                    DT Hya & 0.5679814  & 13.042 & 0.940 & 54583.0190 & 54583.0673 & 54583.0367 & 54583.0872 & --1.43  &  0.10  \\
 6570585628216929408 &                    TY Gru & 0.57006515 & 14.104 & 0.950 & 55820.0694 & 55820.1104 & 55820.0923 & 54690.8307 & --1.99  &  0.10  \\
 5769986338215537280 &                    RV Oct & 0.571178   & 10.954 & 1.130 & 54690.3296 & 54689.8001 & 54689.7797 & 54689.8224 & --1.50  &  0.10  \\
 5412243359495900928 &                    CD Vel & 0.57350788 & 12.000 & 0.870 & 54908.3154 & 54907.7975 & 54907.7630 & 54907.8141 & --1.78  &  0.10  \\
 5461994302138361728 &                    WY Ant & 0.57434364 & 10.773 & 0.850 & 54903.8161 & 58617.5685 & 58617.5474 & 58617.5909 & --1.88  &  0.10  \\
 5806921716937210496 &                    BS Aps & 0.5825659  & 12.155 & 0.680 & 55644.5855 & 55644.0827 & 55644.0232 & 55644.0810 & --1.49  &  0.10  \\
 6787617919184986496 &                     Z Mic & 0.58692775 & 11.489 & 0.640 & 57287.6695 & 58306.0549 & 58306.0075 & 58306.0631 & --1.51  &  0.10  \\
 5773390391856998656 &                    XZ Aps & 0.58726739 & 12.285 & 1.100 & 55018.9886 & 55019.0308 & 55019.0114 & 55019.0660 & --1.78  &  0.10  \\
 4860671839583430912 &                    SX For & 0.6053453  & 11.077 & 0.640 & 56529.0510 & 56529.1188 & 58061.8060 & 58061.8531 & --1.80  &  0.15  \\
 3797319369672686592 &                    SS Leo & 0.62632619 & 11.034 & 1.152 & 58247.1761 & 58246.6024 & 58246.5755 & 58246.6350 & --1.91  &  0.07  \\
 2381771781829913984 &                    DN Aqr & 0.63376712 & 11.139 & 0.720 & 57261.0936 & 57261.1589 & 57261.1112 & 57261.1760 & --1.76  &  0.15  \\
 4709830423483623808 &                     W Tuc & 0.64224028 & 11.429 & 1.178 & 56528.8863 & 56528.9377 & 58062.5835 & 55457.7106 & --1.76  &  0.15  \\
 15489408711727488   &                     X Ari & 0.65117537 &  9.583 & 0.940 & 56531.5437 & 56530.9654 & 58394.5863 & 58394.6450 & --2.52  &  0.17  \\
\multicolumn{11}{c}{---RRab3---} \\   
 1360405567883886720 &       Cl* NGC 6341 SAW V1 & 0.70279828 & 15.059 & 0.986 & 48054.4181 & 48053.7930 &            &            &  --2.38 & 0.07 \\
 4417888542753226112 &                    VY Ser & 0.7141     & 10.065 & 0.675 & 56468.5381 & 54574.8381 & 58654.4207 & 58654.4811 & --1.82  &  0.10  \\
 4454183799545435008 &                    AT Ser & 0.74655408 & 11.463 & 0.890 & 56530.9316 & 56530.2582 & 58326.4245 & 58326.4957 & --2.05  &  0.22  \\
 6701821205809488384 &       ASAS J181215-5206.9 & 0.8375398  & 13.258 & 0.480 & 55328.2311 & 55017.5992 & 55327.4451 & 55327.5812 &  \ldots & \ldots \\
\enddata
\tablenotetext{a}{The iron abundances are all taken from
homogeneous metallicity estimates in \citet{crestani2021a}. The only exception
is Cl* NGC 6341 SAW V1, for which we have adopted the abundance published
in \citet{kraft2003}}
\end{deluxetable*}

\begin{deluxetable}{r c c r r r}
\tablenum{7}
\setlength{\tabcolsep}{4pt}
\tabletypesize{\scriptsize}
\tablecaption{Individual radial velocity measurements for the calibrating RRLs. 
From left to right the different columns list the name of the variable, the 
spectroscopic diagnostic, heliocentric Julian date, the radial velocity, 
the error on the radial velocity and the spectrograph.}
\label{tab:allrvcs}
\tablehead{ \colhead{Name} & \colhead{Species\tablenotemark{a}} & \colhead{HJD} & \colhead{RV\tablenotemark{b}} & \colhead{eRV\tablenotemark{c}} & \colhead{Instrument} \\
  & & (days) & \multicolumn{2}{c}{(km/s)} &  }
\startdata
 CS Eri & Fe & 2456919.6422 & --133.915 & 0.767 & du~Pont \\ 
 CS Eri & Fe & 2456919.6475 & --133.843 & 1.496 & du~Pont \\
 CS Eri & Fe & 2456919.6528 & --133.221 & 0.896 & du~Pont \\
 CS Eri & Fe & 2456919.6582 & --133.492 & 0.884 & du~Pont \\
 CS Eri & Fe & 2456919.6635 & --133.689 & 0.896 & du~Pont \\
 CS Eri & Fe & 2456919.6689 & --134.396 & 0.985 & du~Pont \\
 CS Eri & Fe & 2456919.6761 & --136.157 & 0.888 & du~Pont \\
 CS Eri & Fe & 2456919.6815 & --136.982 & 0.889 & du~Pont \\
 CS Eri & Fe & 2456919.6868 & --138.586 & 0.805 & du~Pont \\
 CS Eri & Fe & 2456919.6921 & --139.538 & 0.517 & du~Pont \\
\enddata
\tablecomments{Only ten lines are listed. The machine-readable version of this 
table is available online on the CDS.\\
\tablenotetext{a}{Fe, Mg and Na indicate the average of [Fe1, Fe2, Fe3, Sr],
[Mg2, Mg3] and [Na1, Na2] lines, respectively (see Section~\ref{section:template_species}). In contrast, Balmer lines 
radial velocity measurements are single.}
\tablenotetext{b}{Velocity plus heliocentric velocity and diurnal velocity correction.}
\tablenotetext{c}{Uncertainty on the radial velocity 
measurements. For the Balmer lines, it is the 
uncertainty from spectroscopic data reduction. For Fe, Mg and Na, it is the standard 
deviation of the RVs from different lines.} 
}
\end{deluxetable}

The RRLs in the TS have well-covered RVCs for all the adopted 
spectroscopic diagnostics. The only exception is a cluster star 
(Cl* NGC 6341 SAW V1) that has good RVCs only for Fe and Mg lines.
The number of
calibrating RRLs adopted in this work is six times larger than the RRL 
sample adopted by S12. Moreover, S12 only included RRab variables 
covering a limited range in pulsation periods (0.56-0.59 days).

To estimate $V_{\gamma}$, $Amp(RV)$, the epoch of mean velocity
on the decreasing branch and the epoch of minimum velocity 
(\tmeanv~ and \tminv, both for Fe
and H$_\beta$ RVCs), we fitted the RVCs with the PLOESS algorithm,
as described in \citet{bono2020}. Then, we derived $V_{\gamma}$ as
the average of the fit and $Amp(RV)$ as the difference between the 
maximum and the minimum of the fit. The estimates of 
$V_{\gamma}$, $Amp(RV)$ and their uncertainties are provided 
in Tables~\ref{tab:rrls_vel_h} and \ref{tab:rrls_vel}.
Note that we provide these estimates for both the Balmer lines 
and for the averaged RVCs of Fe, Na and Mg.
By using $V_{\gamma}$ and $Amp(RV)$, we normalized all 
the RVCs of the TS RRLs and derived the Normalized 
RVCs (NRVCs).

\begin{deluxetable*}{r rr rr rr rr rr rr rr rr}
\tablenum{8}
\setlength{\tabcolsep}{4pt}
\tabletypesize{\scriptsize}
\tablecaption{Barycentric radial velocities and RV amplitudes based on Balmer lines.}
\label{tab:rrls_vel_h}
\tablehead{ Name & \multicolumn{4}{c }{H$_\alpha$} & \multicolumn{4}{c }{H$_\beta$} & \multicolumn{4}{c }{H$_\gamma$} & \multicolumn{4}{c}{H$_\delta$} \\ 
    & $V_{\gamma}$ & e$V_{\gamma}$ & $Amp(RV)$ & e$Amp(RV)$& $V_{\gamma}$ & e$V_{\gamma}$ & $Amp(RV)$ & e$Amp(RV)$& $V_{\gamma}$ & e$V_{\gamma}$ & $Amp(RV)$ & e$Amp(RV)$& $V_{\gamma}$ & e$V_{\gamma}$ & $Amp(RV)$ & e$Amp(RV)$\\
& \multicolumn{4}{c}{(km/s)} & \multicolumn{4}{c}{(km/s)} & \multicolumn{4}{c}{(km/s)} & \multicolumn{4}{c}{(km/s)}
    }
\startdata
              YZ Cap &--107.17 &    1.48&   37.09&    3.16&--109.37&    1.56&   29.56&    2.90&--112.82&    2.09&   29.29&    2.84&--105.05&    1.71&   21.98&    2.47 \\
	          DR Cap &  --1.66 &    1.18&   30.42&    2.94&  --3.25&    1.21&   25.31&    2.54&  --1.83&    1.82&   25.25&    2.94&  --4.42&    1.89&   26.86&    4.80 \\
              CS Eri &--145.11 &    1.02&   48.15&    3.79&--146.63&    1.12&   35.14&    2.96&--146.73&    1.19&   33.42&    2.75&--142.83&    1.24&   28.58&    2.61 \\
              MT Tel &   65.64 &    1.10&   35.65&    2.26&   64.23&    1.15&   27.37&    1.94&   65.26&    1.21&   25.80&    1.91&   66.87&    1.29&   22.03&    1.81 \\
              SV Scl & --14.60 &    1.11&   42.70&    3.19& --15.94&    1.21&   32.64&    2.73& --14.93&    1.34&   29.18&    2.58& --11.09&    1.56&   23.17&    2.48 \\
              AV Peg & --57.29 &    0.94&   92.66&    4.07& --62.00&    0.98&   72.66&    3.27& --66.26&    1.59&   67.53&    3.21& --59.05&    1.55&   59.07&    2.98 \\
              HH Pup &   18.11 &    1.02&  110.28&    4.30&   17.05&    1.02&   86.75&    3.74&   15.65&    1.33&   80.19&    3.29&   20.66&    1.21&   70.89&    2.98 \\
           V0445 Oph & --19.51 &    1.27&   90.33&    7.07& --24.41&    1.18&   68.82&    5.16& --28.84&    1.53&   63.08&    4.79& --21.29&    1.62&   58.15&    4.98 \\
              ST Vir &  --1.26 &    1.40&   95.80&    5.42&  --4.63&    1.54&   73.00&    4.24&  --2.24&    1.61&   75.59&    4.48&  --3.03&    1.57&   61.47&    3.81 \\
               W Crt &   60.86 &    1.21&  101.33&    4.17&   59.32&    1.15&   84.23&    3.55&   56.66&    1.57&   76.17&    3.28&   61.81&    1.38&   66.79&    2.89 \\
              VX Her &--376.05 &    1.72&  105.14&    7.29&--376.92&    1.85&   80.88&    6.88&--379.06&    2.35&   75.14&    5.33&--373.62&    2.01&   57.18&    4.06 \\
              SW Aqr & --49.20 &    1.08&  103.54&    4.13& --49.17&    1.13&   76.46&    3.82& --49.25&    1.36&   80.24&    3.42& --46.23&    1.55&   66.89&    2.99 \\
              DX Del & --56.88 &    0.95&   92.90&    5.42& --60.81&    0.98&   71.37&    3.92& --62.74&    1.31&   66.59&    3.78& --58.20&    1.19&   58.22&    3.24 \\
              UU Vir & --12.08 &    0.89&  110.49&    6.31& --12.51&    0.93&   94.20&    5.79& --11.12&    1.45&   88.31&    4.28&  --9.62&    1.28&   75.87&    3.53 \\
           V0440 Sgr & --64.10 &    1.67&   98.91&    6.24& --64.22&    1.93&   84.25&    6.78& --60.83&    1.96&   74.25&    5.63& --59.86&    1.99&   67.85&    5.09 \\
              ST Leo &  165.14 &    1.72&  114.08&    8.18&  165.08&    1.05&   82.29&    5.35&  163.93&    1.55&   86.00&    5.88&  169.63&    1.68&   69.03&    4.78 \\
               V Ind &  200.88 &    0.95&  101.31&    3.39&  200.31&    0.85&   78.22&    2.58&  201.52&    1.04&   73.88&    2.48&  203.00&    0.95&   61.37&    2.07 \\
              AN Ser & --41.71 &    1.35&   95.50&    6.29& --44.51&    1.50&   73.11&    4.87& --51.79&    1.77&   69.88&    4.65& --42.15&    1.55&   57.87&    3.77 \\
              RR Cet & --77.69 &    1.06&  107.53&    5.23& --77.19&    1.28&   82.88&    4.13& --75.66&    1.72&   77.60&    4.02& --73.85&    1.42&   72.01&    3.67 \\
              DT Hya &   75.15 &    1.30&  110.59&    5.82&   77.66&    1.12&   82.92&    4.23&   77.45&    1.43&   81.03&    4.31&   82.13&    1.46&   71.75&    3.87 \\
	          TY Gru & --12.28 &    1.54&  105.82&    4.14& --11.13&    1.69&   80.50&    3.24&  --6.89&    1.95&   81.11&    3.37&  --8.79&    2.19&   63.10&    3.13 \\
              RV Oct &  137.23 &    1.42&  104.50&    3.63&  139.60&    1.15&   85.93&    3.00&  141.51&    1.35&   85.52&    3.12&  144.78&    1.29&   75.42&    2.86 \\
              CD Vel &  239.70 &    0.95&   96.88&    3.86&  240.15&    0.83&   77.77&    3.02&  241.53&    1.11&   74.63&    2.95&  243.02&    1.07&   62.65&    2.52 \\
              WY Ant &  201.31 &    0.93&  108.83&    5.62&  202.26&    0.78&   84.27&    4.27&  204.58&    1.00&   76.90&    4.12&  206.15&    0.99&   67.96&    3.45 \\
              BS Aps &--106.76 &    1.25&   85.20&    3.01&--108.88&    1.08&   65.30&    2.28&--105.39&    1.22&   61.26&    2.21&--103.51&    1.36&   56.18&    2.22 \\
               Z Mic & --58.64 &    0.94&   92.91&    3.61& --61.08&    0.79&   74.78&    2.88& --58.05&    0.98&   68.05&    2.67& --57.44&    1.05&   60.64&    2.41 \\
              XZ Aps &  192.39 &    1.02&  104.24&    3.14&  195.55&    0.91&   85.62&    2.62&  197.67&    1.46&   85.91&    2.85&  198.92&    1.60&   67.22&    2.49 \\
              SX For &  243.01 &    1.25&   91.87&    5.73&  242.72&    0.99&   69.79&    3.98&  244.01&    1.17&   65.42&    3.86&  245.87&    1.01&   59.78&    3.45 \\
              SS Leo &  160.70 &    0.99&  103.73&    5.54&  161.11&    1.17&   83.63&    4.13&  164.42&    1.56&   80.95&    4.00&  164.07&    1.61&   73.69&    3.62 \\
              DN Aqr &--229.89 &    0.87&  101.94&    5.14&--228.68&    0.84&   83.05&    4.38&--229.00&    1.04&   81.96&    4.14&--225.74&    1.31&   64.72&    3.50 \\
               W Tuc &   56.91 &    1.07&  114.34&    5.36&   59.96&    0.99&   85.68&    4.10&   63.23&    1.20&   82.03&    3.88&   65.48&    1.15&   69.92&    3.41 \\
               X Ari & --41.58 &    1.08&  109.64&    4.40& --39.73&    0.97&   89.55&    3.47& --37.97&    1.27&   77.70&    3.10& --36.22&    1.43&   67.65&    2.86 \\
              VY Ser &--147.44 &    0.95&   97.31&   10.27&--148.75&    0.81&   79.42&    6.68&--146.66&    0.94&   70.92&    5.80&--146.87&    0.87&   64.06&    4.70 \\
              AT Ser & --71.38 &    1.00&  102.45&    7.35& --70.38&    1.16&   79.59&    5.50& --68.68&    1.42&   79.90&    5.60& --63.33&    1.44&   68.66&    4.63 \\
    	   V0384 Tel &  302.85 &    1.45&   87.45&   10.13&  303.03&    0.98&   71.76&    9.27&  303.04&    1.99&   60.79&    8.15&  303.90&    1.75&   55.80&    6.81 \\
\enddata
\tablecomments{Note that Cl* NGC 6341 SAW V1 does not appear because the Balmer RV measurements for this star are not accurate.}
\end{deluxetable*}


\begin{deluxetable*}{r rr rr rr rr rr rr}
\tablenum{9}
\tabletypesize{\scriptsize}
\tablecaption{Barycentric radial velocities and RV amplitudes based on metallic lines.}
\label{tab:rrls_vel}
\tablehead{Gaia DR2 ID & \multicolumn{4}{c }{Fe} & \multicolumn{4}{c }{Mg} & \multicolumn{4}{c}{Na} \\ 
    & $V_{\gamma}$ & e$V_{\gamma}$ & $Amp(RV)$ & e$Amp(RV)$& $V_{\gamma}$ & e$V_{\gamma}$ & $Amp(RV)$ & e$Amp(RV)$& $V_{\gamma}$ & e$V_{\gamma}$ & $Amp(RV)$ & e$Amp(RV)$ \\
& \multicolumn{4}{c}{(km/s)} & \multicolumn{4}{c}{(km/s)} & \multicolumn{4}{c}{(km/s)} }
\startdata
              YZ Cap & --109.60 &     0.16&    26.34&     1.09& --109.05&     0.16&    26.47&     1.55& --109.05&     0.41&    27.76&     2.33 \\
              DR Cap &   --3.34 &     0.18&    22.77&     1.17&   --2.19&     0.20&    23.00&     1.72&   --3.25&     0.48&    27.30&     2.76 \\
              CS Eri & --145.25 &     0.12&    29.08&     1.21& --145.31&     0.17&    29.17&     1.74& --145.60&     0.39&    31.43&     2.61 \\
              MT Tel &    66.43 &     0.14&    22.82&     0.73&    65.98&     1.17&    22.99&     1.90&    65.32&     0.71&    23.19&     1.53 \\
              SV Scl &  --14.14 &     0.19&    26.87&     1.14&  --13.98&     0.17&    26.80&     1.56&  --15.68&     0.30&    26.81&     2.06 \\
              AV Peg &  --60.48 &     0.28&    63.52&     1.41&  --56.63&     0.32&    64.58&     2.00&  --55.89&     0.69&    68.71&     3.06 \\
              HH Pup &    18.59 &     0.21&    69.27&     1.36&    19.65&     0.29&    71.16&     1.95&    19.00&     0.65&    71.42&     2.82 \\
           V0445 Oph &  --21.96 &     0.32&    57.42&     2.25&  --18.57&     0.44&    57.16&     2.98&  --18.36&     0.67&    61.45&     4.70 \\
              ST Vir &   --4.52 &     0.27&    60.60&     1.74&   --1.43&     0.54&    62.44&     2.62&   --2.50&     0.64&    66.75&     3.83 \\
               W Crt &    61.38 &     0.28&    67.49&     1.27&    63.66&     0.50&    68.39&     1.84&    62.69&     0.70&    70.41&     2.76 \\
              VX Her & --375.68 &     0.50&    61.93&     1.86& --374.55&     0.65&    62.76&     2.62& --375.59&     0.98&    62.61&     3.68 \\
              SW Aqr &  --48.64 &     0.23&    62.54&     1.25&  --47.21&     0.30&    62.31&     1.72&  --47.80&     0.48&    64.74&     1.90 \\
              DX Del &  --58.96 &     0.24&    53.90&     1.48&  --55.69&     0.54&    54.27&     2.05&  --55.59&     0.60&    57.46&     3.39 \\
              UU Vir &  --11.15 &     0.20&    67.45&     1.46&   --3.30&     0.45&    69.21&     2.11&   --8.80&     0.62&    69.22&     3.07 \\
           V0440 Sgr &  --60.99 &     0.48&    63.10&     2.60&   \ldots&  \ldots &   \ldots&   \ldots&  --36.16&     1.64&    95.08&     9.03 \\
              ST Leo &   165.87 &     0.29&    65.39&     2.12&   168.07&     0.34&    65.51&     2.94&   167.40&     0.54&    66.96&     4.44 \\
               V Ind &   201.04 &     0.16&    56.09&     0.83&   201.86&     0.25&    56.48&     1.24&   201.17&     0.42&    57.87&     1.39 \\
              AN Ser &  --43.34 &     0.33&    59.90&     1.84&  --39.68&     0.54&    60.80&     2.56&  --38.94&     1.01&    60.08&     3.96 \\
              RR Cet &  --75.49 &     0.18&    61.77&     1.62&  --74.13&     0.37&    61.42&     2.17&  --74.39&     0.55&    63.03&     3.38 \\
              DT Hya &    80.03 &     0.30&    63.78&     1.64&    82.12&     0.41&    63.26&     2.34&    81.90&     0.68&    61.90&     3.09 \\
              TY Gru &   --7.89 &     0.41&    58.61&     1.05&   --6.70&     0.43&    61.07&     1.50&   --7.26&     0.55&    61.90&     2.15 \\
              RV Oct &   141.86 &     0.32&    65.28&     1.10&   142.46&     0.45&    65.63&     1.58&   142.38&     1.01&    69.55&     2.67 \\
              CD Vel &   240.96 &     0.17&    52.13&     0.93&   241.52&     0.22&    53.07&     1.32&   240.73&     0.43&    52.77&     1.88 \\
              WY Ant &   204.31 &     0.17&    59.80&     1.42&   205.28&     0.25&    59.45&     2.01&   204.40&     0.45&    64.67&     3.26 \\
              BS Aps & --106.68 &     0.26&    50.42&     0.82& --105.79&     0.46&    49.27&     1.24& --106.61&     0.59&    49.51&     1.67 \\
               Z Mic &  --59.52 &     0.18&    52.94&     0.98&  --57.99&     0.42&    52.71&     1.45&  --58.17&     0.56&    53.89&     2.06 \\
              XZ Aps &   198.49 &     0.24&    63.52&     0.92&   198.27&     0.29&    63.02&     1.28&   198.34&     0.50&    63.28&     1.88 \\
              SX For &   244.48 &     0.27&    51.57&     1.44&   245.34&     0.38&    52.53&     2.10&   245.31&     0.47&    51.15&     3.01 \\
              SS Leo &   162.46 &     0.23&    61.51&     1.36&   164.25&     0.35&    61.74&     1.99&   163.50&     0.37&    61.84&     1.96 \\
              DN Aqr & --226.83 &     0.20&    54.58&     1.39& --226.24&     0.46&    52.91&     1.96& --227.23&     0.50&    57.35&     2.51 \\
               W Tuc &    64.98 &     0.19&    63.66&     1.35&    64.62&     0.27&    63.19&     1.89&    64.54&     0.75&    65.87&     3.20 \\
               X Ari &  --37.37 &     0.22&    57.66&     1.06&  --36.70&     0.51&    57.26&     1.61&  --36.94&     0.76&    56.32&     2.24 \\
 Cl* NGC 6341 SAW V1 & --125.84 &     1.13&    56.24&     4.54& \ldots  & \ldots  & \ldots  & \ldots  & \ldots  & \ldots  & \ldots  & \ldots   \\
              VY Ser & --147.11 &     0.27&    51.30&     1.85& --145.15&     0.27&    48.82&     2.46& --145.45&     0.43&    50.41&     3.92 \\
              AT Ser &  --67.66 &     0.23&    60.01&     2.57&  --66.03&     0.36&    56.55&     2.98&  --65.78&     0.57&    62.87&     5.45 \\
           V0384 Tel &   305.34 &     0.29&    42.85&     3.24&   304.55&     0.43&    41.72&     4.21&   305.29&     0.63&    46.36&     6.17 \\
\enddata
\end{deluxetable*}

\subsection{Period bins for the RVC templates}\label{chapt_bins}

The shape of both light curves and RVCs of RRLs depends not only on 
the pulsation mode but also on the pulsation period.  
To improve the accuracy of the RVC templates available in the literature, we 
provide independent RVC templates for RRc and RRab variables. Moreover we 
divide, for the first time, the typical period range covered by RRab variables 
into three different period bins. This improvement is strongly required by the 
substantial variation in pulsation amplitudes (roughly a factor of five) when 
moving from the blue to the red edge of the fundamental instability strip 
\citep{bono99d} and in the morphology of both light curves and RVCs \citep{bono2020,braga2020}.

\begin{figure*}[htbp]
\centering
\label{fig:bailey}
\includegraphics[width=15cm]{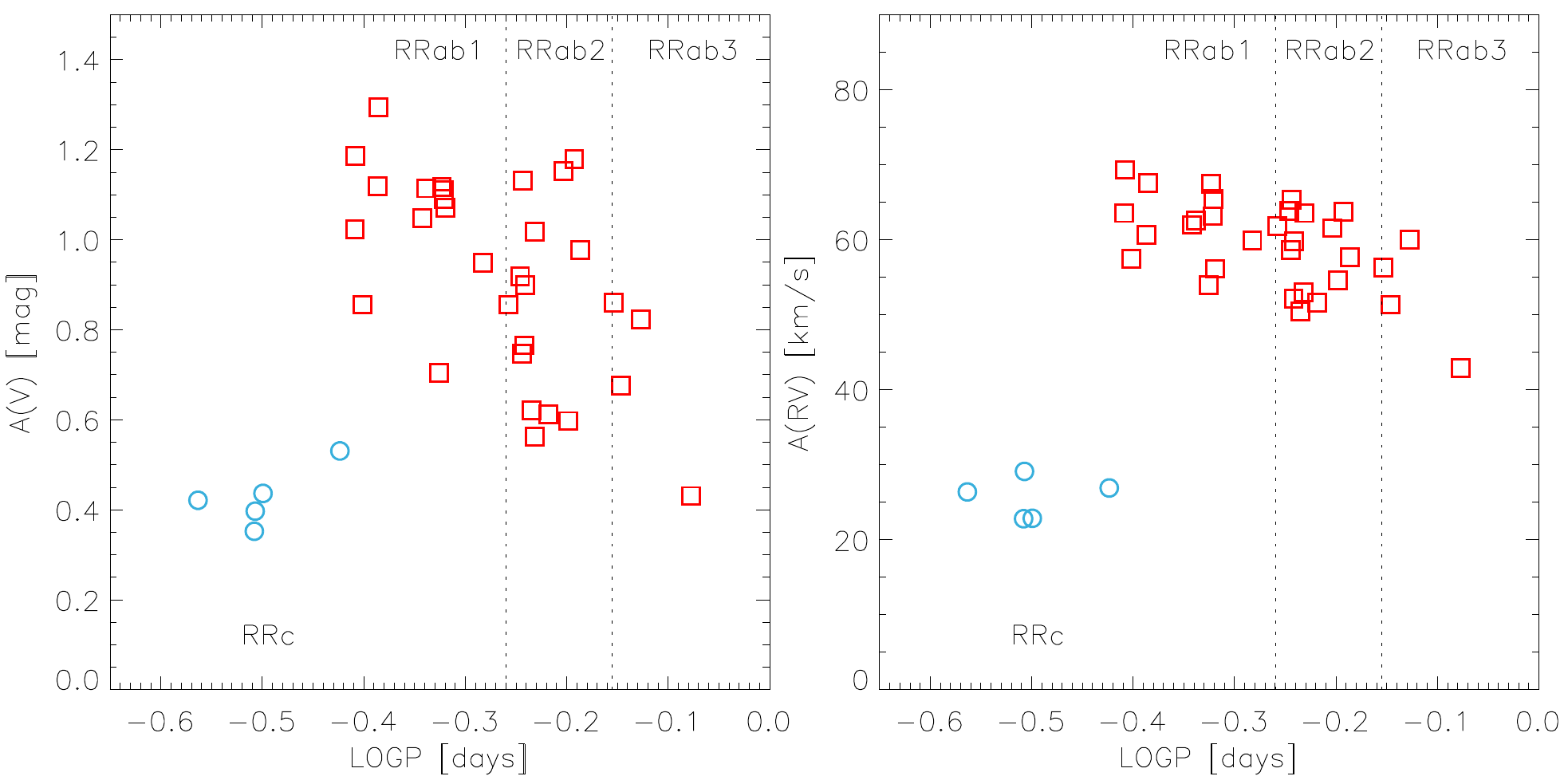}
\caption{Left: Bailey diagram--optical amplitude versus logarithmic 
period--for TS RRLs. Red squares and blue open circles display RRab and RRc 
variables. The vertical dashed line separate the three period bins for 
RRab variables.  
Right: same as the left, but for iron $Amp(V)$.}
\end{figure*}

With this in mind, we adopted the same period bins that 
were used for the NIR light curve templates in \citep{braga2019}
and for the optical light curve templates in Section~\ref{sect:template_vband}.
The reasons for the selection of these specific thresholds 
were already discussed in \citet{braga2019}: they are basically 
to maximize the number of points per bin without disregarding the 
change of the curve shape with period, and to separate RRLs 
with/without Blazhko modulations. The same arguments
hold for the current investigation, with the additional advantage that,
by adopting the same bins, the whole set of RVC templates and 
NIR light curves templates are rooted on homologous sub-samples of 
RRL variables. The Bailey diagram and its velocity amplitude counterpart 
in Fig.~\ref{fig:bailey} show their $Amp(V)$, $Amp(RV)$, pulsation periods, 
and adopted period bins.

We provide 28 RVC templates in total, by considering the combination of seven
different spectroscopic diagnostics (H$_\alpha$, H$_\beta$, 
H$_\gamma$, H$_\delta$, Mg, Na, and Fe+Sr, see Table~\ref{tab:wav}) and four period. 
%

\subsection{The reference epochs of the RVC templates}\label{chapt_phase}

The photometric data available in the literature for the TS RRLs were not 
collected close in time with our spectroscopic data. Therefore, we cannot 
anchor the RVC templates to the photometric reference epochs (e.g., \tmeano).
Small period variations and/or random phase shifts might significantly 
increase the dispersion of the points in the cumulative RVCs.
The phase coverage of the TS RRLs is good enough to provide 
independent estimates of both the pulsation period and of the reference epoch. 
Moreover, \tmeano~ matches \tmeanvfe~ within 5\% of the pulsation cycle 
(see Section~\ref{sect:deltaphi_opt_rv} for more details), therefore we can 
adopt the latter to compute the RVC templates of the metallic lines
(Fe, Mg, Na). This choice allows anyone to adopt 
\tmeano~ to phase the RV measurements and then use our templates 
(see Appendix~\ref{chapt_howto} for detailed instructions). 
Note that, to compute the RVC templates of the Balmer lines, 
we use \tmeanvhb because there is a well defined difference in phase 
between \tmeanvhb and \tmeanvfe. To provide a solid proof of our assumptions, 
we performed the same test discussed in Section~\ref{sect:t0tmax}.

We derived \tmeanv~ and \tminv from the 
average of the RVCs for both the Fe group lines and the H$_\beta$ line, 
taken as representative of the Balmer lines. As expected, \tminv~ matches 
in first approximation \tmaxo. Indeed, the latter was adopted by 
\citet{liu1991} and by \citet{sesar2012} to anchor their RVC templates.
The working hypothesis behind this assumption is that the minimum 
in the RVC of the metallic lines takes place at the 
same phases at which the RVCs based on the Balmer lines 
attain their minimum, i.e. that \tminvfe~ matches \tminvhb. However, we 
checked this assumption and found that the mean difference in phase 
between the two epochs is 0.036$\pm$0.051, with the minimum in the radial 
velocity curves of Fe group lines leading the H$_\beta$ minimum. 
Although the difference is consistent with being zero, its standard 
deviation is not negligible: a 
systematic phase drift of one twentieth around the minimum
of the RVC might lead to offsets in the estimate of $V_\gamma$ of 
the order of 10~km/s.

We point out that TS RRLs have multiple estimates 
$V_{\gamma}$ and of $Amp(RV)$---three from the metallic lines plus four from individual 
Balmer lines---but they only have two reference epochs: one for the Fe group 
lines, representative of the metallic RVCs and one for H$_\beta$. 
The individual estimates of the reference epochs for the two groups of lines 
are listed in Table~\ref{tab:rrls}.

The basic idea is to have the different RVC templates phased at reference 
epochs originating from similar physical conditions. Weak metallic lines 
and strong Balmer lines display a well-defined RV gradient 
and their RVCs are also affected by a phase shift since the former form 
at high optical depths, and the latter at low optical depths 
\citep{liujanes90a,carney1992,bono94c}.
We verified that the reference epochs for the weak metallic lines (Fe, Mg, Na) 
are the same within $\sim$3\% of the pulsation cycle, while those 
for the Balmer lines, anchored to the H$_\beta$ RVC, 
are the same within $\sim$5\% of the pulsation cycle.

\begin{figure*}[!htbp]
\centering
\includegraphics[width=8.5cm]{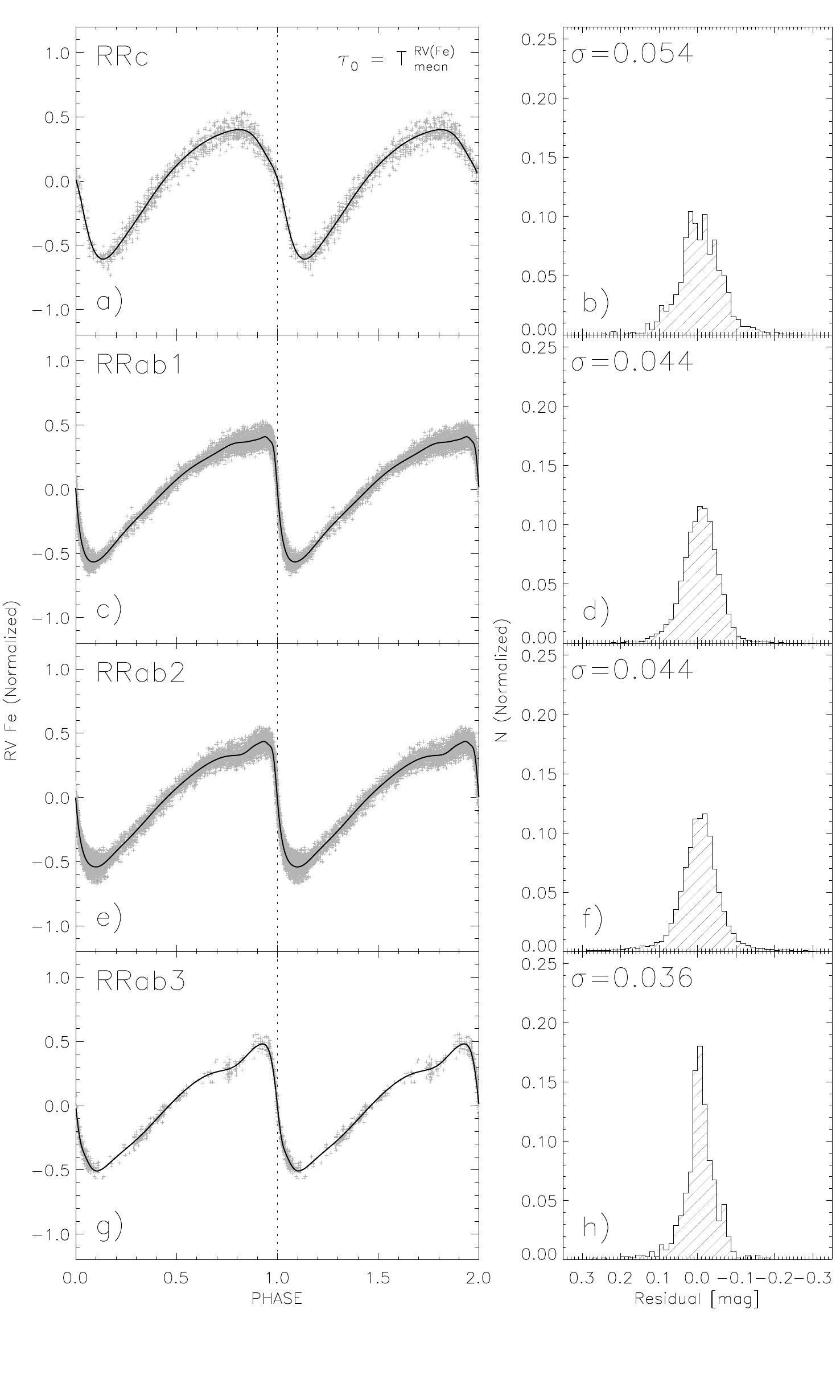}\quad\quad\quad
\includegraphics[width=8.5cm]{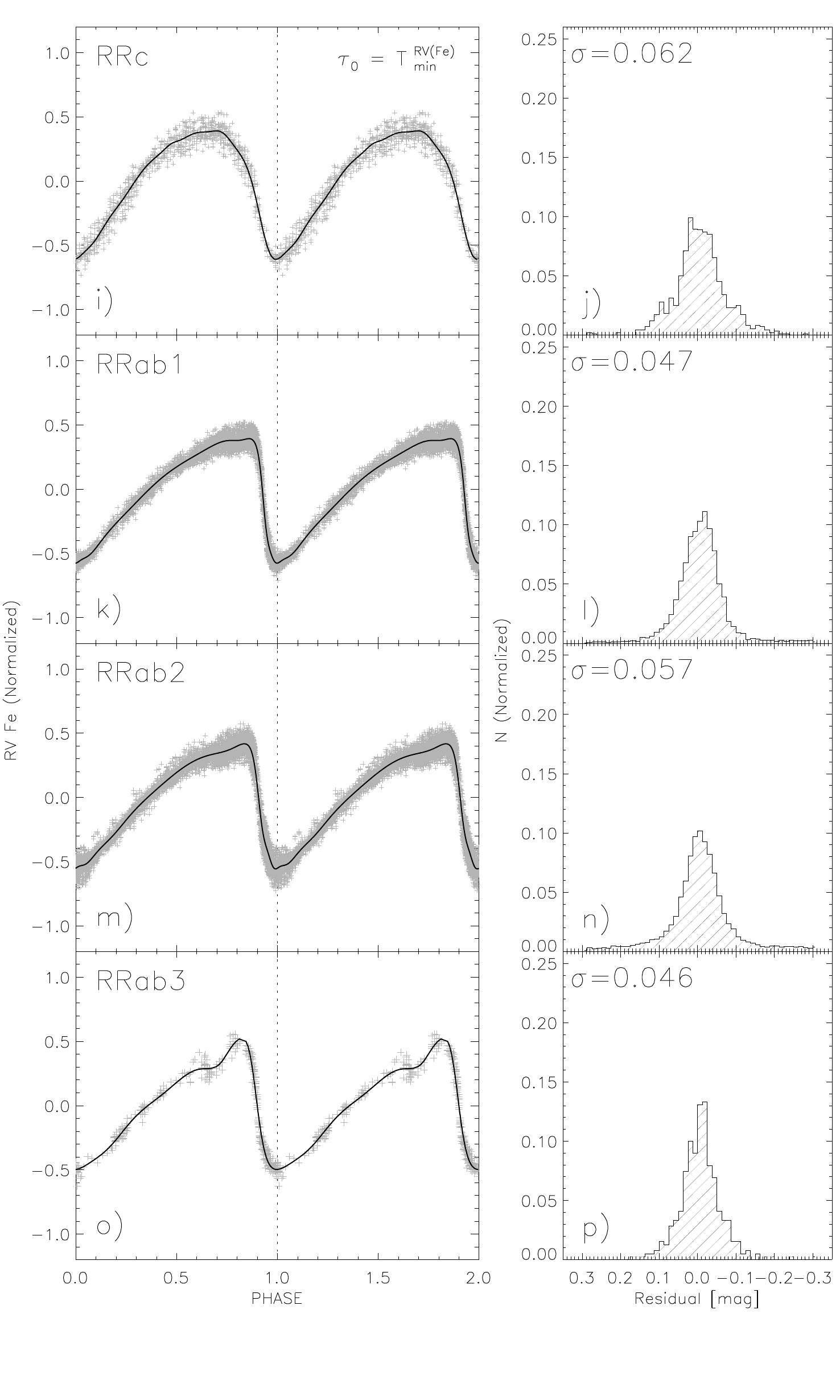}
\vspace{-1truecm}
\caption{Same as Figure~\ref{fig:vbandtemplate}, but for the cumulative and 
normalized RVCs based on the Fe group lines. a), c), e) and g) panels
versus i), k), m), and o) panels 
show the difference between RVCs phased by assuming as a reference epoch 
$\tau_0$=\tmeanvfe~ and $\tau_0$=\tminvfe, respectively. b), d), f), h),
j), l), n) and p) panels
display the residuals of the observations from the analytical fits.}
\label{fig:tmeantmax_fe}
\end{figure*}

\begin{figure*}[!htbp]
\centering
\includegraphics[width=8.5cm]{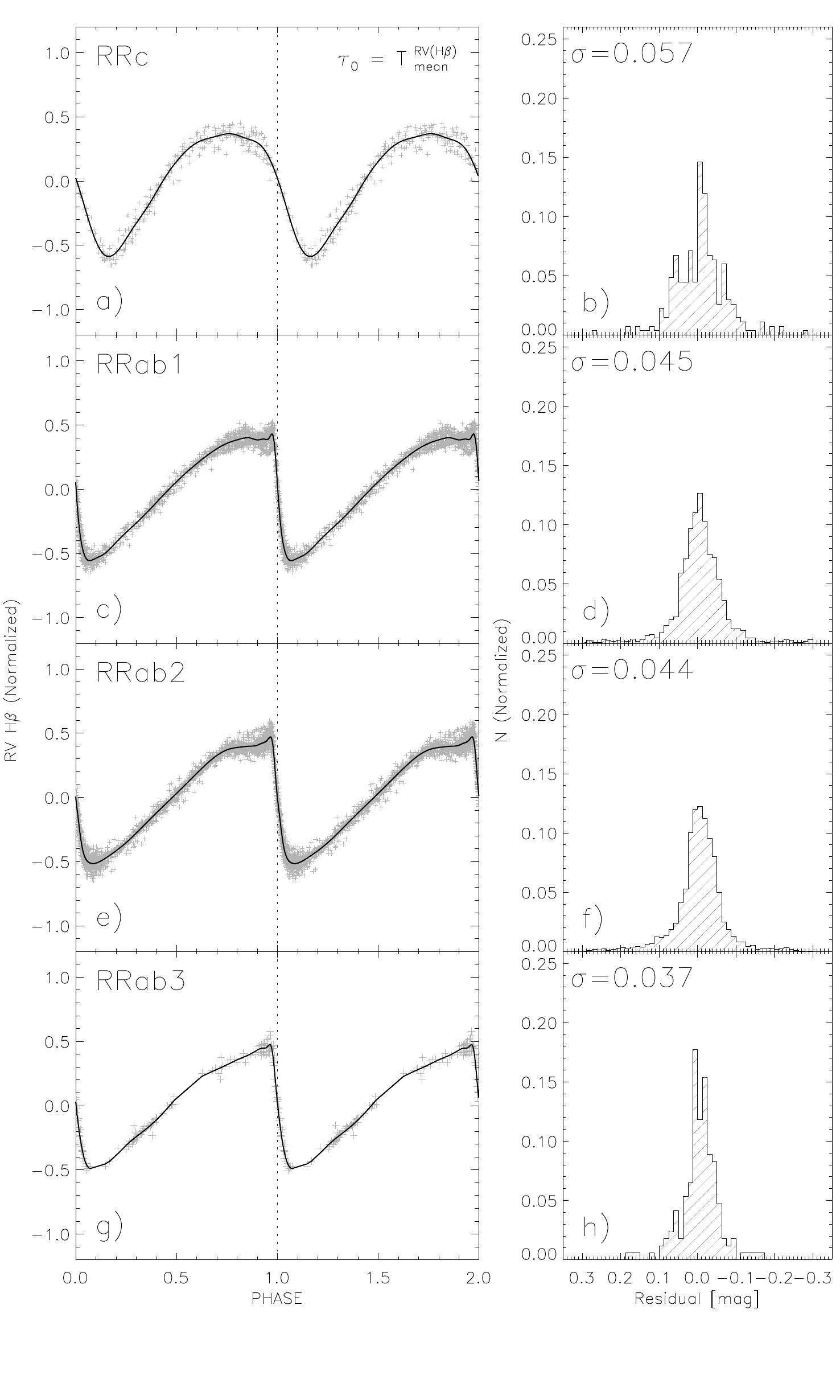}\quad\quad\quad
\includegraphics[width=8.5cm]{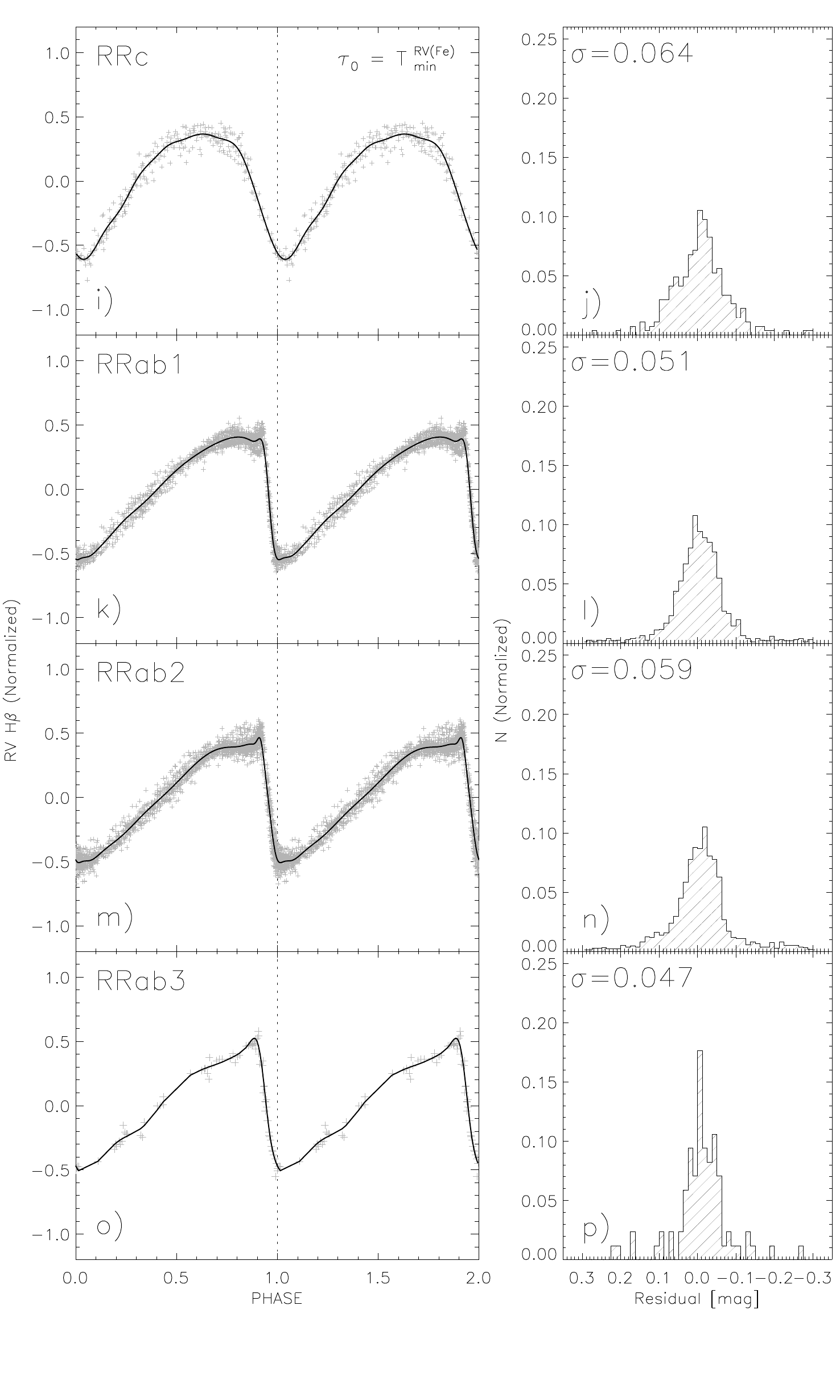}
\vspace{-1truecm}
\caption{Same as Figure~\ref{fig:tmeantmax_fe}, but for cumulative and 
normalized RVCs based on the H$_{\beta}$ line. Note that the a) to h) panels 
are anchored to $\tau_0$=\tmeanvhb~ and the i) to p) panels are anchored to 
$\tau_0$=\tminvfe.}
\label{fig:tmeantmax_hb}
\end{figure*}

Figures~\ref{fig:tmeantmax_fe} and \ref{fig:tmeantmax_hb} display the 
cumulative and normalized RVCs (CNRVCs) based on the Fe group lines and 
on the H$_\beta$ line, respectively. Data plotted in these figures display two 
interesting features worth being discussed in detail. 
{\em i)}--The residuals between observations and analytical fits for the CNRVCs 
based on the Fe group lines and phased using the reference epoch anchored to 
\tmeanvfe~ are systematically smaller than the residuals of the same CNRVCs 
anchored to \tminvfe. The difference ranges from $\sim$30\% for TS RRLs in 
the period bin RRab1 to $\sim$40\% for TS RRLs in the period bin RRab3.  
{\em ii)}--The impact of the two different reference epochs is even more 
evident for the Balmer lines (Fig.~\ref{fig:tmeantmax_hb}). Indeed, the 
difference in the standard deviation ranges from $\sim$15\% in the period bin 
RRc to $\sim$45\% for the period bin RRab2. 
Moreover, the CNRVCs for the RRc and the RRab3 period bin show quite 
clearly that the reference epoch (vertical dotted line) anchored to 
$\tau_0$=\tmeanvfe~ takes place at phases that are slightly earlier than 
the actual minimum (see right panels). 
This difficulty is associated with the shape of both light curves and RVCs
and it causes larger and asymmetrical residuals when compared with 
the CNRVCs of the same period bin anchored to the mean magnitude/systemic 
velocity (see the histograms plotted in the panels to the right of the CNRVCs).

The current circumstantial evidence indicates that RVC templates based on 
Fe group RV measurements and anchored to $\tau_0$=\tmeanv~ provide 
$V_{\gamma}$ that are on average $\sim$30\% more accurate than the same RVCs  
anchored to $\tau_0$=\tminv. The improvement in using \tmeanv~ compared with 
\tminv~ becomes even more relevant in dealing with the RVCs based 
on Balmer lines. Indeed, uncertainties are smaller by up to a 
factor of three (see Section~\ref{chapt_validation}). This further 
supports the use of \tmeanv~ as the optimal reference epoch to construct 
RVC templates. 


%

\section{Radial velocity curve templates}\label{chapt_template}

Before deriving the RVC templates, two pending issues need to be 
addressed: are the RVCs for the different lines 
in the Fe group, in the Mg Ib triplet and in the Na doublet, within the errors, the same? 
Are the {\em mean} RVCs of these three groups of lines 
affected by possible systematics?

\subsection{Mean RVC templates for the three different groups of metallic lines}\label{section:template_species}

Our dataset is large enough to derive RVC 
templates for each single absorption line listed in Table~\ref{tab:wav}.
However, our goal is to provide RVC templates that can be adopted
as widely as possible. Therefore, we aim to provide
one RVC template for each of the Balmer lines and one for 
each of the metallic groups (Fe, Na, Mg), making a total of 
seven different sets of RVC templates. This is feasible only if the 
RVCs of the lines belonging to the same group have, within the errors,  
the same intrinsic features, i.e. the same shape, amplitude and phasing.

In principle, the RVC derived with a specific absorption line 
is different with respect to the one derived from any another line,
because different lines may form in different physical conditions of 
the moving atmosphere. This is quite obvious for the Balmer series, 
with $Amp(RV)$ progressively increasing by $\sim$60-70\%
from H$_\delta$ to H$_\alpha$ \citep[S12,][]{bono2020}. 
This is the reason why independent RVC templates have to be provided 
for each Balmer line. However, for the Fe, Mg and Na groups, it is not 
a priori obvious whether different lines of the same group
(e.g., Fe1 and Fe2) display, within the uncertainties,
similar $Amp(RV)$ and RVC shapes. Therefore, we verified 
whether the RVCs within the Fe, Mg and Na groups agree within uncertainties.

\begin{figure*}[htbp]
\centering
\includegraphics[width=18cm]{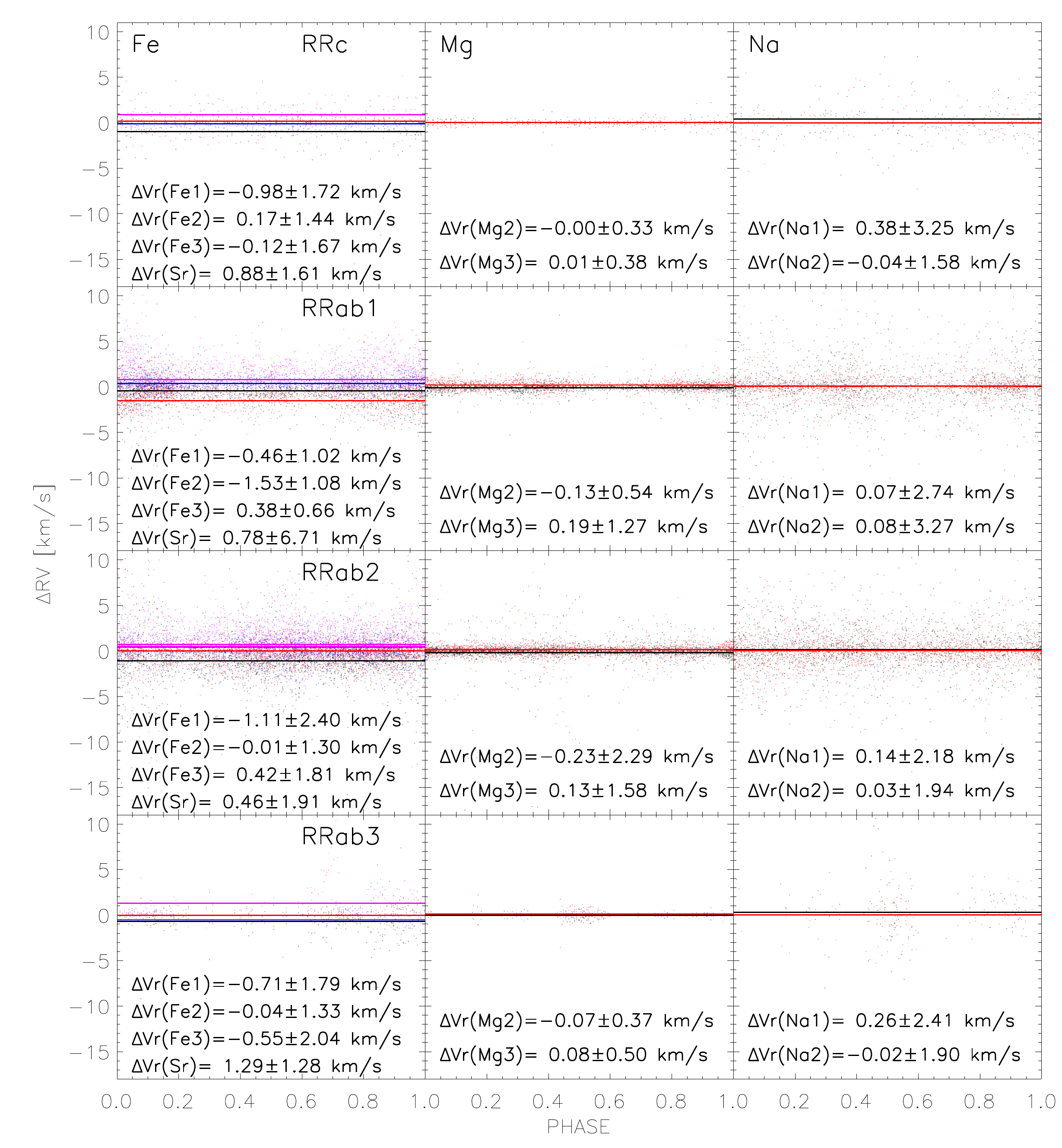}
\caption{Left: Residuals of single-line Fe group RV measurements with respect 
to the mean RVC. Black (Fe1-mean), red (Fe2-mean), blue (Fe3-mean) and 
magenta (Sr-mean) symbols display the difference for the individual lines. 
The mean of the residuals are displayed as solid lines of the same color of 
the symbols and are labeled together with their standard deviations.
From top to bottom, residuals are displayed for the four period bins.
Middle: Same as the left, but for Mg RV measurements.
Right: Same as the left, but for Na RV measurements.}
\label{fig:delta_check_fe}
\end{figure*}

To investigate on a quantitative basis the difference, we inspected the 
residuals of the RVCs of each line with respect to the average RVC of the 
group. The left panels in Fig.~\ref{fig:delta_check_fe} display
the residuals of the RV measurements based on Fe1, Fe2, Fe3 
and Sr lines with respect to the average of the four lines. The middle and right 
panels are the same, but for the two Mg and the two Na lines. Note that we discarded all 
the Mg b1 RV measurements because this line is blended with an 
Fe~I line (5167.50 \AA). This iron line can be as strong as the Mg b1 line 
itself or even stronger depending on the Mg abundance and on the effective 
temperature. Therefore, even using high-resolution 
spectra the velocity measurements with the Mg b1 line refer to an 
absorption feature with a center that changes across the pulsation cycle.

Figure~\ref{fig:delta_check_fe} displays, both quantitatively and qualitatively,
that, for Mg and Na, there is no clear trend within the dispersion 
of the residuals. The maximum absolute offset is vanishingly small, being 
always smaller than 0.50 km/s, which is also smaller than
our RV uncertainty. This means that these lines trace the dynamics 
of the same atmospheric layer and they can be averaged in order to derive 
a single set of RVC templates for both Mg and Na groups.

The same outcome does not apply to Fe and Sr lines. Indeed, although the average
offsets are all smaller than the dispersions, they seem to follow 
a trend. More specifically, the average offset of the Fe1 curve
is always smaller than the other, while the average offset of the 
Sr curve is typically larger. This is mostly due to the interplay of a 
small difference in $Amp(RV)$ (generally increasing from Fe1 to Sr)
and a mild trend in the average velocity (generally increasing 
from Fe1 to Sr). However, all these offsets are smaller than the 
dispersions ($\le$ 2.5 km/s) and similar to the intrinsic dispersion 
of the RVC templates (see Section~\ref{chapt_template_sub}).  
We also note that the standard deviation of the points around 
the offsets is, within the uncertainties, constant along the pulsation 
cycle. Indeed, a minimal increase around phase zero is only present for 
RV measurements based on iron in the period bin RRab1.

\begin{figure*}[!htbp]
\centering
\includegraphics[width=15cm]{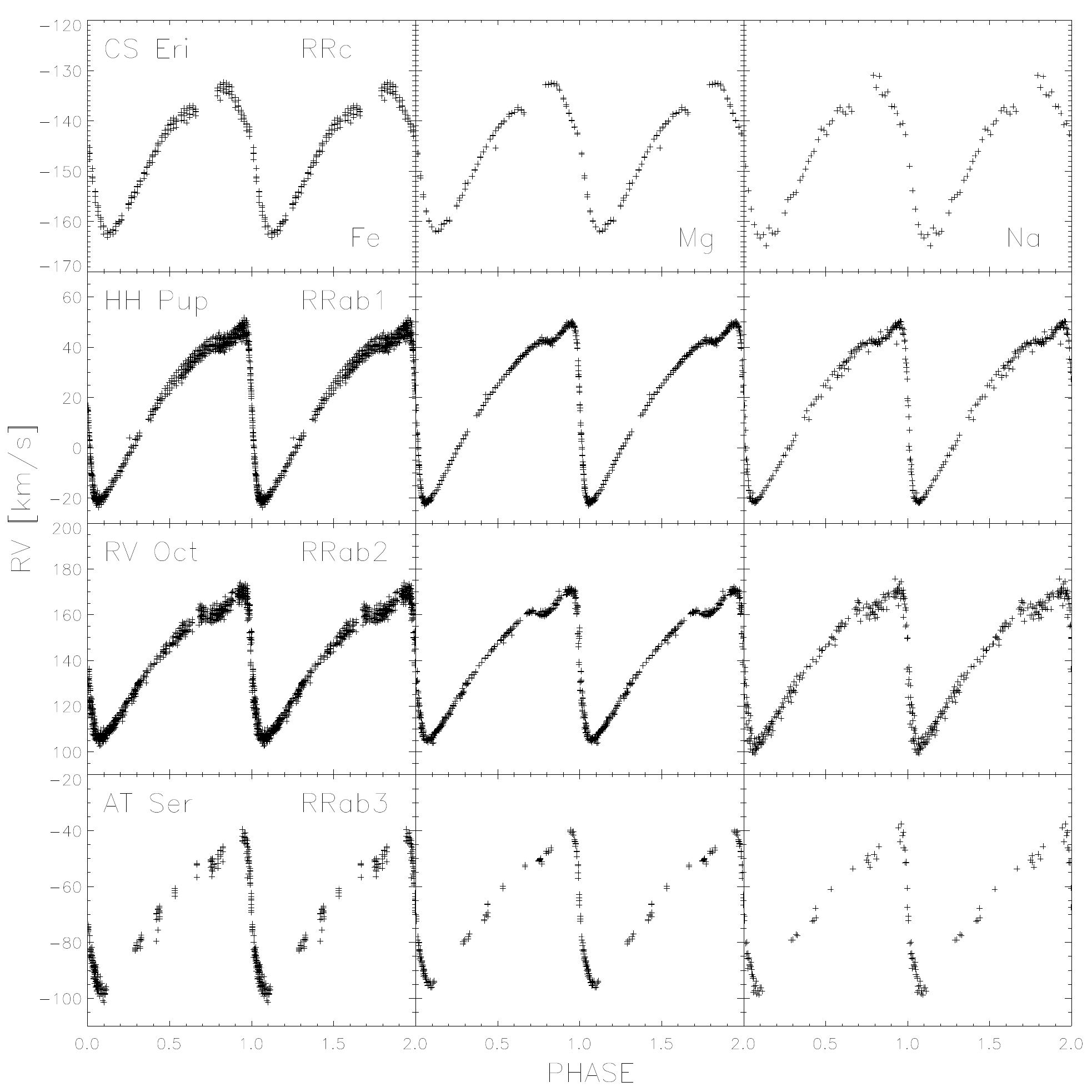}
\caption{First row: From left to right radial velocity curves of the 
RRc variable CS Eri for the average Fe, Mg and Na groups of lines.
Second row: Same as the top, but for the RRab1 variable HH Pup.
Third row: Same as the top, but for the RRab2 variable RV Oct.
Fourth row: Same as the top, but for the RRab3 variable AT Ser.}
\label{fig:rvcurves_metal}
\end{figure*}

\begin{figure*}[!htbp]
\centering
\includegraphics[width=15cm]{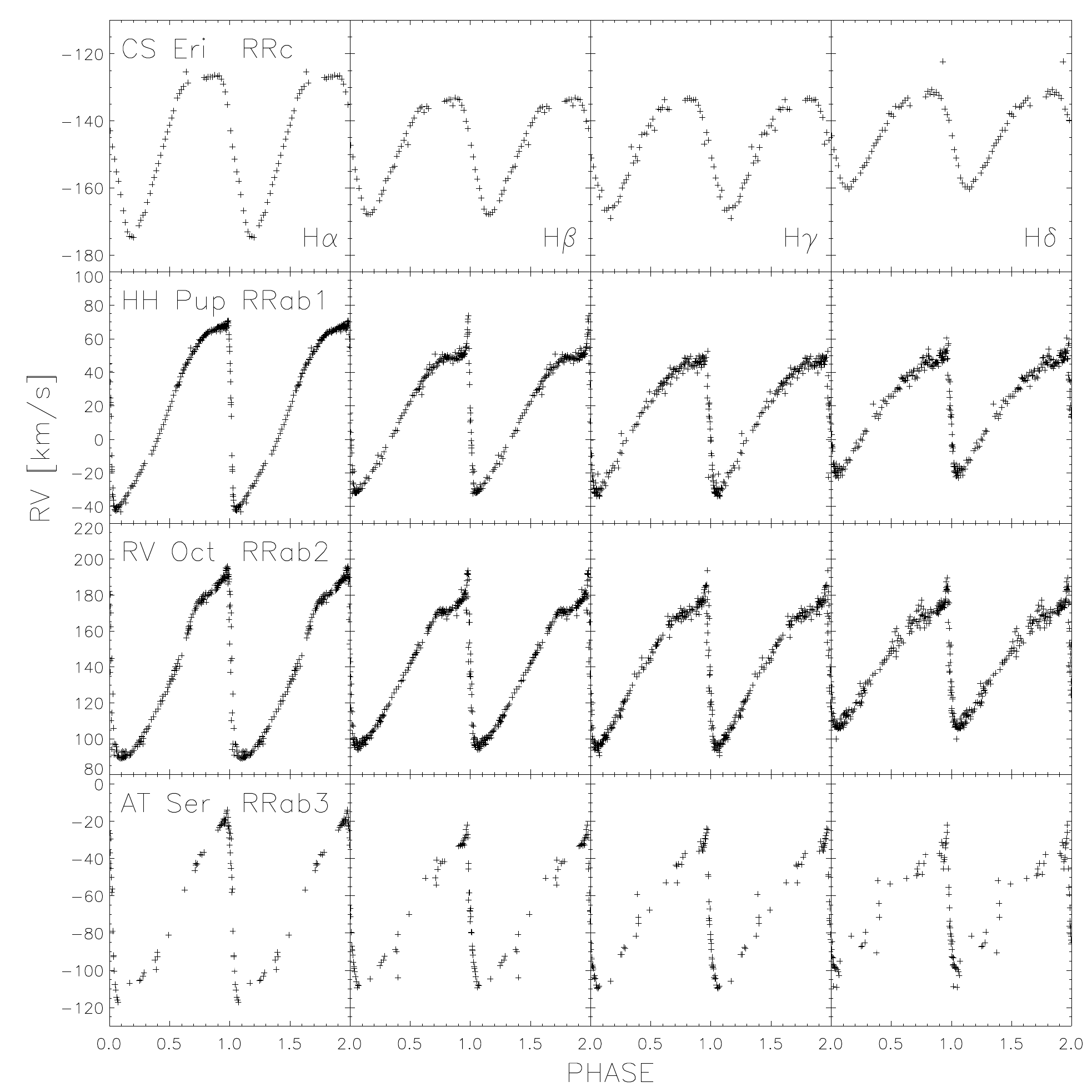}
\caption{First row: From left to right radial velocity curves of the 
RRc variable CS Eri for individual Balmer lines 
(H$_{\alpha}$, H$_{\beta}$, H$_{\gamma}$ and H$_{\delta}$).
Second row: Same as the top, but for the RRab1 variable HH Pup.
Third row: Same as the top, but for the RRab2 variable RV Oct.
Fourth row: Same as the top, but for the RRab3 variable AT Ser.}
\label{fig:rvcurves_balmer}
\end{figure*}

In the light of these results, we opted to derive three independent mean 
RVCs for the Fe, the Mg and the Na groups of lines. Selected RVCs  
for these three metallic diagnostics and for the four individual Balmer 
lines, anchored to  $\tau_0$=\tminvfe~ and  $\tau_0$=\tminvhb, are shown 
in Figures~\ref{fig:rvcurves_metal} and \ref{fig:rvcurves_balmer}.

\subsection{Analytical fits of the RVC templates}\label{chapt_template_sub}

To construct the RVC templates, we normalized the RVCs by subtracting 
$V_{\gamma}$ and dividing by $Amp(V_r)$. Once normalized, we stacked 
the RVCs within the same period bin of RVC template, thus obtaining 
the CNRVCs (see Appendix~\ref{chapt_cnrvc}).

To provide robust RVC templates we decided to fit the CNRVCs with an analytical 
function. We discarded the Fourier series as a fitting curve 
because the number of phase points is not large enough (at least in the 
{\it RRab3} period bin) to avoid non-physical bumps and spurious secondary 
features in the fits. We adopted the PEGASUS fit as for the 
light curve templates in Section~\ref{sect:template_vband}.



\begin{longrotatetable}
\begin{deluxetable*}{ll r c ccc ccc ccc ccc ccc c}
\tablenum{10}
 \tabletypesize{\scriptsize}  
 \tablecaption{Coefficients and standard deviations of the PEGASUS fits to the CNRVCs.}
\tablehead{\colhead{Template} & \colhead{Bin} & \colhead{N} & \colhead{$A_0$} & \colhead{$A_1$} & \colhead{$\phi_1$} & \colhead{$\sigma_1$} & \colhead{$A_2$} & \colhead{$\phi_2$} & \colhead{$\sigma_2$} & \colhead{$A_3$} & \colhead{$\phi_3$} & \colhead{$\sigma_3$} & \colhead{$A_4$} & \colhead{$\phi_4$} & \colhead{$\sigma_4$} & \colhead{$A_5$} & \colhead{$\phi_5$} & \colhead{$\sigma_5$} & \\
\multicolumn{4}{c }{} & \colhead{$A_6$} & \colhead{$\phi_6$} & \colhead{$\sigma_6$} & \colhead{$A_7$} & \colhead{$\phi_7$} & \colhead{$\sigma_7$} & \colhead{$A_8$} & \colhead{$\phi_8$} & \colhead{$\sigma_8$} & \colhead{$A_9$} & \colhead{$\phi_9$} & \colhead{$\sigma_9$} & & & & \colhead{$\sigma$}}
 \startdata 
\enddata
\tablecomments{We will provide this table only after the paper will be officially published on ApJ.}
\label{tab:coeff}
 \end{deluxetable*} 
\end{longrotatetable}

Table~\ref{tab:coeff} provides the coefficients of the PEGASUS functions obtained 
from the fitting procedure. When possible, 
we favored the lowest possible order, especially for the RRc 
template, as first overtone pulsators have more sinusoidal RVCs. These are also the 
coefficients for the analytical form of the RVC templates. 
Figures~\ref{fig:templates_metal} and ~\ref{fig:templates_balmer} display 
the analytical fits of the RVC templates together with the observed 
RV measurements. 

The largest standard deviations are those for 
the H$_\gamma$ and H$_\delta$ RVC templates for the RRc period bin
($\sim$0.08 and 0.13, respectively). To convert the $\sigma$ into the uncertainty 
on $V_{\gamma}$, one has simply to factor in $Amp(RV)$. Since the 
typical $Amp(RV)$ for RRc variables in H$_\gamma$ and H$_\delta$ range between
10 and 30 km/s \citep{bono2020b}, the largest possible uncertainty 
introduced by the RVC template is of $\sim$4 km/s. However, the largest 
absolute uncertainties are associated with 
the H$_\alpha$ and H$_\beta$ RVC templates 
for RRab1 period bin ($\sigma \sim$0.05). Since 
$Amp(RV)$ is much larger for these diagnostics, 
the absolute uncertainties on $V_{\gamma}$ based on these RVC templates 
are of $\sim$7 km/s. For metallic lines, all these 
uncertainties are on average smaller than $\sim$3 km/s. These results 
concerning both metallic and Balmer lines indicate that the current 
RVC templates can provide V$_\gamma$ for typical Halo RRLs 
with an accuracy better than 1-3\%.

\section{Reference epoch to apply the radial velocity curve templates}\label{sect:deltaphi_opt_rv}

A crucial aspect of templates is that they are used especially when 
the number of RV measurements is small. A first consequence is that, in a realistic
scenario, the RV data is insufficient to accurately estimate 
\tmeanv~ for either metal or Balmer RVCs. 
However, optical photometry usually 
is conducted before spectroscopic observations, and a good knowledge of the pulsation 
period and of $Amp(V)$ are required to apply the RVC template. This means 
that we are typically dealing with a fairly well sampled $V$-band light curve, 
and in turn, with an accurate estimate of \tmeano. With this in mind, 
it is necessary to check whether \tmeanv~ and \tmeano~ 
in RRLs take place, within the errors, at the same 
phase along the pulsation cycle. If this is the case, we could safely 
use of \tmeano~ to phase the spectroscopic data and to apply the RVC template.

We phased a subset of RRLs with a good sampling of the 
pulsation cycle both in the $V$-band and in metallic RVCs.
Fortunately, among the RRLs of the TS, there are several 
objects that were used for the Baade-Wesselink (BW) analysis and
for which there are available $V$-band light curves and RVCs 
collected at relatively close epochs (at most within $\sim$3 years). 
This is an important advantage for this consistency test 
because possible changes in phases (phase drifts) and the 
effect of period derivatives are small. We derived both \tmeanvfe~ 
and \tmeano~ and they are listed in Table~\ref{tbl:deltaphase}.

\begin{deluxetable*}{l r r r r r l}
\tabletypesize{\scriptsize}
\tablenum{11}
\caption{Difference in the reference epoch between light and radial velocity curves.
From left to right the columns give the ID, the pulsation mode, the pulsation period, 
\tmeanv, \tmeano, the difference in phase and the reference.}
\label{tbl:deltaphase}
\tablehead{ \colhead{Name} & \colhead{Type} & \colhead{Period} & \colhead{\tmeanvfe} & \colhead{\tmeano} & \colhead{$\Delta\Phi$} & \colhead{Ref.\tablenotemark{a}} \\ 
 & & (days) & \multicolumn{2}{c}{HJD-2,400,000 (days)} &  &}
\startdata
     DH Peg & RRc  & 0.25551624 & 46667.5938 & 46684.6796 &   0.0239 &  0 \\
     TV Boo & RRc  & 0.31256    & 47220.7066 & 47227.5768 &   0.0260 &   1 \\
      T Sex & RRc  & 0.32468493 & 47129.6898 & 47226.4140 &   0.1229 &   1 \\
     RS Boo & RRab & 0.37736549 & 46985.6766 & 46949.4482 &   0.0109 &  2 \\
     AV Peg & RRab & 0.3903809  & 47130.3415 & 47123.3206 & --0.0427 &   1 \\
  V0445 Oph & RRab & 0.397026   & 45842.2860 & 46980.9462 & --0.0111 &   3 \\
     RR Gem & RRab & 0.3973     & 47220.6576 & 47226.5974 &   0.0498 &   1 \\
     TW Her & RRab & 0.39959577 & 46925.8161 & 46947.3878 &   0.0168 &  2 \\
     AR Per & RRab & 0.42556048 & 47128.3468 & 47123.6650 &   0.0019 &   1 \\
      V Ind & RRab & 0.47959915 & 57619.5115 & 47814.0720 & --0.0061 &   4 \\
     BB Pup & RRab & 0.48055043 & 46136.1969 & 47192.9099 &   0.0052 &  5+6 \\
     TU UMa & RRab & 0.5569     & 47129.8698 & 47227.4521 &   0.0088 &   1 \\
     SW Dra & RRab & 0.56967009 & 46519.6350 & 46496.2706 &   0.0140 &  7 \\
     WY Ant & RRab & 0.57434364 & 46135.3591 & 47193.2805 & --0.0091 &  5+6 \\
     RX Eri & RRab & 0.58725159 & 47130.7581 & 47226.4736 &   0.0096 &   1 \\
     RV Phe & RRab & 0.59641862 & 47054.1546 & 46305.0578 & --0.0075 &   8 \\
     TT Lyn & RRab & 0.59744301 & 47129.6929 & 47123.1210 &   0.0003 &   1 \\
     UU Cet & RRab & 0.60610163 & 47055.3286 & 47055.9229 &   0.0195 &   4 \\
      W Tuc & RRab & 0.64224028 & 47057.8121 & 47493.2459 &   0.0046 &   4 \\
      X Ari & RRab & 0.65117537 & 45640.8262 & 45639.5058 &   0.0278 &  9 \\
     SU Dra & RRab & 0.66041178 & 47129.4805 & 47227.8778 &   0.0081 &   1 \\
     VY Ser & RRab & 0.7141     & 44743.7769 & 47655.1533 & --0.0120 &   3 \\
\enddata
\tablecomments{In column 7, the references for the RVC and light curves 
adopted to derive $\Delta\Phi$ are listed in the following order:
\tablenotetext{a}{0: \citet{jones88a}; 1: \citet{liujanes89}; 2: \citet{jones88b}; 
3: \citet{fernley90b}; 4: \citet{clementini90};  5+6: \citet{skillen93a,skillen93b}; 
7: \citet{jones87b}; 8: \citet{cacciari87}; 9: \citet{jones87a}
}}
\end{deluxetable*}

Data listed in this table clearly show that \tmeanvfe~ and \tmeano~ trace, within 
the errors, the same phase along the pulsation cycle. Indeed the average phase difference 
is 0.007$\pm$0.019 and always smaller than 
0.05 (see column 6 in Table~\ref{tbl:deltaphase})\footnote{There is 
only one exception to this trend: the RRc variable 
T Sex for which the phase difference is $\sim$0.12. This large offset might 
be explained by the fact that this variable is multiperiodic, the shape of its 
light curve and its luminosity amplitude change night by night \citep{hobart1991}. 
It is worth noticing that recent photometric surveys 
\citep[e.g., ASAS, TESS,][]{pojmanski1997,benko2021} display a very narrow 
light curve for T Sex, meaning that the multiperiodic behavior might have 
been a transient phenomenon.} This means the two reference 
epochs provide the same phasing. As a consequence, the photometric \tmeano~ can be 
safely adopted to anchor the RVC templates. 

\begin{figure*}[!htbp]
\centering
\includegraphics[width=8.5cm]{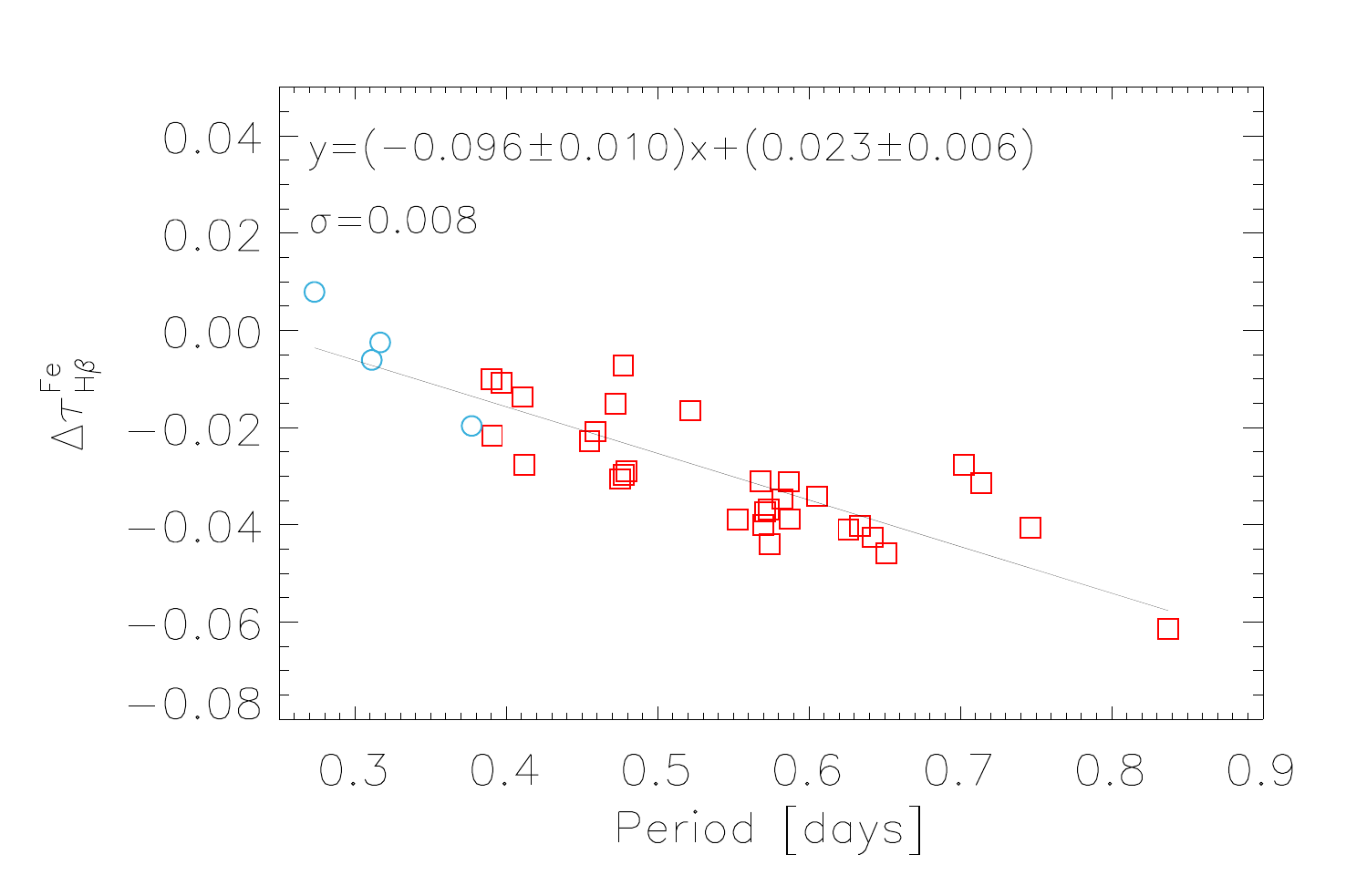}
\caption{$\Delta\tau^{Fe}_{H\beta}$ versus period for the TS RRLs. The linear 
relation fitting the data is displayed as a solid line and labeled at the top, 
together with the standard deviation of the relation.}
\label{fig:deltatau_fehb}
\end{figure*}

We already mentioned that there is a difference in phase between 
\tmeanvfe~ and \tmeanvhb. This means that, when adopting \tmeano~ to use the 
template on Balmer lines, it is necessary to first shift the phases by an offset 
$\Delta\tau^{Fe}_{H\beta}$=$\Phi$(\tmeanvfe-\tmeanvhb). For this reason, 
we adopted the data listed in Table~\ref{tab:rrls} and found a 
linear trend of $\Delta\tau^{Fe}_{H\beta}$ as a function of the pulsation 
period (see Fig.~\ref{fig:deltatau_fehb}). The plausibility of the phase 
difference between metallic and Balmer lines is further supported by the 
empirical evidence that the standard deviation of the relation is vanishing 
(0.008). Indeed, it is almost one order of magnitude smaller than the standard deviation 
of the phase offset between \tminvfe~ and \tminvhb~ 
(see Section~\ref{chapt_phase}).

Large photometric surveys, however, often provide \tmaxo~ but not \tmeano. 
To overcome this limitation and to facilitate the use of the RVC templates,
we provide relations for $\Delta\Phi$ in Section~\ref{chapt_deltaphi} that allow the
template user to easily convert the phases anchored on $\tau_0$=\tmaxo~ can into
phases anchored on $\tau_0$=\tmeano.

\section{Validation of the radial velocity curve templates}\label{chapt_validation}

A solid validation of the RVC templates requires that photometric and radial 
velocity measurements be as close as possible in time. This methodological 
approach provides the unique opportunity to derive accurate photometric 
(pulsation period, $V$-band mean magnitude, $Amp(V)$, reference epoch) and 
spectroscopic ($V_{\gamma}$) properties of the RRLs adopted for the validation. 
We have already mentioned in Section~\ref{chapt_phase} that the temporal 
proximity of both photometric and spectroscopic data is only available for a 
small number of field RRLs. To overcome this limitation we decided to select 
from the calibrating sample one RRL per period bin: YZ Cap for the RRc, 
V Ind for the RRab1, W Crt for the RRab2 and AT Ser for the RRab3.    
We label these four RRLs as the Template Validation Sample (TVS) and they 
are the benchmark for the RVC template validation. 

Ideally, the validation should be done with an RRL sample independent from 
the one adopted to construct the RVC templates. However, we have verified that, 
by removing  the four TVS RRLs from the calibrating RRLs, the coefficients 
of the analytical fits are only minimally affected.
Note that the selection of the validating variable might bias the uncertainties 
of the result because there is a small degree of internal variation in the shape 
of the RVCs of the different period bins. To investigate on a quantitative basis 
the dependence on the validating variable, we performed several tests by using 
as validating variable one after the other all the RRLs included in each period 
bin. Interestingly enough, we found that the medians (see Tables 12 and 13) agree 
within 1$\sigma$, while the variation of the dispersions is on average smaller 
than 20\%. This means that the selection of the validating variable has a minimal 
impact on the accuracy of the validation.

We adopted the $V_{\gamma}$ of the TVS RRLs obtained by directly 
fitting their RVCs (see Tables~\ref{tab:rrls_vel_h} and ~\ref{tab:rrls_vel}) 
and  assumed these as the best estimates for the systemic velocity ($V_{\gamma(best)}$)
to be used in the validation process. However, to use the RVC template 
we need to convert $Amp(V)$ into $Amp(RV)$ and then use the 
latter to rescale the normalized analytical function. 
The ratio between $Amp(V)$ and $Amp(RV)$ together with the equations for the 
conversion are thoroughly discussed in Appendix~\ref{chapt_amplratio}. 

The analytical form of the RVC templates can be used both as curves to be 
anchored to a single phase point and as functions to fit a small number 
of phase points (three or more). Therefore, we followed two different 
paths for the validation process, based either on a single phase point 
approach (Section~\ref{wcen_1punto}) or on  multiple phase points 
(Section~\ref{wcen_3punti}).
The key idea is to estimate the accuracy of the light-curve templates
from the difference $\Delta V_{\gamma}$ between $V_{\gamma(best)}$ 
and the systemic velocities estimated by adopting the RVC template 
($V_{\gamma(templ)}$).  
Furthermore, the accuracy of the current RVC templates is quantitatively 
compared, using TVS RRLs, with similar RVC templates available in the 
literature (S12).

\subsection{Single phase point}\label{wcen_1punto}

As a first step, we generated a 
grid of 100 phase points for the seven RVCs (Fe, Mg, Na, H$_\alpha$, 
H$_\beta$, H$_\gamma$, H$_\delta$, that we label, respectively, with 
a $j$ index running from 1 to 7). We interpolated the analytical 
fits of the RVCs for the TVS RRLs on an evenly-spaced grid of phases 
($\phi_i$=[0.00, 0.01, ... 0.99], where $i$ runs from 1 to 100).
For each $\phi_i$ and each RVC($j$), we generated a RV
measurement with the sum $RV_{ij} = RV(fit(\phi_i))_j + r\sigma$.
The two addenda are: 
{\it i)} $RV(fit(\phi_i))_j$, which is the value of the fit of the RVC($j$) 
interpolated at the phase $\phi_i$; 
{\it ii)}  $r\sigma$, which simulates random noise: $\sigma$ is the standard 
deviation of the phase points around the fit and $r$ is a random number 
extracted from a standard normal distribution.

\begin{figure*}[!htbp]
\centering
\includegraphics[width=5cm]{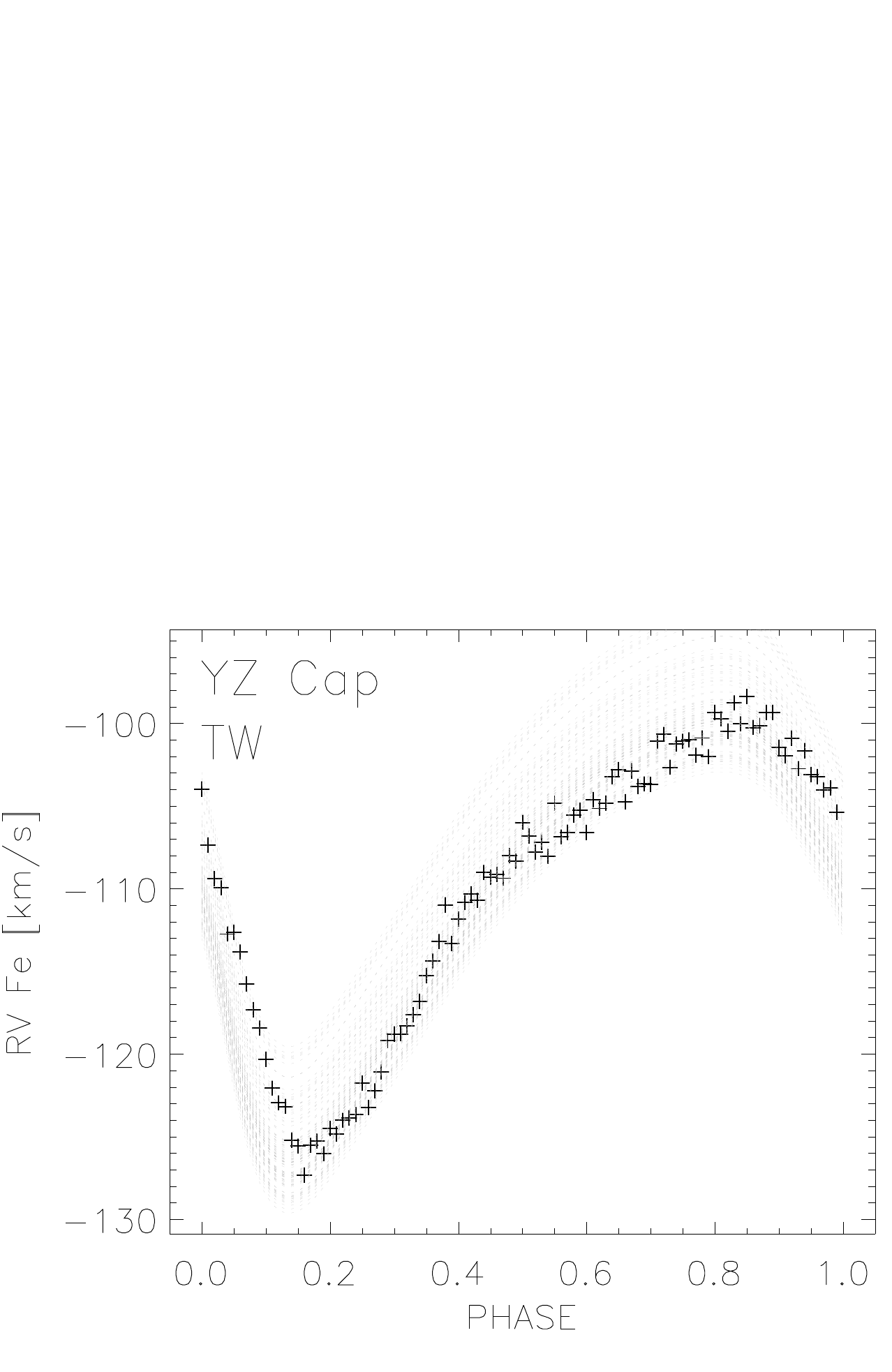}
\includegraphics[width=5cm]{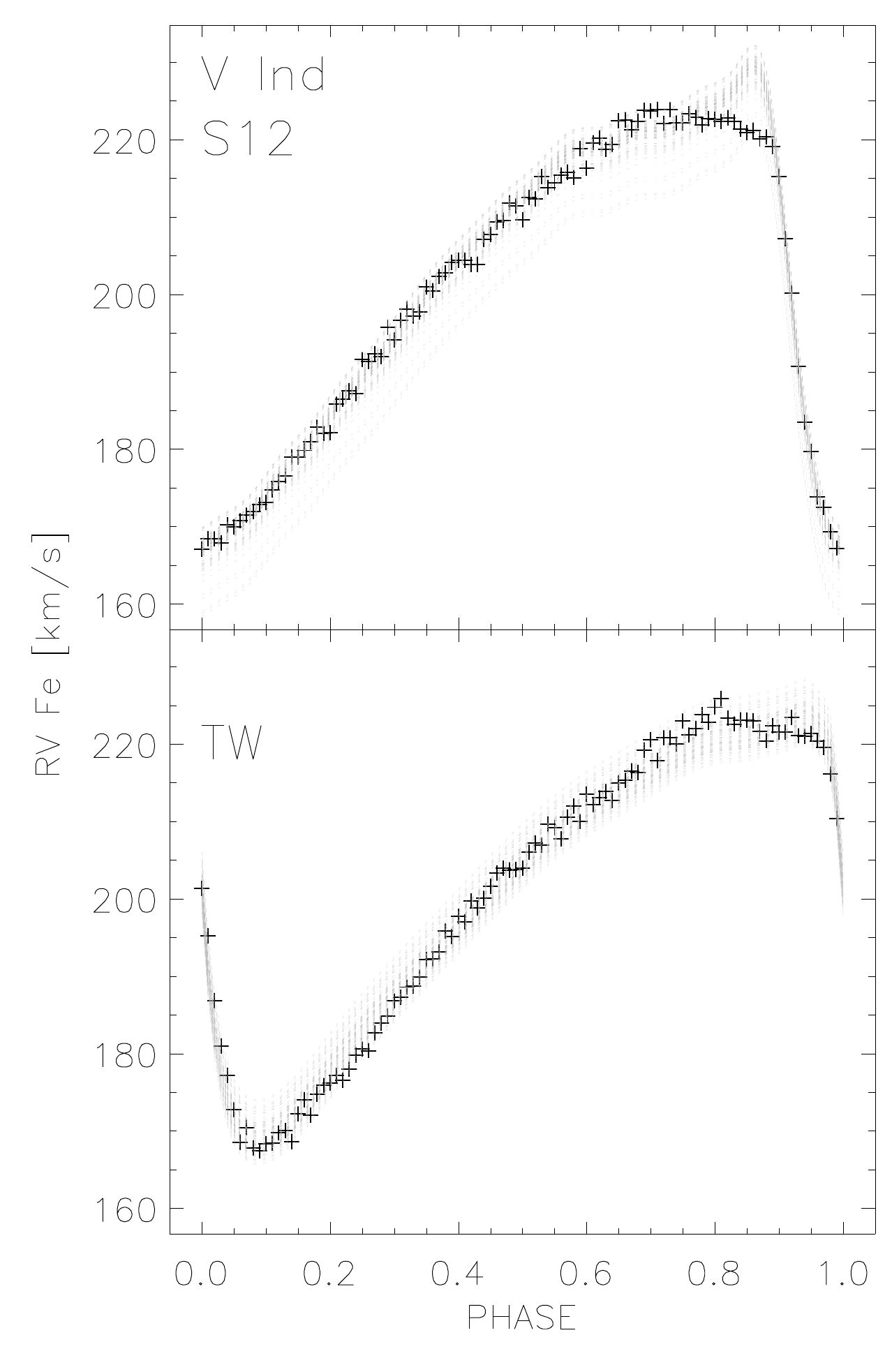}
\includegraphics[width=5cm]{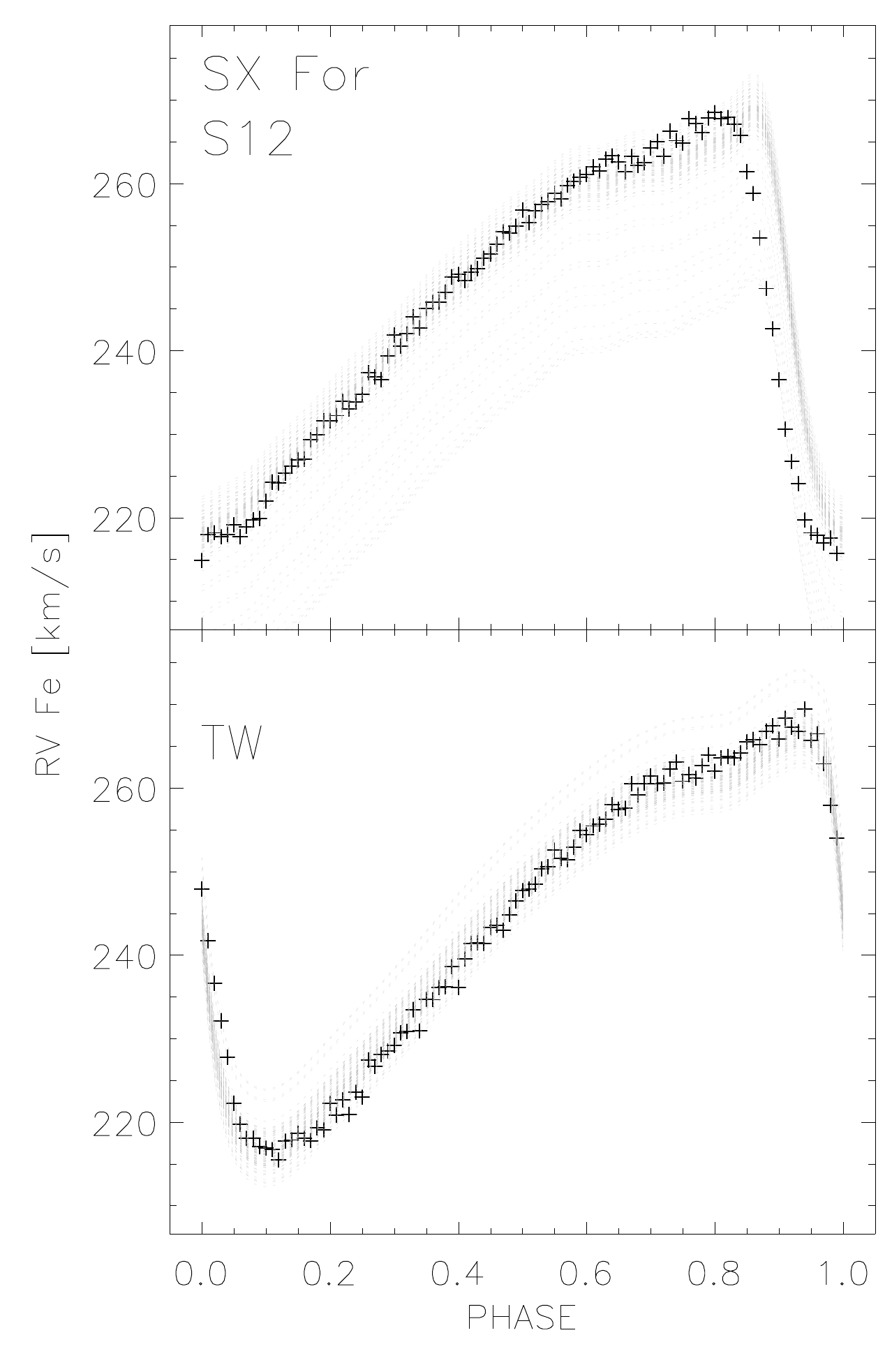}
\caption{Left: Black crosses show the Fe RV measurements that 
we generated from the grid for the RRc variable YZ Cap. Gray dashed lines 
display the RVC templates from this work (labeled as TW) 
associated with each phase point. The ID of the RRL is labeled. 
Note that the comparison with the RVC template provided 
by S12 is missing because YZ Cap is an RRc variable.
Middle: Same as the left, but for the RRab variable V Ind belonging to the 
RRab1 period bin. Top: Radial velocity curves based on the RVC templates
provided by S12. Bottom: same as the top, but based on our RVC templates. 
Right: Same as the middle, but for the RRab variable SX For belonging to the 
RRab2 period bin.}
\label{fig:check_fe1p}
\end{figure*}

We applied the RVC templates on each of the simulated 
points, thus deriving 100 estimates of the systemic velocity ($V_{\gamma(i,j)}^{templ}$)
for each RVC($j$). To provide a quantitative comparison, these estimates were 
performed using both our own and S12 RVC templates. The S12 RVC templates were 
not applied to YZ Cap, since they are valid only for RRab variables. 
In discussing the difference between our own RVC templates and those provided 
by S12, there are three key points worth being mentioned: 
{\em i)}-- the comparison for Fe, Mg and Na RVCs was performed by using the 
generic metallic RVC template from S12, because they do not provide 
specific-element RVC templates; 
{\em ii)}-- the RVC templates by S12 were anchored to \tmaxo, while our own 
were anchored to \tmeano; 
{\em iii)}-- the RVC templates by S12 were rescaled, for internal consistency, 
by using their conversion equations from $Amp(V)$ to $Amp(RV)$ and not our 
own equation for the amplitude ratio.

Note that, due to the paucity of RRab3 variables,
it was not possible to find one with both a reliable estimate of the 
reference epoch and with RV measurements close in epoch to the optical 
light curve. Therefore, the RVC templates for the RRab3 period bin were
not validated with the single phase point approach. Nonetheless, we 
anticipate that we successfully validated the RRab3 RVC templates with 
the three-point approach (see section~\ref{wcen_3punti}). 
Figure~\ref{fig:check_fe1p}, displays the simulated RV(Fe) phase points and 
the RVC templates applied to them. Finally, we derived the offsets 
$\Delta V_{\gamma(i,j)} = V_{\gamma(i,j)}^{templ} - V_{\gamma(j)}^{best}$ 
for each point and each RVC template. 
Table~\ref{tab:validation_1p} gives the median and standard deviations 
of $\Delta V_{\gamma}$ for each RVC template.

\begin{deluxetable*}{l cc cc cc cc cc cc cc}
\tablenum{12}
\tabletypesize{\footnotesize}
\caption{Validation of the RVC templates based on the single phase point approach.}
\label{tab:validation_1p}

\tablehead{  & \multicolumn{2}{c }{$\Delta V_{\gamma}$(Fe) } & \multicolumn{2}{c }{$\Delta V_{\gamma}$(Mg) } & \multicolumn{2}{c }{$\Delta V_{\gamma}$(Na) } & \multicolumn{2}{c }{$\Delta V_{\gamma}$(H$_\alpha$) } & \multicolumn{2}{c }{$\Delta V_{\gamma}$(H$_\beta$) } & \multicolumn{2}{c }{$\Delta V_{\gamma}$(H$_\gamma$) } & \multicolumn{2}{c}{$\Delta V_{\gamma}$(H$_\delta$) } \\
\colhead{Name} & \colhead{mdn} & \colhead{$\sigma$} & \colhead{mdn} & \colhead{$\sigma$} & \colhead{mdn} & \colhead{$\sigma$} & \colhead{mdn} & \colhead{$\sigma$} & \colhead{mdn} & \colhead{$\sigma$} & \colhead{mdn} & \colhead{$\sigma$} & \colhead{mdn} & \colhead{$\sigma$} \\
& \multicolumn{2}{c}{(km/s)}& \multicolumn{2}{c}{(km/s)}& \multicolumn{2}{c}{(km/s)}& \multicolumn{2}{c}{(km/s)}& \multicolumn{2}{c}{(km/s)}& \multicolumn{2}{c}{(km/s)}& \multicolumn{2}{c}{(km/s)}}
\startdata
\multicolumn{15}{c}{---Our RVC templates---}\\
\hline
    YZ Cap &   --0.082 & 2.270 & --0.548 &  2.755 &  --0.970 &   2.600 &  --0.815 & 3.027 &  --1.116 &  3.131 &   0.202 & 2.566 &   0.062 &   2.906  \\
    V Ind  &   --0.301 & 1.925 & --0.173 &  2.271 &    0.469 &   1.729 &  --0.940 & 2.847 &    0.055 &  2.998 &   0.203 & 2.357 & --0.194 &   2.505  \\
    SX For &     0.066 & 2.138 & --0.257 &  2.670 &  --0.241 &   1.887 &    1.063 & 3.181 &    0.527 &  3.110 &   0.725 & 4.293 &   1.007 &   3.347  \\
\hline
\multicolumn{15}{c}{---S12---}\\
\hline
    V Ind  &    0.049  &  2.977 &--0.583  &  3.287 &  0.440  &  3.200 & 0.685  &  4.356 & 0.877  &  6.601 &  0.644 &   6.895 &  \ldots &  \ldots \\
    SX For &    1.488  &  5.418 &  1.115  &  5.583 &  1.467  &  5.634 & 3.863  &  8.812 & 3.820  & 10.926 &  4.154 &  10.899 &  \ldots &  \ldots \\
\enddata
\tablecomments{Medians (mdn) and standard deviations ($\sigma$) of for the $\Delta
V_{\gamma}$ based on a single phase point validation, for both our and S12 RVC 
templates.}
\end{deluxetable*}

We note that that the median of the $\Delta$V  is 
always smaller, in absolute value, than 1.0 km/s and 1.5 km/s  
for metallic and Balmer RVC templates.  In all cases, the standard deviations 
are larger than the residuals, meaning that the latter can be considered 
vanishing within the dispersion. The largest standard deviations are found 
for the H$_\alpha$ and H$_\beta$ RVC templates and progressively decrease when moving to 
H$_\gamma$, H$_\delta$ and metallic lines. 

The comparison between the current and the RVC templates provided by S12 
brings forward that the standard deviations of the former ones are systematically 
smaller by a factor ranging from $\sim$1.5 to $\sim$3. The higher accuracy 
of the current RVC templates is due to an interplay of a more robust 
estimate of T$_{mean}$ with respect to T$_{max}$ and a more accurate 
optical-to-RV amplitude conversion (note e.g., in the upper-right panel 
of Fig.~\ref{fig:check_fe1p}, $Amp(RV)$ is clearly overestimated for 
the S12 RVC template).

\subsection{Multiple phase points}\label{wcen_3punti}

To apply the RVC templates to single phase points, it is necessary to know
four parameters with great accuracy: {\it i)} the pulsation period, 
{\it ii)} the pulsation mode, {\it iii)} $Amp(V)$ and 
{\it iv)} the reference epoch of the anchor point (\tmeano).
The last one is particularly delicate, not only because a good
sampling of the light curve is needed but also because, when the
spectroscopic data were collected several years before/after the 
photometric data, even small phase shifts or period changes can 
affect the phasing of the RV measurements. 
Note that for RRLs affected by large Blazhko modulations, these 
parameters---especially $Amp(V)$ and reference epoch---cannot 
be accurately estimated. Therefore we suggest to avoid the application of the 
templates to RRLs with evident Blazhko modulations.

To overcome this limitation, it is possible to use the RVC templates 
as fitting functions if at least three RV measurements are available. 
In this empirical framework, only three parameters are required in order
to apply the RVC template, namely the pulsation period, the pulsation mode and 
$Amp(V)$. We performed a number of tests by assuming that 
three independent RV measurements were available. In this approach, 
the $Amp(V)$ has to be converted into the $Amp(RV)$ and then perform a 
least-squares fit of the RV measurements by adopting the RVC template as the
fitting function. The minimization of the $\chi^2$ is based on two parameters: 
$\Delta\phi$ (a horizontal shift) and $\Delta V_{\gamma}$ (a vertical shift). 
The function to be minimized is: 

\begin{align}\label{eq_pegasusfit}
P(\phi; \Delta\phi,\Delta V_{\gamma}) = \Delta V_{\gamma} + Amp(RV) \cdot 
\Big(A_0^P + \Sigma_i A_i^P \exp{\Big(-\sin{\Big(\dfrac{\pi (\phi - \phi_i - \Delta\phi)}{\sigma_i^P}\Big)^2}\Big)\Big)}
\end{align}

To simulate three RV measurements, we extracted three random
phases over the pulsation cycle and generated three RV measurements 
following the same approach adopted in Section~\ref{wcen_1punto}. 
To rely on a wide statistical sample, the process was repeated 5000 times 
to generate 5000 different triplets of RV measurements for each 
RVC template. Henceforth, we label each of the triplets with a subscript $k$.

We estimated the systemic velocity $V_{\gamma(k,j)}^{templ}$ by fitting 
each of the triplets with both S12 and our own analytical form of the 
RVC templates. Analogously to Section~\ref{wcen_1punto}, we derived the 
offsets $\Delta V_{\gamma(k,j)}$ and their median and standard deviations 
and they are listed in Table~\ref{tab:validation_3p}.

We note that the uncertainties are larger for the three-point approach
with respect to the single point one discussed in the former section. 
This happens because, when the three points are randomly extracted, 
it may happen that two (or all) of them are too close in phase, and the 
fitting procedure provides less solid estimates. This is a realistic 
case in which one might not have control over the sampling of the 
spectroscopic data. Also, the very fact that we are fitting the shift in 
phase, and not anchoring the RV data to the true reference epoch of the 
variable, means that the phasing of the template is not fixed, and 
this naturally leads to larger uncertainties. We have verified that, 
if the reference epochs were available, we would obtain standard deviations 
that are $\sim$10-30\% smaller and medians that are up to $\sim$50\% 
smaller than in the case of the single-epoch approach.

\begin{deluxetable*}{l cc cc cc cc cc cc cc}
\tablenum{13}
\tabletypesize{\footnotesize}
\tablecaption{Results of the validation for the multiple phase point approach.}
\label{tab:validation_3p}
\tablehead{  & \multicolumn{2}{c }{$\Delta V_{\gamma}$(Fe) } & \multicolumn{2}{c }{$\Delta V_{\gamma}$(Mg) } & \multicolumn{2}{c }{$\Delta V_{\gamma}$(Na) } & \multicolumn{2}{c }{$\Delta V_{\gamma}$(H$_\alpha$) } & \multicolumn{2}{c }{$\Delta V_{\gamma}$(H$_\beta$) } & \multicolumn{2}{c }{$\Delta V_{\gamma}$(H$_\gamma$) } & \multicolumn{2}{c}{$\Delta V_{\gamma}$(H$_\delta$) } \\
\colhead{Name} & \colhead{mdn} & \colhead{$\sigma$} & \colhead{mdn} & \colhead{$\sigma$} & \colhead{mdn} & \colhead{$\sigma$} & \colhead{mdn} & \colhead{$\sigma$} & \colhead{mdn} & \colhead{$\sigma$} & \colhead{mdn} & \colhead{$\sigma$} & \colhead{mdn} & \colhead{$\sigma$}\\
& \multicolumn{2}{c}{(km/s)}& \multicolumn{2}{c}{(km/s)}& \multicolumn{2}{c}{(km/s)}& \multicolumn{2}{c}{(km/s)}& \multicolumn{2}{c}{(km/s)}& \multicolumn{2}{c}{(km/s)}& \multicolumn{2}{c}{(km/s)}}
\startdata
\multicolumn{15}{c}{---Our RVC templates---}\\
\hline
    YZ Cap &  0.028  &  3.049 &  1.682 &   2.857 &  0.873  &  3.666 &  4.147 &   5.813 &  0.742 &   3.863 &--1.702 &   5.413&  4.839 &   4.799 \\   
     V Ind &--0.472  &  5.658 &  0.808 &   4.649 &--0.621  &  6.230 &  6.268 &  17.383 &  3.227 &  11.646 &  2.620 &   9.894&  2.978 &   8.356 \\   
    SX For &  1.436  &  8.031 &  2.943 &   8.068 &  1.840  &  9.490 &  4.373 &  16.630 &  1.700 &  12.225 &  3.050 &  10.995&  4.196 &   9.721 \\   
    AT Ser &  0.219  &  9.340 &  2.311 &  11.129 &  0.363  & 12.087 &--0.583 &  17.990 &--0.844 &  15.311 &--0.371 &  13.432&  4.477 &  12.727 \\   
\hline
\multicolumn{15}{c}{---S12---}\\
\hline
     V Ind & --0.273 &  7.822 &  0.770 &  9.314 & -0.777 &  9.805 & 4.817 & 19.250 &  2.676 & 13.259 &   2.263 & 12.345 & \ldots & \ldots \\
    SX For &   2.297 &  7.673 &  3.953 &  7.502 &  2.437 &  9.102 & 3.988 & 16.418 &  2.207 & 11.638 &   3.522 & 10.763 & \ldots & \ldots \\
    AT Ser &   0.454 &  9.522 &  2.412 & 10.869 &  0.528 & 11.037 & 1.251 & 19.826 &  0.989 & 16.109 &   1.289 & 13.684 & \ldots & \ldots \\
\enddata
\tablecomments{Medians (mdn) and standard deviations ($\sigma$) for the $\Delta
V_{\gamma}$ obtained from the three-points validation, for both our and 
S12 RVC templates.}
\end{deluxetable*}

Note that we did not put any restriction on which RV measurements were chosen
to form a triplet. More specifically, we did not remove triplets in which two phase
points were very close in phase, thus mimicking a realistic situation 
where either only two RV independent measurements were obtained or 
where the points are really close in phase due to the sampling of 
the spectroscopic data. Although this scenario usually leads
to flawed estimates of $V_{\gamma}$, they are a minority, with less than 
$\sim$10\% of the triplets having points closer than 0.05 in phase. 
Moreover, it is not only the difference in phase that matters but also the 
distribution along the pulsation cycle. Indeed, phase points close in phase 
located along the decreasing branch of RRab variables produce larger 
uncertainties when compared with other phase intervals. The decision to keep
even these troublesome triplets in our tests 
allowed the computed errors to take into account
all the possible drawbacks of real observations, including the scenario of
spectroscopic surveys that scan the sky by taking multiple consecutive measurements.

Data listed in Table~\ref{tab:validation_3p} show that, with the only exception 
of the RRab3, for which the standard deviations are similar, our RVC templates 
provide smaller standard deviations of the $\Delta$V compared to S12 templates. 
This is true even for 
the three phase points validation, where the median offsets are smaller than the 
standard deviations. This means that the RVC template provides $V_{\gamma}$ 
estimates that are more accurate than the simple average of the three 
RV measurements.

\section{RR Lyrae in NGC~3201 as a test case}\label{chapt_3201}

The Galactic globular cluster (GGC) NGC~3201 is a nearby \citep[4.7 kpc,][]{vasiliev2021} 
metal intermediate  ([Fe/H]$\sim$ --1.4 \citealt{mucciarelli14,magurno2018}) 
stellar system hosting a sizeable number of RRLs 
\citep[77 RRab, 8 RRc, 1 candidate RRd,][]{piersimoni02,layden2003,arellanoferro2014}
This cluster has a very high RV: 494.5$\pm$0.4 km/s based on 
giant and subgiant stars and 496.47$\pm$0.11 km/s based on turn-off, subgiant and
red giant stars \citep[][respectively]{ferraro2018,wan2021}.  
Note that \citet{magurno2018} adopted the S12 RVC templates and applied them to 
eleven RRLs in NGC~3201 obtaining a cluster RV of 494$\pm$2$\pm$8 km/s which is 
within 1$\sigma$ of literature estimates.
Interestingly, by taking advantage of accurate proper motion measurements based 
on Gaia DR2 \citet{massari2019} suggested that NGC~3201 is likely an object 
accreted during either the Sequoia or the Gaia-Enceladus merger. 

\begin{figure*}[!htbp]
\centering
\includegraphics[width=12cm]{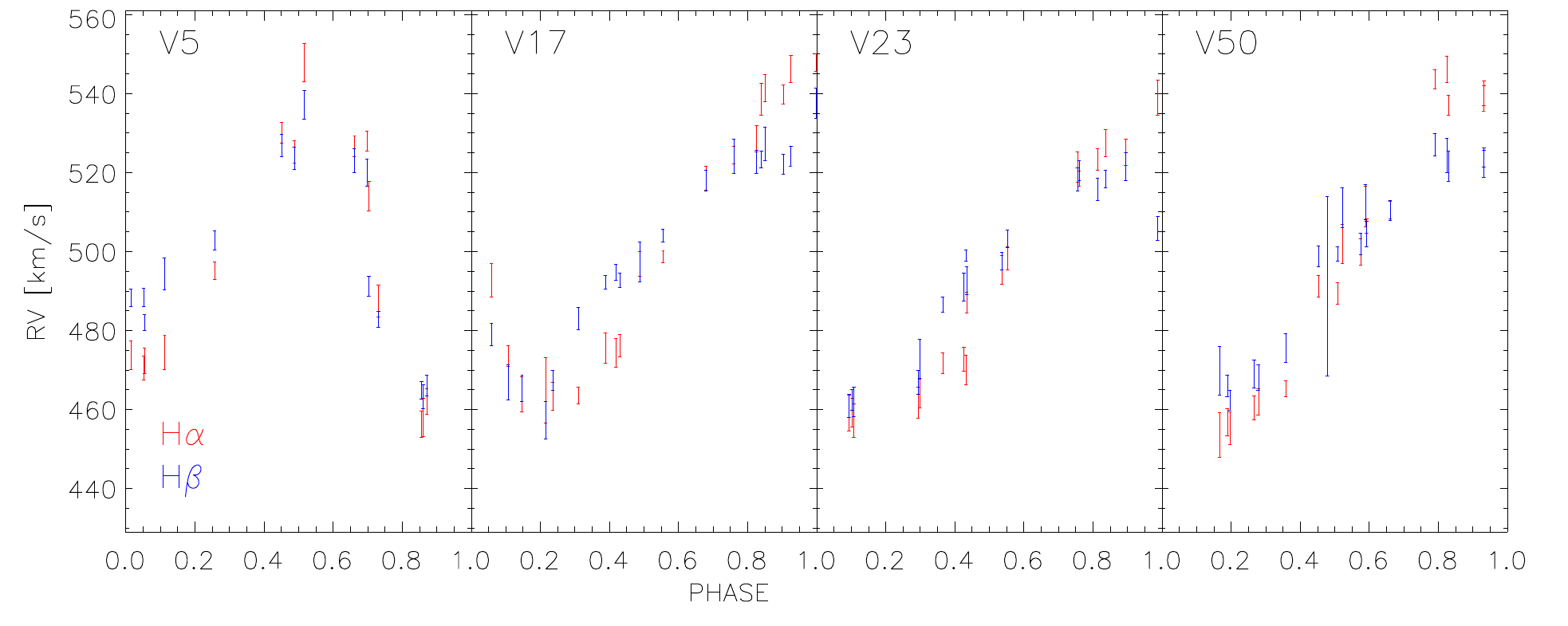}\\
\includegraphics[width=12cm]{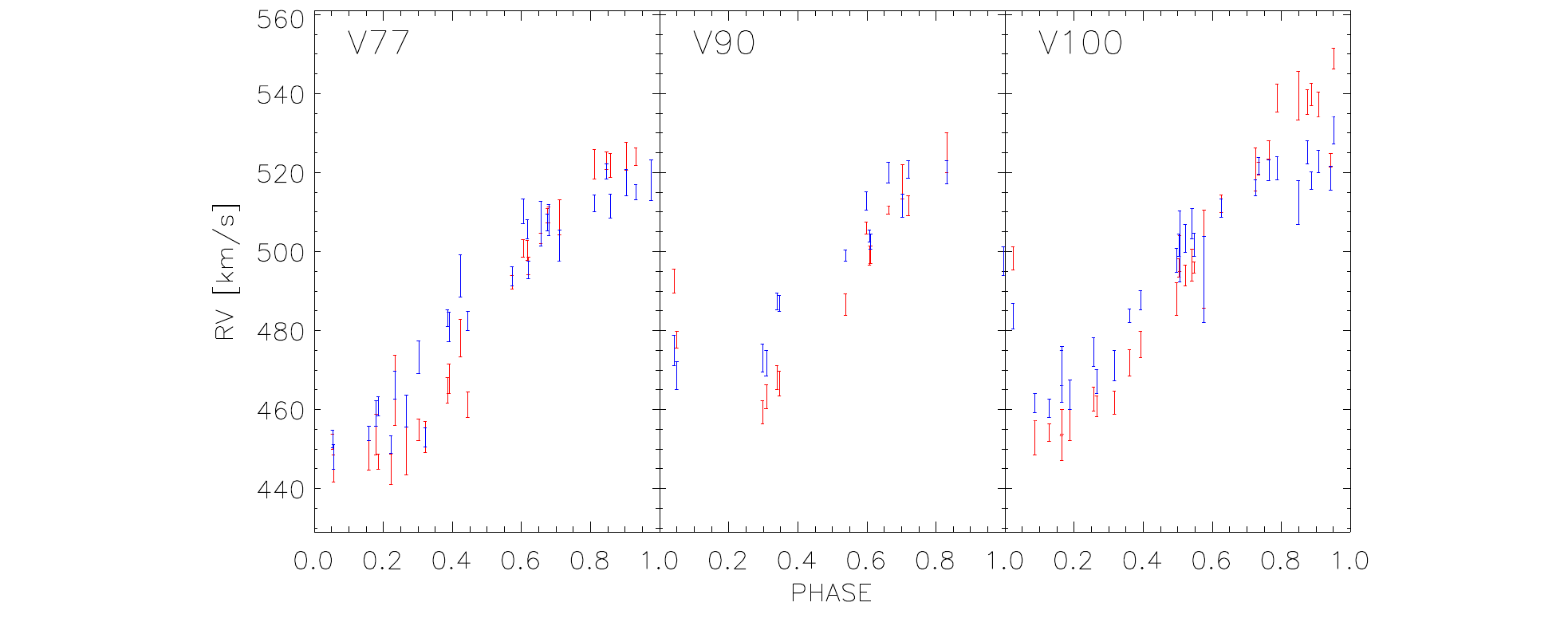}
\caption{Red: H$_\alpha$ RV measurements; Blue: H$_\beta$ RV measurements. The phases
in the figures were derived by adopting the period and the reference epoch
(\tmeano) displayed in Table~\ref{tab:3201}. For V5 and V100, we have
assumed an arbitrary reference epoch, since we could not derive an accurate 
estimate from the optical light curves.}
\label{fig:3201}
\end{figure*}

To investigate on a more quantitative basis the cluster properties, we used  
the MUSE@VLT spectra collected by \citet{giesers2019} for seven cluster RRab 
variables. From these spectra, we measured H$_\alpha$ and H$_\beta$ RVs 
(see data listed in Table~\ref{tab:rvc3201}) and the RVCs are displayed 
in Figure~\ref{fig:3201}.

\begin{deluxetable*}{l c c c c c c c c}
\tablenum{14}
\tabletypesize{\footnotesize}
\tablecaption{Properties of the RRLs observed by MUSE in NGC~3201}
\label{tab:3201}
\tablehead{ \colhead{Name} & \colhead{Type} & \colhead{Period} & \colhead{$V$} & \colhead{$Amp(V)$} & \colhead{\tmeano} & \colhead{\tmaxo} & \colhead{$V_{\gamma}$(H$_\alpha$)} & \colhead{$V_{\gamma}$(H$_\beta$)} \\
  & & (days) & \multicolumn{2}{c}{(mag)} & \multicolumn{2}{c}{HJD-2,400,000 (days)} & \multicolumn{2}{c}{(km/s)} }
\startdata
  V5 & RRab & 0.501527 & 14.786 & \ldots&  \ldots     &  \ldots     & 497.9$\pm$3.5 & 500.1$\pm$2.8 \\
 V17 & RRab & 0.565590 & 14.794 & 0.824 & 56374.33072 & 56373.80043 & 501.0$\pm$3.3 & 501.6$\pm$3.0 \\
 V23 & RRab & 0.586775 & 14.795 & 0.742 & 56374.44344 & 56374.50391 & 496.4$\pm$3.8 & 493.1$\pm$3.0 \\
 V50 & RRab & 0.542178 & 14.795 & 0.876 & 56374.15802 & 56374.19955 & 499.0$\pm$3.2 & 498.6$\pm$3.4 \\
 V77 & RRab & 0.567644 & 14.672 & 0.829 & 56374.05485 & 56374.10073 & 483.2$\pm$3.3 & 488.0$\pm$3.2 \\
 V90 & RRab & 0.606105 & 14.706 & 1.048 & 56374.21846 & 56374.27385 & 489.8$\pm$3.1 & 493.0$\pm$3.0 \\
V100 & RRab & 0.548920 & 14.786 & 0.863 & \ldots      &  \ldots     & 495.6$\pm$3.2 & 496.1$\pm$3.1 \\
\enddata
\end{deluxetable*}

\begin{deluxetable*}{r c c r r}
\tablenum{15}
\setlength{\tabcolsep}{4pt}
\tabletypesize{\scriptsize}
\tablecaption{Radial velocity measurements for RRLs in NGC~3201.}
\label{tab:rvc3201}
\tablehead{ \colhead{Name} & \colhead{Species} & \colhead{HJD} & \colhead{RV\tablenotemark{a}} & \colhead{eRV\tablenotemark{b}} \\
  & & (days) & \multicolumn{2}{c}{(km/s)}}
\startdata
  V5 &     Halpha & 2456978.8484 &  487.562 &    4.023 \\
  V5 &     Halpha & 2456989.8680 &  513.958 &    3.719 \\
  V5 &     Halpha & 2457008.8324 &  547.947 &    4.861 \\
  V5 &     Halpha & 2457009.8025 &  530.079 &    2.620 \\
  V5 &     Halpha & 2457131.4746 &  472.353 &    3.208 \\
  V5 &     Halpha & 2457134.4825 &  470.448 &    2.996 \\
  V5 &     Halpha & 2457138.4761 &  473.754 &    3.604 \\
  V5 &     Halpha & 2457419.7608 &  462.131 &    3.225 \\
  V5 &     Halpha & 2457421.7614 &  458.032 &    4.901 \\
  V5 &     Halpha & 2457787.8737 &  456.275 &    3.388 \\
\enddata
\tablecomments{Only ten lines are listed. The machine-readable version of this 
table is available online on the CDS.\\
\tablenotetext{a}{Velocity plus heliocentric velocity and diurnal velocity correction.}
\tablenotetext{b}{Uncertainty on the RV measurements. For the Balmer lines, 
it is the uncertainty from spectroscopic data reduction. For Fe, Mg and Na, 
it is the standard deviation of the RV measurements from different lines.} }
\end{deluxetable*}

Unfortunately, the optical light curves that we adopted from \citet{arellanoferro2014} do not
fully cover the pulsation for all the cluster RRLs. More specifically, we could not derive a reliable
estimate of the reference epochs for V5 and V100. Due to this, we are not able to apply the templates
to these two stars. In principle, we could apply the three-point method, but we wanted to keep 
our V$_\gamma$ estimates as homogeneous as possible.

After deriving---with the PLOESS algorithm---the best 
estimate for the systemic velocity ($V_{\gamma(best)}$, displayed in 
Table~\ref{tab:3201}) by fitting the RVCs, we applied the RVC template by using the single phase
point approach. Note that, in this case, we do not generate a grid of synthetic points, 
but we simply extract the points one by one from the RVC. Table~\ref{tab:3201templ} 
shows the average $V_{\gamma(templ)}$ estimates for all the lines, together with the
standard deviation and the median of the offsets. The results are similar to those found in 
Section~\ref{wcen_1punto}, meaning that all the medians are smaller than the standard deviations
and, on average, the standard deviations obtained by using our own RVC templates are smaller than
those coming from the use of the S12 RVC templates.

Our final estimate for the velocity of the whole system, by using both H$_{\alpha}$ and 
H$_{\beta}$ is 496.89$\pm$8.37 (error) $\pm$3.43 (standard deviation) km/s by using the RVC templates
and 495.21$\pm$3.23 (error) $\pm$4.32 (standard deviation) km/s by simply fitting the RVCs, which
agree quite well, within 1$\sigma$, with similar estimates available in the literature.




\begin{deluxetable*}{l ccc ccc}
\tablenum{16}
\tablecaption{Systemic velocities of the RRLs in NGC~3201 based on the RVC templates.}
\label{tab:3201templ}
\tabletypesize{\footnotesize}
\tablehead{  & \multicolumn{3}{c}{H$_\alpha$} & \multicolumn{3}{c}{H$_\beta$} \\
 \colhead{Name} & \colhead{$V_{\gamma(templ)}$} & \colhead{$\sigma V_{\gamma(templ)}$} & \colhead{mdn$\Delta$V}  & \colhead{$V_{\gamma(templ)}$} & \colhead{$\sigma V_{\gamma(templ)}$} & \colhead{mdn$\Delta$V} \\
  & \multicolumn{3}{c}{(km/s)} & \multicolumn{3}{c}{(km/s)}}
\startdata
& \multicolumn{6}{c}{---Our RVC templates---}\\
V17 & 500.4 &  6.0 & --0.6 & 499.4 &  5.6 & --2.3 \\
V23 & 495.7 &  6.1 & --0.6 & 498.1 &  8.8 &   4.9 \\
V50 & 498.0 &  6.3 & --1.0 & 498.0 &  5.5 & --0.6 \\
V77 & 486.9 & 12.2 &   3.6 & 489.4 &  8.0 &   1.4 \\
V90 & 491.1 & 10.3 &   1.3 & 494.8 & 10.7 &   1.8 \\
& \multicolumn{6}{c}{---S12---}\\
V17 & 496.2 &  15.3 & --4.8 & 498.5 &   9.4 & --3.2 \\
V23 & 498.6 &   9.2 &   2.2 & 497.5 &  10.2 &   4.3 \\
V50 & 499.3 &   5.3 &   0.3 & 497.1 &   7.3 & --1.5 \\
V77 & 488.3 &  12.1 &   5.1 & 488.3 &  11.7 &   0.3 \\
V90 & 492.3 &  18.3 &   2.5 & 494.1 &  22.3 &   1.1 \\
\enddata
\end{deluxetable*}

\section{Summary and final remarks}\label{chapt_discussion}

We provide accurate and homogeneous RVC templates for 
both RRab and RRc variables using for the first time spectroscopic 
diagnostics based on well-identified metallic (Fe, Mg, Na) lines and 
on Balmer (H$_\alpha$, H$_\beta$, H$_\gamma$, H$_\delta$) lines. 
In the 
following we discuss the approach adopted to construct the 
RVC templates and summarize the most relevant results.

{\it $V$-band light curve templates ---} To demonstrate on a quantitative 
basis the difference between the reference epoch anchored to the maximum 
brightness and to the mean magnitude along the rising branch, we collected 
homogeneous $V$-band photometry for cluster (\wcen, M4) and field RRLs. 
We have grouped them into four period bins (the same that we have used for 
the RVC templates) and found that the anchoring to the epoch of the mean 
magnitude on the rising branch (\tmeano) provides smaller standard deviations
on the light curve templates than the anchoring to the maximum brightness
(\tmaxo). The decrease is of the order of 35\% for the period bins with cuspy 
light curves (RRab1, RRab2) and of 45\% for those with flat-toopped light 
curves. These finding strongly supports the results obtained by \citet{inno15} 
and by \citet{braga2019} in using \tmeano~ to phase the NIR light curves of 
both classical Cepheids and RRLs.

{\it Spectroscopic catalog ---} In this work, we present 
the largest collection of RRL spectra ever compiled in the literature. Altogether, we 
collected 23,865 spectra for 7,070 RRab and 3,343 RRc variables. 
These measurements were secured using eleven different spectrographs, 
ranging from low (2,000) to very high spectral resolution (115,000). To build 
the RVC templates, the most important dataset is the du~Pont (see Table~\ref{tab:spectra}). 
Spectroscopic observations at this telescope were specifically planned to cover  
the entire pulsation cycle of several bright RRLs ($<V> \sim$10-11 mag
for the majority of them). We also collected RV measurements and $V$-band 
time series from the literature (the Baade-Wesselink [BW] sample), which 
were crucial to investigate the reference epoch and the ratio between RV 
and optical amplitudes.

{\it Amplitude ratio ---} To apply the RVC templates, it is necessary to have
prior knowledge on the optical amplitude of the variable, to correctly 
rescale the RVC template itself and to optimally match the true RVC. We  
provide new equations for the ratio of $Amp(RV)$ over $Amp(V)$. 
Those available in the literature (S12), for RVCs based on metallic 
and Balmer (H$_\alpha$, H$_\beta$, H$_\gamma$) lines, were constructed 
using six RRab variables covering a very narrow range in pulsation period 
(0.56-0.59 days). In this investigation we provide different RVC templates 
for both RRc and RRab variables based on metallic and Balmer (H$_\alpha$, 
H$_\beta$, H$_\gamma$, H$_\delta$) lines. Even more importantly, 
our relations are based on three dozen variables covering a wide range in 
pulsation periods (0.27-0.83 days) and metal content (--2.6$\lesssim$[Fe/H]$\lesssim$0.0).

{\it Reference epoch ---} When applying the RVC template to single 
RV measurements, it is necessary to anchor the RVC template at the same epoch 
of the observations. The RVC templates are applied to RRLs with well sampled 
optical light curves and a few spectroscopic measurements. Therefore, the only 
pragmatical possibility to phase the spectroscopic data is to derive a reference 
epoch from the light curve. By using the light curves and the RV curves of our 
BW sample, we demonstrated that \tmeano~ is, within 5\% of the pulsation cycle, 
identical to the time of mean velocity on the decreasing branch of the Fe RVC 
(\tmeanvfe). This means that RVC templates based on Fe, Mg and Na lines can be 
safely anchored to \tmeanvfe, and this approach does not introduce any 
systematic error when \tmeano~ is adopted to apply the RVC template. 
We also found that RVCs based on Balmer lines display a well defined 
phase lag across the phases of mean RV. Therefore, we decided 
to anchor the current H$_\alpha$, H$_\beta$, H$_\gamma$ and H$_\delta$ 
RVC templates to the time of mean velocity on the decreasing branch of 
the H$_\beta$ RVC (\tmeanvhb), taken as representative of the Balmer lines.  
Additionally, we found a clear trend in the phase difference between \tmeanvfe~ 
and \tmeanvhb~ as a function of the pulsation period. The new analytical 
relation gives the phase difference required to use the Balmer 
RVC templates, because \tmeano~ does not match \tmeanvhb.

To discuss the concerning pros and cons of the reference epochs 
anchored to \tmeanv~ and to \tminv~ into a more quantitative framework, we 
performed a series of numerical experiments. Interestingly enough, we found 
that RVC templates based on metallic RV measurements and anchored to 
$\tau_0$=\tmeanv~ provide $V_{\gamma}$ that are on average 
$\sim$30\% more accurate than the same RVCs anchored to $\tau_0$=\tminv.
Even more importantly, we found that the improvement 
in using \tmeanv~ compared with \tminv~ becomes even more relevant in 
dealing with RVCs based on Balmer lines. Indeed, the uncertainties decrease 
up to a factor of three (see Section~\ref{chapt_validation}). 
The current circumstantial evidence further supports not only 
the use of two independent reference epochs for metallic and 
Balmer lines, but also the use of \tmeanv~ as the optimal reference 
epoch to construct RVC templates.

Finally, we investigated the correlation in phase between 
\tmeano~ ($\Phi_{mean}$) and \tmaxo~ ($\Phi_{max}$). This search 
was motivated by the fact that the current photometric surveys only 
provide reference epochs anchored to the time of maximum light (\tmaxo). 
We found that the difference between the two 
($\Delta \Phi_{max-min} =  \Phi_{max} - \Phi_{min}$) is 
constant---within the dispersion---for RRc variables and it follows 
a linear trend for RRab variables. We provide these relations for those 
interested in using the current RVC templates to RRLs for which only 
\tmaxo~ reference epochs are available.

{\it Radial velocity curve templates ---} We provide a total of 28 RVC 
templates of RRLs: these are divided into four different period bins 
(one for the RRc and three for the RRab variables) and seven diagnostics (Fe, Mg, Na, 
H$_\alpha$, H$_\beta$, H$_\gamma$ and H$_\delta$). The analytical form of the 
templates is provided in the form of a PEGASUS series (fifth to ninth order).
Analytical fits based on PEGASUS series, when compared with Fourier series, 
do not display spurious ripples when there are neither gaps in the phase coverage 
nor in the case of uneven sampling. The RVC templates have intrinsic
dispersions that lead to errors smaller than 10 km/s in the worst case (H$_\alpha$ 
and H$_\beta$ for high amplitude RRLs) and one order of magnitude smaller for 
the RVC templates with the smaller intrinsic dispersion (metallic lines RVC templates).
To maximize the reach of the results of this work, we provide, in 
Appendix~\ref{chapt_howto}, clear instructions
on how to apply the RVC templates in different real-life observational scenarios.

{\it Template validation ---} To validate the current RVC templates we performed 
a detailed comparison with the RVC templates provided by S12 for RRab variables and 
based on metallic and Balmer (H$_\alpha$, H$_\beta$, H$_\gamma$) RVCs.
We performed these tests on a sub-sample of variables that were used to build 
the RVC template (YZ Cap, V Ind, W Tuc, AT Ser). The validation process was performed 
using both a single phase point approach, where the knowledge of the reference epoch
is mandatory, and with a three phase points approach, where the RVC template is used 
as a fitting function. We found that the median offset of the $V_{\gamma}$ estimates
from the RVC templates is always smaller than 1.5 km/s (one point approach) and 6 km/s
(three point approach). The medians are smaller than the standard deviations, 
meaning that systematic errors are negligible with respect to the statistical errors. 
We also found that our RVC templates provide $V_{\gamma}$ estimates that 
have a dispersion smaller by a factor of 1.5-3 than those based on the RVC templates 
provided by S12.

{\it RRLs in NGC~3201 ---} We reduced the MUSE spectra already presented in \citet{giesers2019}
and obtained H$_\alpha$ and H$_\beta$ RVCs. We derived V$_{\gamma}$ both by 
fitting the RVC and by extracting the measurements one by one, and by adopting the RVC templates.
Their results based on these RV measurements are very similar to those 
of the validation process, with offsets smaller than 6 km/s and standard deviations 
that are smaller than those on the S12 RVC templates. Our estimate of the $V_{\gamma}$
of the cluster is 496.9$\pm$8.4 (error) $\pm$3.4 (standard deviation) km/s from the 
RVC templates and 495.2$\pm$3.2 (error) $\pm$4.3 (standard deviation) km/s from the 
fit with the RVC templates. They both agree, within 1$\sigma$, with literature estimates.

In the next coming years the ongoing (OGLE, ASAS-SN, ZTF, sky-mapper, PanSTARRS) 
and near future (VRO) ground-based and space-based (Gaia, WISE, WFIRST) observing facilities 
will provide a complete census of evolved variable stars associated with old stellar
tracers in the Milky Way and in Local Group galaxies. This will open new paths in the 
analysis of the early formation and evolution of the Galactic spheroid. However, 
firm constraints on the formation mechanism, namely, the dissipative collapse \citep{eggen1962}, 
dissipation-less mechanism \citep{searle78} and Cold Dark Matter 
cosmological models \citep{monachesi2019} require detailed information concerning 
the kinematics and the metallicity distribution of the adopted stellar tracers. 
This is the approach already adopted to fully characterize the stellar streams and the 
merging history of the Galactic Halo \citep{helmi2018,vasiliev2019,prudil2021}.

Upcoming and ongoing low- 
(LAMOST-LR, \citealt{lamost}; SDSS \citealt{sdss_dr15}), 
medium- (4MOST, \citealt{4most}; SEGUE, \citealt{yanny2009}; GALAH,
\citealt{galah_dr2}; HERBS, \citealt{duong2019}; LAMOST-MR, \citealt{lamost};
RAVE, \citealt{steinmetz2006}; WEAVE, \citealt{weave}) and 
high-resolution (APOGEE, \citealt{apogee}; MOONS, \citealt{moons}; 
PFS, \citealt{pfs}) spectroscopic surveys will provide 
a wealth of new data for large samples of dwarf and giant field stars. 
In this context, old variable stars play a crucial role because their 
individual distances can be estimated with a precision of the order of 3-5\% 
within the Local Group. Recent spectroscopic investigations based on high 
resolution spectroscopy \citep{for11,sneden2017,crestani2021a}
indicate that detailed abundance analysis can be performed with spectra collected 
at random phases. Unfortunately, the typical limiting magnitudes for spectroscopic investigations 
are roughly five magnitudes brighter than photometric ones, with current spectrographs 
available at the 8-10m class telescope allowing us to reach 
limiting magnitudes of the order of V$\sim$20-21 mag. However, even with the usage of large
telescopes, the estimate of V$_{\gamma}$ is 
time consuming, because it requires a spectroscopic time series of
at least a dozen points. The 
RVC templates developed in this investigation provide new solid 
diagnostics to provide accurate V$_{\gamma}$ determinations by using 
a small number (three or less) of RV measurements, based on low-resolution spectra.  
Highly accurate estimates of line-of-sight velocities of stream stars are imperative
for constraining the dark matter distribution \citep{bonaca2018}; these RVC 
templates will provide just that for the numerous RRL harbored by MW streams.
The current diagnostics are focused on spectroscopic features located either 
in the blue or in the visual wavelength regime. A further extension into the 
red and the near-infrared regime is mandatory to fully exploit the most 
advanced space ({\it Gaia}, \citealt{gaia_alldr}; JWST, \citealt{jwst_main}; 
{\it Roman} \citealt{wfirst1}) and ground-based (GMT, ELT, TMT) observing 
facilities.

\section*{Acknowledgements}
Based on observations made with ESO Telescopes at the La Silla Paranal Observatory under ESO programmes 
0100.D-0339, 0101.D-0697, 0102.D-0281, 076.B-0055, 077.B-0359, 077.D-0633, 079.A-9015, 079.D-0262, 
079.D-0462, 079.D-0567, 082.C-0617, 083.B-0281, 083.C-0244, 094.B-0409, 095.B-0744, 097.A-9032, 098.D-0230, 
189.B-0925, 267.C-5719, 297.D-5047, 67.D-0321, 67.D-0554, 69.C-0423, 71.C-0097, 0100.D-0273, 083.C-0244, 098.D-0230,
095.D-0629, 096.D-0175, 097.D-0295, 098.D-0148, 0100.D-0161, 0101.D-0268, 0102.D0270, and 0103.D-0204.

Based on observations made with the Italian Telescopio Nazionale Galileo (TNG) operated on the island of La Palma 
by the Fundaci\'{o}n Galileo Galilei of the INAF (Istituto Nazionale di Astrofisica) at the Spanish Observatorio del 
Roque de los Muchachos of the Instituto de Astrofisica de Canarias.

Some of the observations reported in this paper were obtained with the Southern African Large Telescope (SALT).

Funding for the SDSS and SDSS-II has been provided by the Alfred P. Sloan Foundation,
the Participating Institutions, the National Science Foundation, the U.S. Department of
Energy, the National Aeronautics and Space Administration, the Japanese Monbukagakusho,
the Max Planck Society, and the Higher Education Funding Council for England. The SDSS
Web Site is \url{http://www.sdss.org/.}
The SDSS is managed by the Astrophysical Research Consortium for the Participating
Institutions. The Participating Institutions are the American Museum of Natural History,
Astrophysical Institute Potsdam, University of Basel, University of Cambridge, Case Western
Reserve University, University of Chicago, Drexel University, Fermilab, the Institute for Advanced Study, 
the Japan Participation Group, Johns Hopkins University, the Joint Institute
for Nuclear Astrophysics, the Kavli Institute for Particle Astrophysics and Cosmology, the
Korean Scientist Group, the Chinese Academy of Sciences (LAMOST), Los Alamos National
Laboratory, the Max-Planck-Institute for Astronomy (MPIA), the Max-Planck-Institute for
Astrophysics (MPA), New Mexico State University, Ohio State University, University of
Pittsburgh, University of Portsmouth, Princeton University, the United States Naval Observatory, 
and the University of Washington.

V.F.B. acknowledges the financial support of the Istituto Nazionale di 
Astrofisica (INAF), Osservatorio Astronomico di Roma, and 
Agenzia Spaziale Italiana (ASI) under contract to INAF: 
ASI 2014-049-R.0 dedicated to SSDC.

We acknowledge financial support from US NSF under Grants AST-1714534 (M.M., J.P.M.) and AST1616040 (C.S.). 
B.L., Z.P., and H.L. were supported by the Deutsche Forschungsgemeinschaft 
(DFG, German Research Foundation) - Project-ID 138713538 - SFB 881 
(``The Milky Way System'', subprojects A03, A05, A11). E.V. acknowledges 
the Excellence Cluster ORIGINS Funded by the Deutsche Forschungsgemeinschaft (DFG, German Research Foundation) 
under Germany's Excellence Strategy \-- EXC \-- 2094 \--390783311. Z.P. gratefully acknowledges funding 
by the Deutsche Forschungsgemeinschaft (DFG, German Research Foundation) -- Project-ID 138713538 -- SFB 881 
(``The Milky Way System''), subprojects A03, A05, A11.
G.F. has been supported by the Futuro in Ricerca 2013 (grant RBFR13J716).   
M.L. acknowledges funding from the Deutsche Forschungsgemeinschaft (grant DR 281/35-1).
N.M. is grateful to JSPS Japan Society for the Promotion of Science (JSPS)
for the Grant-in-Aid for Scientific Research, KAKENHI 18H01248.

\begin{appendix}

\section{Amplitude ratios}\label{chapt_amplratio}

The fundamental piece of information that is needed to use the RVC template 
RVCs is the relation of $Amp(RV)$ vs $Amp(V)$. 
These relations should be adopted to re-scale the normalized 
templates to the true $Amp(RV)$ of the star, with the 
knowledge of $Amp(V)$. The leading
argument is always that, since there is far more 
availability of light curves than of RVCs, the
amplitudes of the former should be adopted to derive 
those of the latter.

For their RVC templates, S12
adopted $Amp(RV)$ vs $Amp(V)$ relations
based on six RRLs for metallic lines, H$_{\alpha}$, 
H$_{\beta}$ and H$_{\gamma}$. We provide 
the same relations, but separating RRc and RRab variables.
We have considered the use of $Amp(RV)$ vs $Amp([3.6])$ 
relations, since the [3.6] band traces radius variations and 
is not affected by temperature variations as the $V$ band.
However, there are two arguments against this 
option: $i)$--- there is paucity of MIR light 
curves respect to optical light curves; $ii)$--- There is
mounting empirical evidence that amplitude ratios 
using NIR and MIR bands follow a mild trend with 
metallicity \citep{mullen2021}.

\begin{figure*}[htbp]
\centering
\includegraphics[width=8cm]{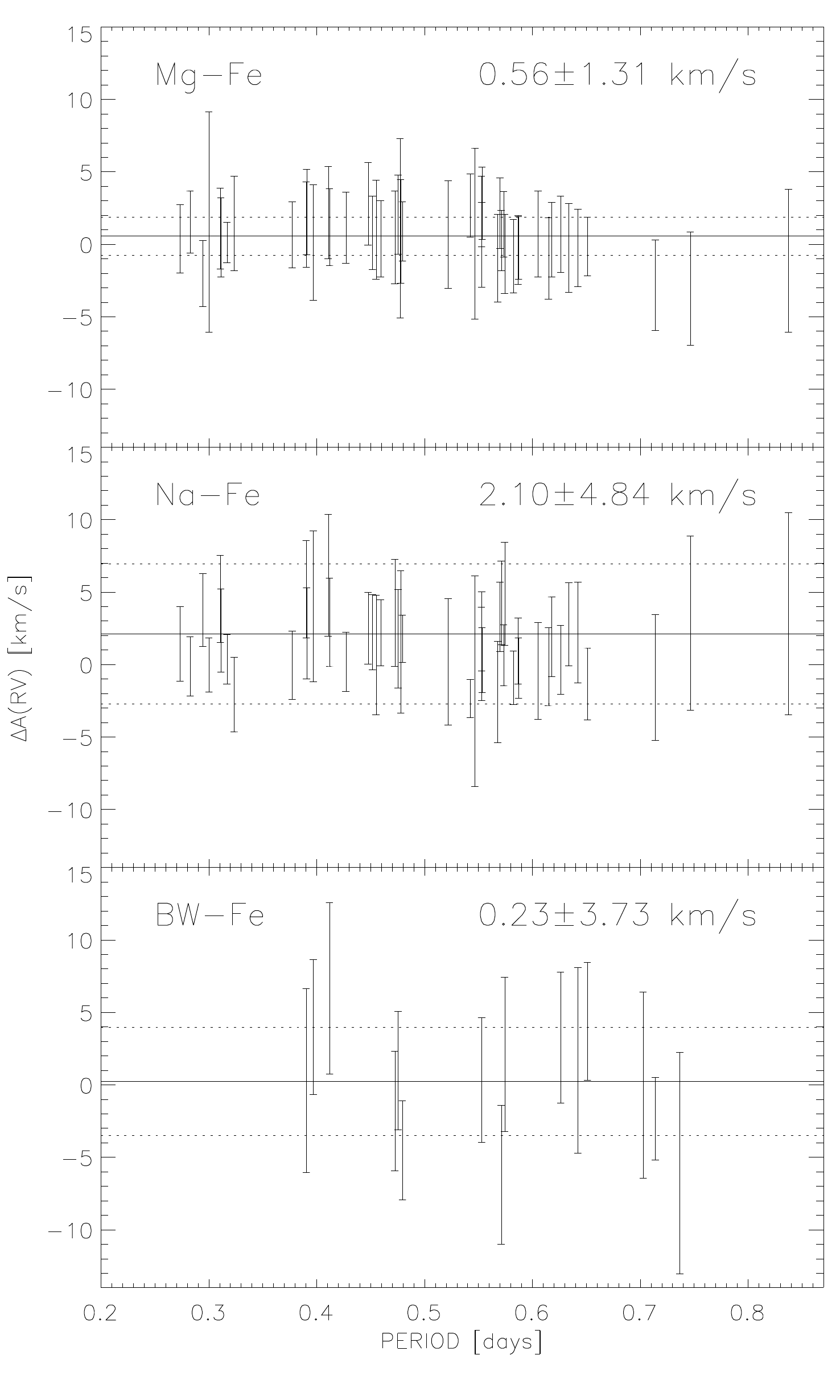}
\includegraphics[width=8cm]{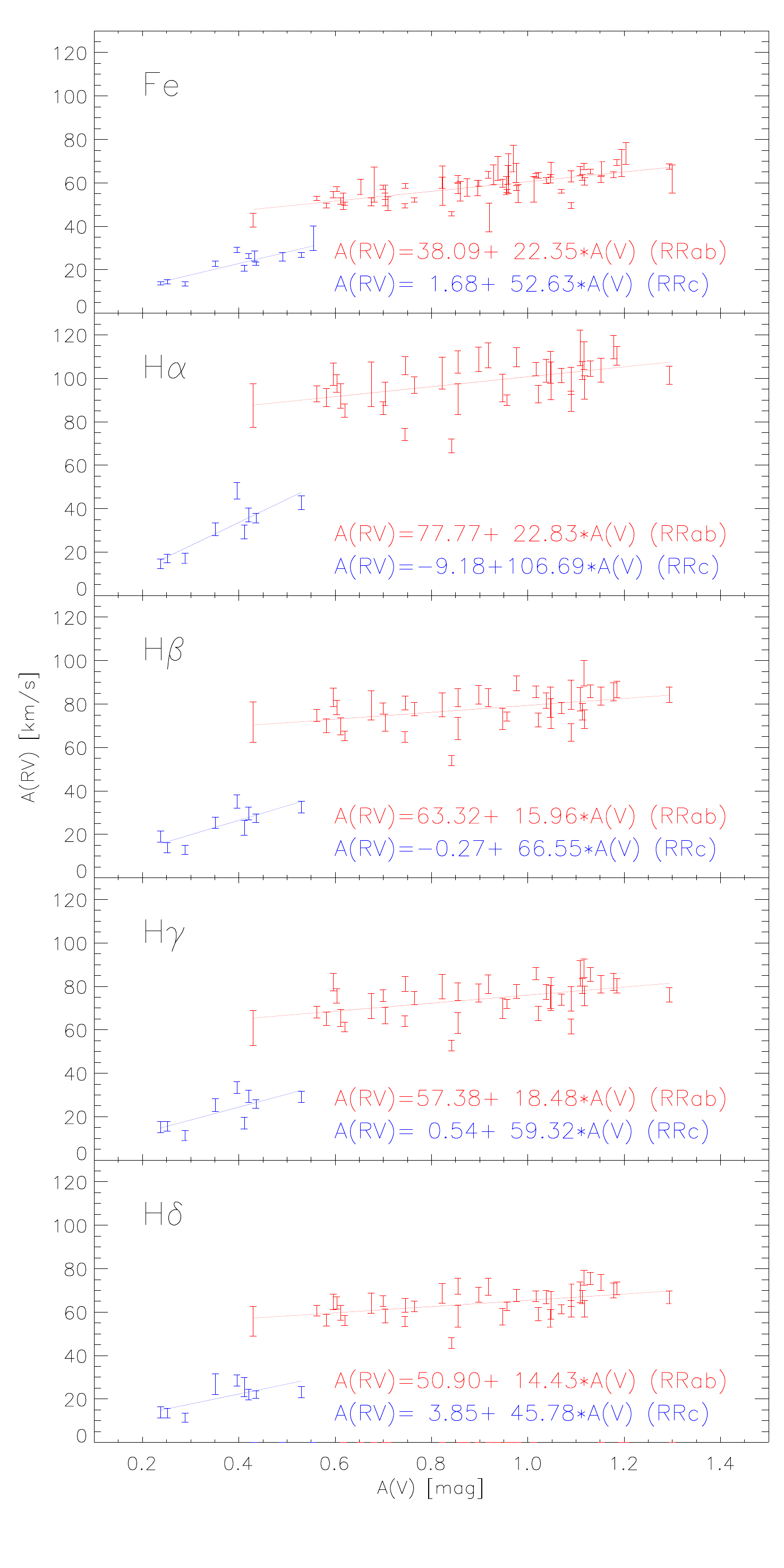}
\caption{Left-top: difference between radial velocity amplitudes 
based on Mg and Fe group RVC templates. The mean difference is labeled 
at the top-right corner. Left-Middle: same as left-top, but for Na and Fe group 
RVC templates. Left-Bottom: same as the left-top, but between the BW sample  and the 
Fe group RVC templates. 
Right panels: From top to bottom, visual amplitude ($Amp(V)$) versus radial 
velocity amplitudes ($Amp(RV)$) based on  Fe, H$_\alpha$, H$_\beta$, H$_\gamma$ 
and H$_\delta$ lines. RRab and RRc variables are marked with red and blue 
symbols. Uncertainties on $Amp(RV)$ are displayed. Solid lines show the
linear relations between $Amp(RV)$ and $Amp(V)$. The analytical form of 
the relations are labeled at the bottom.}
\label{fig:ampldiff}
\end{figure*}

Before deriving the relation between $Amp(RV)$ and $Amp(V)$, we 
checked whether we could assume the Fe, Mg and Na amplitudes 
as equivalent. Figure~\ref{fig:ampldiff} shows the residuals of 
Na and Mg $Amp(RV)$ versus the 
Fe $Amp(RV)$. It is clear that there is no trend with period 
and also the offset is well within the $Amp(RV)$ uncertainty. This
means that we can provide only two $Amp(RV)$ vs $Amp(V)$
relations (one for RRc and one for RRab variables) that hold
for both Fe, Mg and Na. We also collected RV curves 
of RRLs from literature, all derived from 
metallic lines (BW sample, see caption of Table~\ref{tbl:deltaphase}).
We found that the BW $Amp(RV)$ displays a very small difference
with our Fe $Amp(RV)$ (0.23$\pm$3.73 km/s). This allowed us to 
merge the two sets of $Amp(RV)$ and derive a more solid relation 
for the metallic $Amp(RV)$, based on a larger number of 
objects (12 RRc and 60 RRab).

The right panels of Figure~\ref{fig:ampldiff} display the trend of 
Fe and of the Balmer $Amp(RV)$ with $Amp(V)$. A steady increase of the 
slopes for the Balmer $Amp(RV)$, from $\delta$ to $\alpha$ is clear enough,
both visually in the figure and quantitatively from the coefficients 
listed in Table~\ref{tab:amplratio}. This is also the first time
that different trends have been found for RRc and RRab.

\begin{deluxetable*}{l ccccc ccccc}
\tablenum{17}
\tablecaption{Coefficients of the $Amp(RV)$-$Amp(V)$ relations.}
\label{tab:amplratio}
\tabletypesize{\footnotesize}
\tablehead{& \multicolumn{5}{c }{RRc} & \multicolumn{5}{c}{RRab} \\
\colhead{Line} & \colhead{a} & \colhead{$e$a} & \colhead{b} & \colhead{$e$b} & \colhead{$\sigma$} & \colhead{a} & \colhead{$e$a} & \colhead{b} & \colhead{$e$b} & \colhead{$\sigma$} \\
 & \multicolumn{2}{c}{km/s} & \multicolumn{2}{c}{km/s/mag} & & \multicolumn{2}{c}{km/s} & \multicolumn{2}{c}{km/s/mag} & }
\startdata
 Fe &    1.68 & 4.92 &  52.63 & 12.94 & 3.50 & 38.09 & 2.90 & 22.35 & 3.12 & 4.20 \\
 H${\alpha}$ &  --9.18 & 9.36 & 106.69 & 24.61 & 6.66 & 77.77 & 6.17 & 22.83 & 6.64 & 8.92 \\
 H${\beta}$  &  --0.27 & 6.92 &  66.55 & 18.20 & 4.93 & 63.32 & 4.96 & 15.96 & 5.34 & 7.17 \\
 H${\gamma}$ &    0.54 & 7.99 &  59.32 & 21.02 & 5.69 & 57.38 & 5.01 & 18.48 & 5.39 & 7.24 \\
 H${\delta}$ &    3.85 & 6.80 &  45.78 & 17.88 & 4.84 & 50.90 & 3.84 & 14.43 & 4.14 & 5.55 \\
\enddata
\tablecomments{Coefficients (a,b), uncertainties ($e$a, $e$b) and standard
deviations ($\sigma$) of the $Amp(RV)$ vs $Amp(V)$ relations for RRc and
RRab variables.}
\end{deluxetable*}

\section{The cumulated radial velocity curves}\label{chapt_cnrvc}

\begin{figure*}[!htbp]
\centering
\includegraphics[width=20cm]{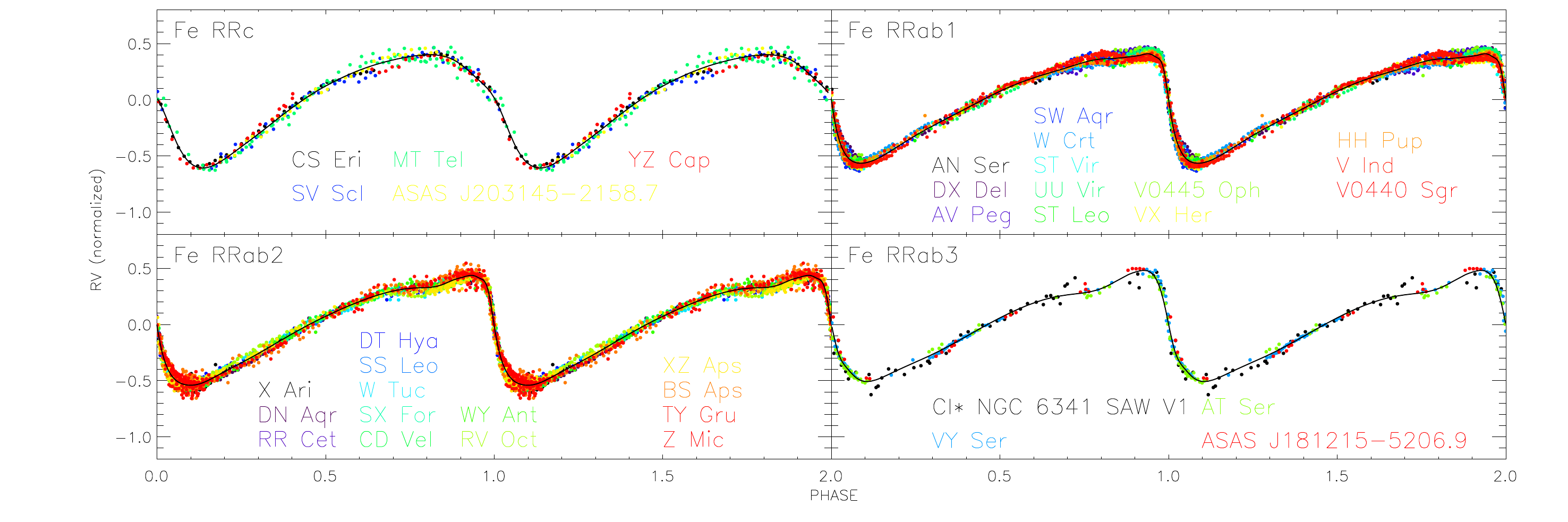}
\includegraphics[width=20cm]{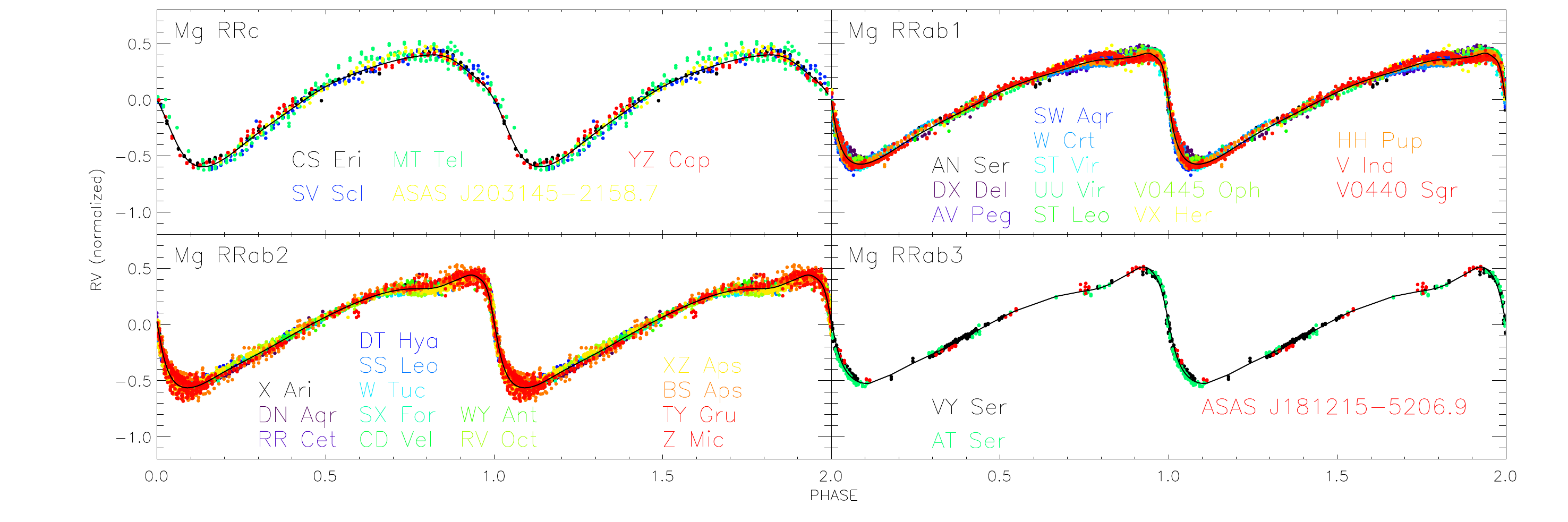}
\includegraphics[width=20cm]{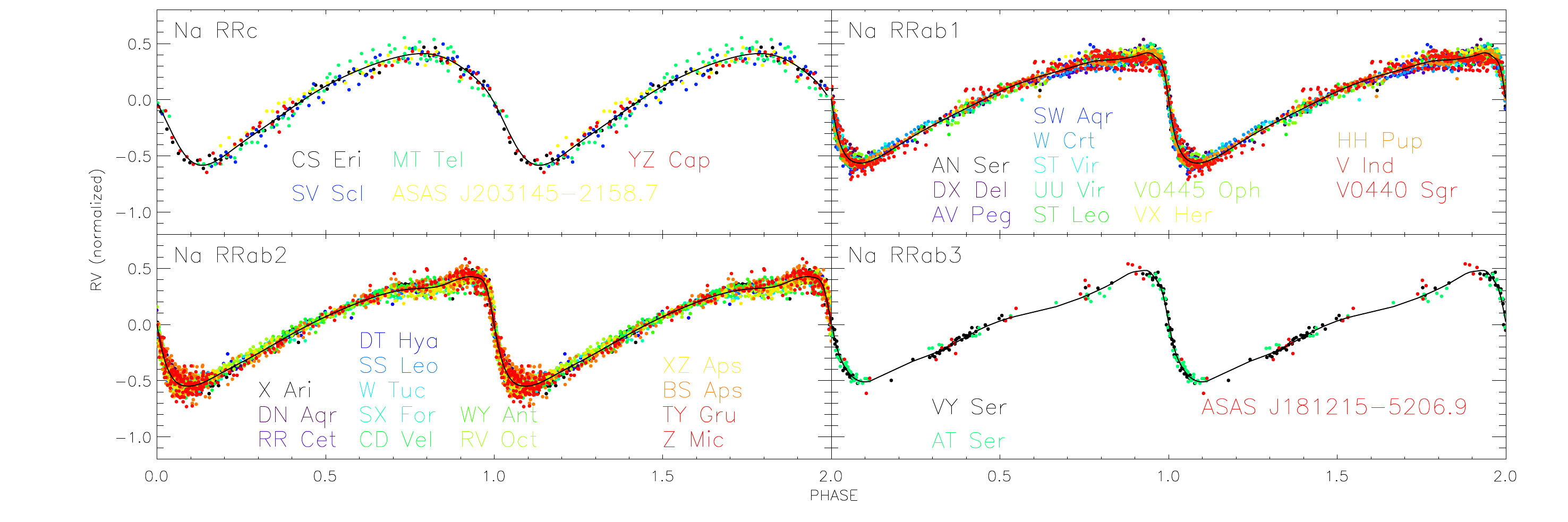}
\caption{From top to bottom: cumulative and normalized radial 
velocity curves based on metallic (Fe, Mg, Na) lines. 
The period bin of the RVC template is labeled on the top left corner. 
Small circles are color-coded by variables and their names are  
labeled at the bottom. The solid line displays the analytical 
form of the RVC templates.}
\label{fig:templates_metal}
\end{figure*}

\begin{figure*}[!htbp]
\centering
\includegraphics[width=18cm]{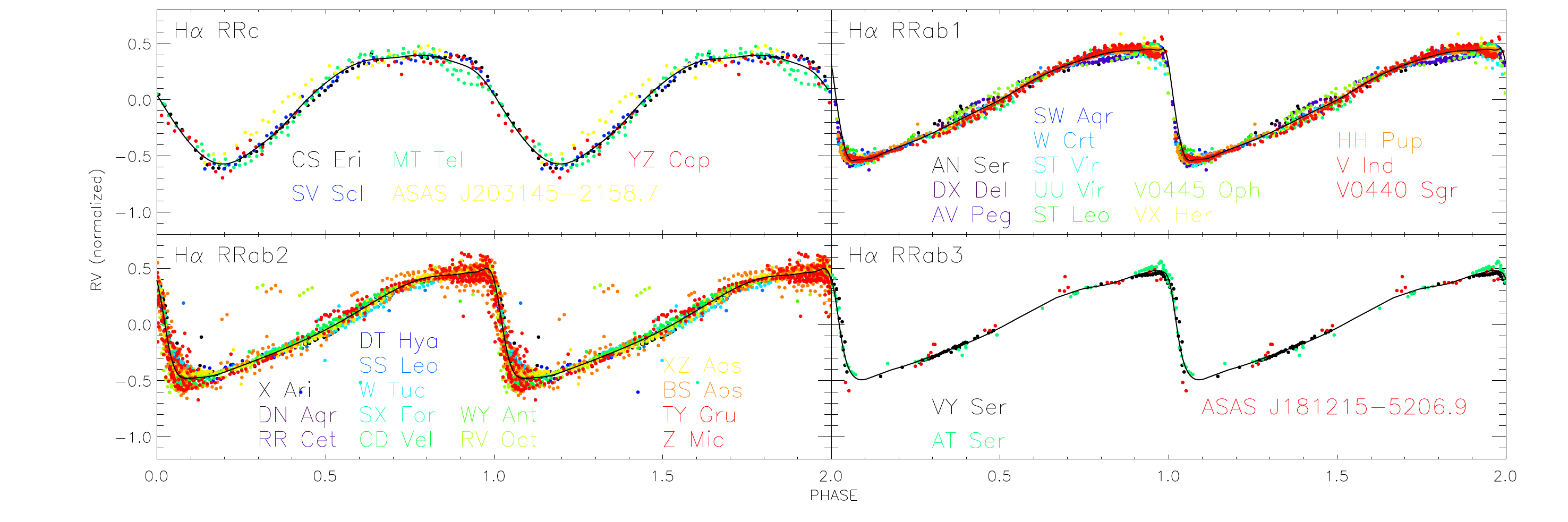}
\includegraphics[width=18cm]{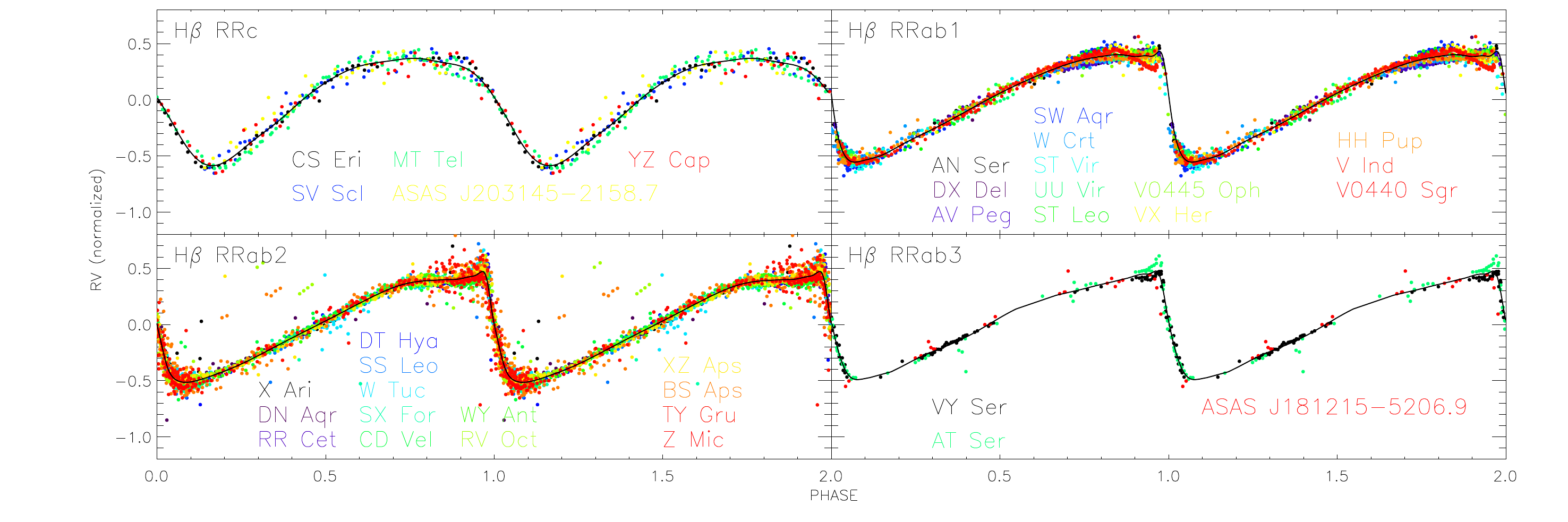}\\
\includegraphics[width=18cm]{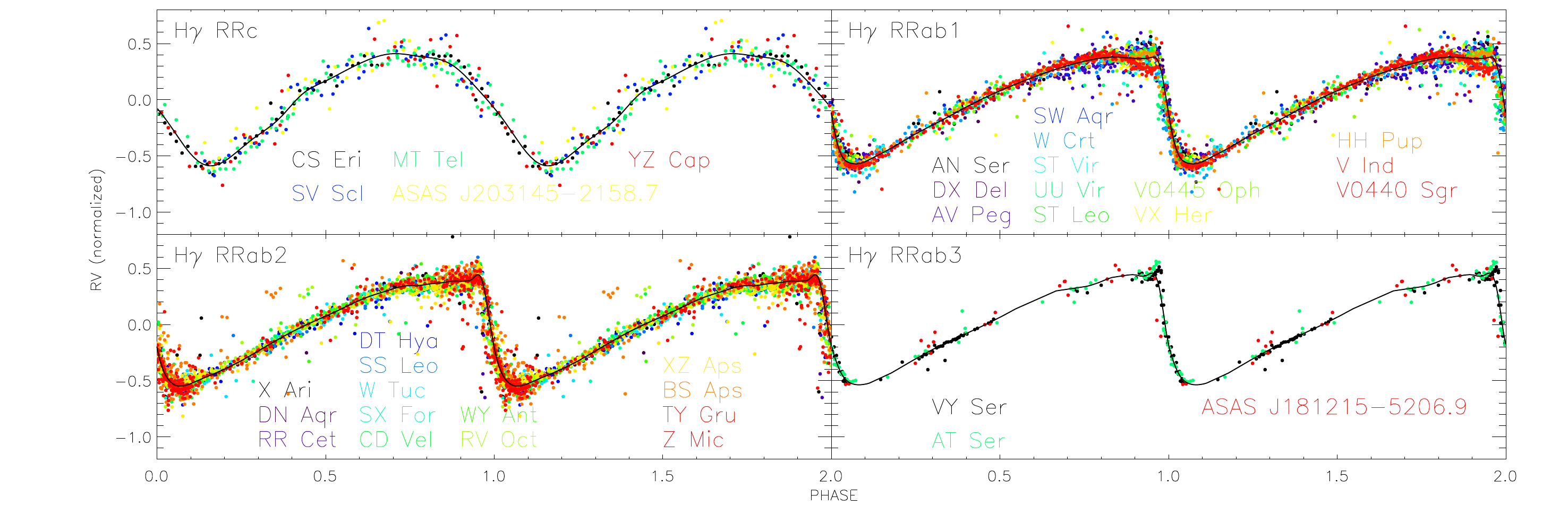}
\includegraphics[width=18cm]{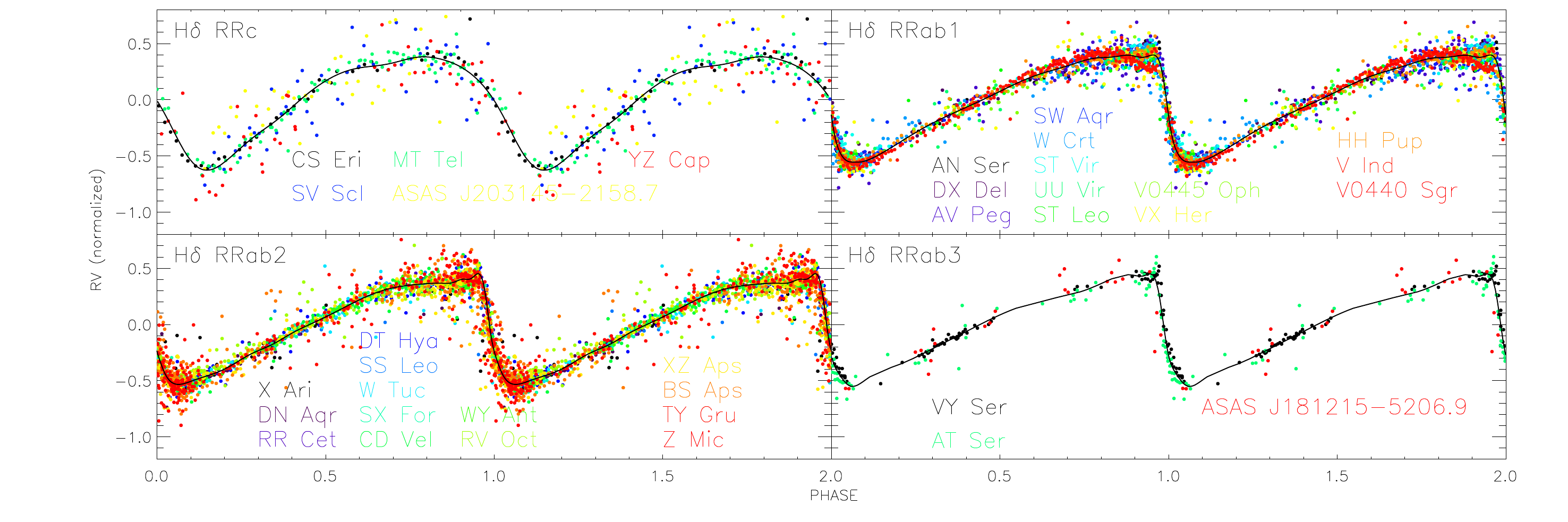}
\caption{From top to bottom: cumulative and normalized radial 
velocity curves based on the Balmer (H$_\alpha$, H$_\beta$, H$_\gamma$, 
H$_\delta$) lines.  The period bin of the RVC template bin is labeled 
on the top left corner. Small circles are color-coded by variables 
and their names are labeled at the bottom. The solid line displays the 
analytical form of the RVC templates.}
\label{fig:templates_balmer}
\end{figure*}

Figures~\ref{fig:templates_metal} and \ref{fig:templates_balmer} display the CNRVCs 
for all the RVC templates (colored small circles), together with the analytical form of the 
template (solid line). 

By inspecting the CNRVCs it is clear that, 
within the bins {\it RRab1} and {\it RRab2}, the morphology of the RVCs is, 
on a first approximation, dichotomic. More specifically, 
within the {\it RRab1} bin, SW Aqr, ST Leo, 
VX Her, V Ind and V0440 Sgr display a local maximum 
around the phase 0.70-0.75, instead of the more or less
steady rising behavior of the other {\it RRab1} RRLs.
Moreover, TY Gru, CD Vel and SX For ({\it RRab2} bin), do not
display a local minimum around phase 0.7-0.8, as the other 
{\it RRab2} variables do. This happens for all the diagnostics, although it is 
more evident for the Balmer lines. We remark that these features are 
also present in the optical light curves of these stars.

We checked whether these features can be associated with either pulsation
or physical properties of the stars. While it is true that the {\it RRab1} variables
with a more prominent local maximum  are located in the High-Amplitude Short-Period
\citep[HASP,][]{fiorentino15a} 
region, they are not the only HASP variables in our sample. Their iron abundances 
range between --1.9 and --1.4, which
is around the peak of the distribution of field RRLs \citep{crestani2021a}
and it was not even possible to constrain a morphological class of RVCs 
based on the Fourier parameters R2, R31, $\phi_{21}$ and $\phi_{31}$ of 
their light curve.

To sum up, there is no quantitative way to predict, either from the light curve or
from the physical properties, which is the RVC morphology of {\it RRab1} and 
{\it RRab2}. This has several consequences: the first and most obvious is that
we cannot split these bins and provide more RVC templates because we cannot provide
criteria for using one or the other. This may seem to be a disadvantage, but luckily, this dichotomy
introduces a $\lesssim$5 km/s 
offset in the estimate of $V_{\gamma}$ with a probability of $\sim$10-15\% 
(i.e., the fraction of pulsation cycle where the RVCs have a different behavior).

\section{How to use the new RVC templates}\label{chapt_howto}

In this section of the Appendices, we provide precise instructions on how
to use the new RVC templates with the aim of estimating V$_\gamma$ 
in different realistic cases: $i)$--- when only one RV measurement and a well-sampled 
light curve that allows the estimate of the reference epoch are available; $ii)$--- when 
a few (we assume three) RV measurement and a well-sampled light 
curve that allows the estimate of the reference epoch are available; $iii)$--- when 
a few (we assume three) RV measurement and a light 
curve that does not allow an accurate estimate of the reference epoch are available.

\subsection{Estimate of \tmeano}\label{chapt_howto_tmeano}

Before describing how to use the templates, we want to instruct the reader on
how to derive \tmeano, which is not as common as \tmaxo, therefore it might not
be straightforward to estimate.

\begin{figure*}[!htbp]
\centering
\includegraphics[width=10cm]{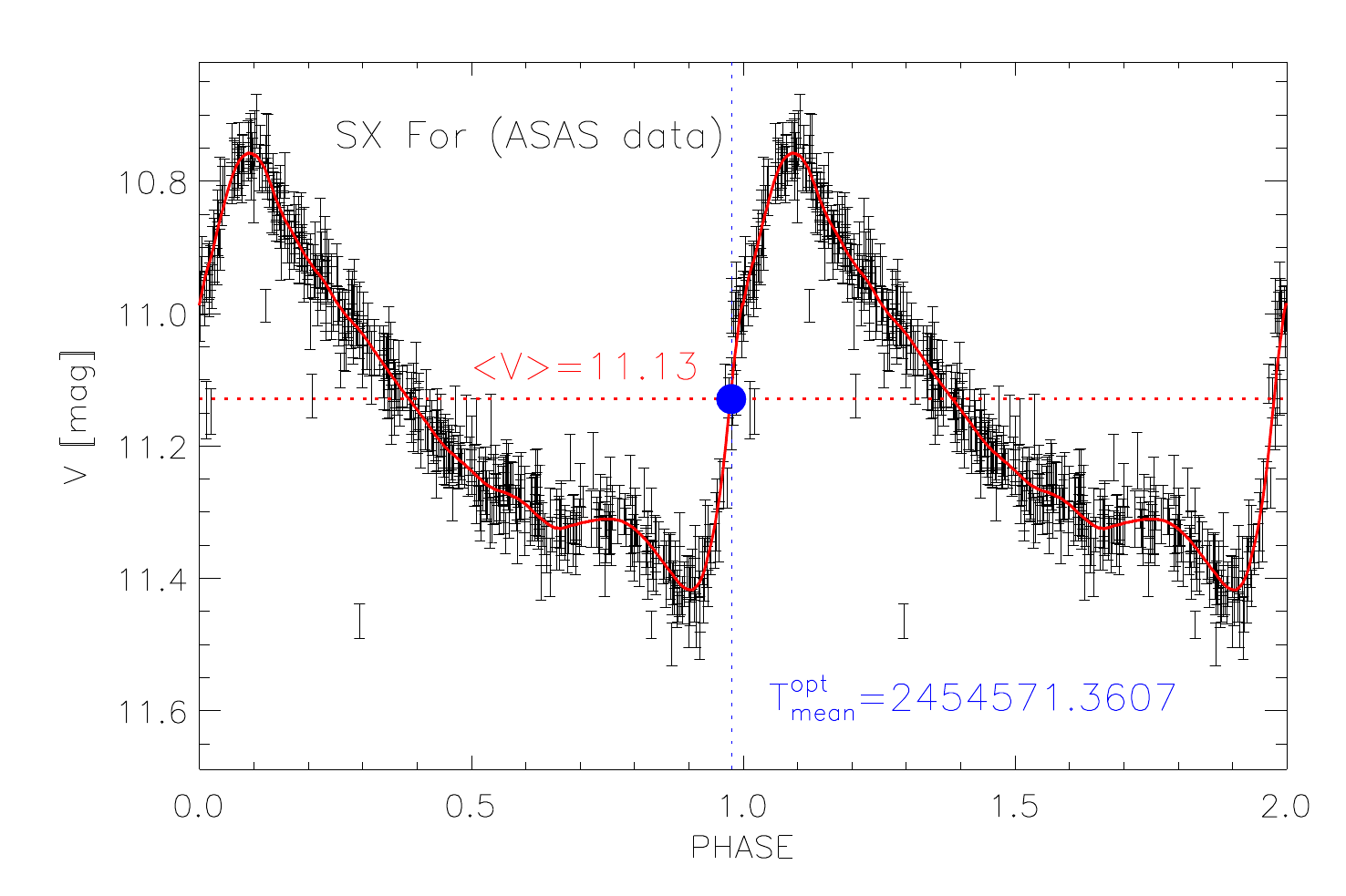}
\caption{$V$-band ASAS light curve of SX For. The solid red line displays the fitting model.
The dashed red line represents the mean magnitude and the blue circle displays the 
position of \tmeano. The values of the mean magnitude and \tmeano are labeled in
red and blue, respectively.}
\label{fig:tmeanopt}
\end{figure*}

First of all, the light curve must be phased to an arbitrary 
reference epoch, e.g., HJD=0, as in Figure~\ref{fig:tmeanopt}. Then, the phased light curve 
must be fit with a model. It is crucial that the model fits well the rising branch.
After that, $<V>$ is derived by converting each point of the model into arbitrary 
flux ($F=10^{-0.4*mag}$), integrating the flux model and finding the mean 
flux $<F>$ and converting back to magnitude ($<V>=-2.5\cdot\log_{10}(<F>) $).

If the model is analytical (e.g., Fourier or PEGASUS), one can easily find 
the phase on the rising branch at which the model intersects $<V>$, namely, 
$\phi_{mean}$. If the model
is not analytical (e.g., spline or PLOESS), it is necessary to interpolate $<V>$ with 
the model sampled on an even grid of phases. A convenient choice for the step of
the grid is between 0.001 and 0.01.

Once $\phi_{mean}$ is known, the next step is to select any phase point of the light curve. This will 
be characterized by an epoch of observation ($t_{V(i)}$, where $i$ indicates the $i$-th
point of the light curve) and a phase ($\phi_{V(i)}$). Finally, 
\tmeano = $t_{V(i)} - (\phi_{V(i)}-\phi_{mean})\cdot P$ can be derived.

\subsection{Single RV measurement}\label{chapt_howto1}

This is the most classical situation when using any type of template. In this case, only one 
RV measurement is available. Note that, if the spectral range of the instrument is
large enough, it is possible to have one RV measurement 
per diagnostic (e.g., Fe, Mg and H$_\gamma$)
but still no more than one epoch is available. This means that any RVC template can be
applied to only one point.

The quoted RV measurement consists of an epoch of observation ($t$),
a velocity ($RV$) and its uncertainty ($eRV$). In this case, a decent sampling
of the light curve is needed, in order to estimate its period ($P$), 
amplitude ($Amp(V)$), mean magnitude ($<V>$)
and the reference epoch \tmaxo~ or \tmeano.

The first step is to anchor the RV measurement to the same reference epoch as the
template. As demonstrated in the body of the paper, it is possible to assume 
\tmeano = \tmeanvfe~ and derive the phase as 
$\phi=\dfrac{t-T_{mean}^{opt}}{P} \bmod{1}$.
If only \tmaxo~ estimates are available, as is often the case of data 
releases of large surveys, it is necessary to derive the phase anchored 
to \tmaxo~ and then apply the offsets provided in Section~\ref{chapt_deltaphi}: 
$\phi_{max}=\dfrac{t-T_{max}^{opt}}{P} \bmod{1}$
and then $\phi = \phi_{max}+0.223$ for RRc variables or  $\phi = \phi_{max}+0.043+0.099\cdot P$ 
for RRab variables. In this case, an uncertainty must be associated with $\phi$, namely,
the $\sigma$ of the quoted relations: 0.036 for RRc and 0.024 for RRab.
By using $\phi$, the RV measurement is now anchored to the same reference
of metallic (Fe, Mg, Na) templates. If, however, the RV measurement is from a Balmer line, 
it is necessary to convert $\phi$ by using the relation provided in Figure~\ref{fig:deltatau_fehb}, 
that is $\phi_{H\beta} = \phi + 0.023 -0.096*P$.

The second step is to rescale of the normalized template. For this end, it 
suffices to convert $Amp(V)$ into $Amp(RV)$ by using the relations provided in 
Table~\ref{tab:amplratio}.

The third step is to derive the analytical form of the template both rescaled by $Amp(RV)$
and shifted in zero-point to pass through the RV measurement. For this step, the right coefficients 
for the  RVC template must be selected from Table~\ref{tab:coeff}. Afterwards, these 
coefficients are substituted into Equation~\ref{eq_pegasus} to calculate the value of the template
(T($\phi$)) at the phase $\phi$. Finally, V$_\gamma$ is obtained as V$_\gamma$=$RV-Amp(RV)\cdot T(\phi)$.

Of course, if RV measurements are available for more than one diagnostic, 
the quoted steps can be separately applied to 
all the diagnostics, providing multiple V$_\gamma$ estimates that can be averaged.

\subsection{Multiple RV measurement with reference epoch}\label{chapt_howto2}

In this case, more than one RV measurement per diagnostic and   
a light curve enough well-sampled are available. The procedure is qualitatively
identical to that described 
in~\ref{chapt_howto1}, but having more than one RV measurement per diagnostic
allows the averaging of the V$_\gamma$ estimates for the same diagnostic.
Also in this case, if RV measurements are available for more than one diagnostic,
these can be averaged to constrain V$_\gamma$ on a broader statistical basis.

In principle, this method could be applied to any number of RV measurements.
However, when these are ten-twelve (or more) and they are more or less evenly sampled, 
with a good knowledge of the period, it is possible to just fit the points and directly 
derive a V$_\gamma$ estimate as accurate as the template itself, or even 
more if the variable has experienced some phase drift or period change during
the time elapsed from the collection of light curve and RV data. 

\subsection{Multiple RV measurement without reference epoch}\label{chapt_howto3}

In this case, more than two RV measurements per diagnostic are available, but 
the light curve is only modestly sampled. This means that period and $Amp(V)$ 
can be estimated from photometric data, but the reference epoch cannot. In this
situation, the templates can be used as fitting curves.

This approach is qualitatively different from the other two because one does not have to anchor 
the template but to fit it to the data. The steps are the following

First of all, phases must be derived by adopting an arbitrary reference epoch
$T_0$ (e.g., $T_0$=0 or $T_0$=2,400,000) $\phi = \dfrac{t-T_0}{P} \bmod{1}$.
Still, if one wants to use the Balmer templates, the conversion
$\phi_{H\beta} = \phi + 0.023 -0.096*P$ must be applied.

Secondly, $Amp(V)$ has to be rescaled into $Amp(RV)$ by using the relations provided in 
Table~\ref{tab:amplratio}. This step is analogous to that described in~\ref{chapt_howto1}.

The third step, is the selection of the RVC template coefficients from
Table~\ref{tab:coeff} and perform a $\chi^2$ minimization when fitting
the RV data with Equation~\ref{eq_pegasusfit}. The minimization must 
be performed on the two free paramaters $\Delta\phi$ 
(a horizontal shift) and $\Delta V_{\gamma}$, while all the others 
remain fixed.

Finally, V$_\gamma$ is simply derived by integrating the final 
analytical form of Equation~\ref{eq_pegasusfit}. Also in this case,
if RV measurements are available for more than one diagnostic,
these can be averaged to constrain V$_\gamma$ on a broader
statistical basis.


\section{Phase-gridded templates}\label{chapt_gridded}

The anonymous referee suggested that reader might be interested in using analytical 
functions different from PEGASUS series to apply to RVC templates. To facilitate 
the independent readers we list, in Table~\ref{tab:phasegridded}, the phase-gridded values of the 
RVC templates by using the coefficients of the PEGASUS series given in Table~\ref{tab:coeff}.  
the reader can fit these gridded values with any analytical function and use them 
one to apply the RVC template. We adopted a step of 0.01, This means 101 points 
per template. We have checked that, for a Fourier series of tenth-to-fifteenth 
order, this step in phase provides a very good fit of the gridded points. In passing, 
we stress that these phase-gridded values should not be used in substitution of the 
analytical templates, but to derive alternative analytical forms of the RVC templates. 
This is recommended for the single phase point approach and necessary for the multiple 
phase point approach.

\begin{deluxetable*}{l l c c}
\tablenum{18}
\setlength{\tabcolsep}{4pt}
\tabletypesize{\scriptsize}
\tablecaption{Phase-gridded RVC templates.}
\label{tab:phasegridded}
\tablehead{ \colhead{Species} & \colhead{Template bin} & \colhead{Phase} & \colhead{Template value\tablenotemark{a}}}
\startdata
Fe & 0 & 0.000 &   0.0176 \\
Fe & 0 & 0.001 &   0.0124 \\
Fe & 0 & 0.002 &   0.0072 \\
Fe & 0 & 0.003 &   0.0019 \\
Fe & 0 & 0.004 & --0.0036 \\
Fe & 0 & 0.005 & --0.0092 \\
Fe & 0 & 0.006 & --0.0148 \\
Fe & 0 & 0.007 & --0.0206 \\
Fe & 0 & 0.008 & --0.0265 \\
Fe & 0 & 0.009 & --0.0325 \\ 
\enddata
\tablecomments{Only ten lines are listed. The machine-readable version of this 
table is available online on the CDS.\\
\tablenotetext{a}{Calculated at the given phase by using Equation~\ref{eq_pegasus} and the coefficients 
provided in Table~\ref{tab:coeff}}}
\end{deluxetable*}

\end{appendix}

\end{document}